\documentclass[11pt]{article}
\usepackage{epsfig,amsmath,latexsym,amssymb}
\usepackage{amscd,pdflscape,amsthm}
\usepackage{multirow, array}
\usepackage{fancyvrb, bm, hyperref}
\usepackage{graphicx, subcaption, caption, listings, url}
\usepackage[tableposition=top, labelfont=footnotesize]{caption}
\usepackage{enumitem, framed, tcolorbox,leftidx}
\usepackage[section]{placeins}
\usepackage[top=2.8cm,bottom=2.3cm,textwidth=16cm]{geometry}
\usepackage{amsmath}

\def\R{{\mathbb R}}
\def\N{{\mathbb N}}

\def\x{{\boldsymbol x}}

\def\p{{\mathbb P}}

\def\G{{\mathcal G}}
\def\A{{\mathcal A}}
\def\T{{\mathcal T}}
\def\K{{\mathcal K}}
\def\X{{\mathcal X}}

\def\I{{\mathcal I}}

\newcommand{\Remm}[1]{}
\newtheorem{theo}{Theorem}[section]
\newtheorem{lemma}[theo]{Lemma}

\newtheorem{model ass}[theo]{Model Assumptions}

\theoremstyle{definition}

\newtheorem{remark}[theo]{Remark}

\numberwithin{equation}{section}

\makeatletter
\newcommand{\eqnum}{\refstepcounter{equation}\textup{\tagform@{\theequation}}}
\makeatother

  {\begin{tcolorbox}[colback=gray!15,colframe=gray!15, fonttitle=\bfseries,
 				 left=1mm, right=1mm, top=.01mm, bottom=.1mm, middle=1mm]
	\begin{lemma}}%
  {\end{lemma}\end{tcolorbox}}

  {\begin{tcolorbox}[colback=gray!15,colframe=gray!15, fonttitle=\bfseries,
 				 left=1mm, right=1mm, top=.01mm, bottom=.1mm, middle=1mm]
	\begin{theo}}%
  {\end{theo}\end{tcolorbox}}

\begin{document}

\author{
Philippe Deprez\footnote{RiskLab, Department of Mathematics,
ETH Zurich, Switzerland} \qquad
Pavel V.~Shevchenko\footnote{Department of Applied Finance and Actuarial Studies,
Macquarie University, Australia} \qquad
Mario V.~W\"uthrich$^\ast$\footnote{Swiss Finance Institute SFI Professor}}

\date{February 20, 2017}
\title{Machine Learning Techniques for Mortality Modeling}

\maketitle

\begin{abstract}
Various stochastic models have been proposed to estimate mortality rates.
In this paper we illustrate how machine learning techniques
allow us to analyze the quality of such mortality models.
In addition, we present how these techniques can be used
for differentiating the different causes of death in mortality modeling.
\end{abstract}



\section{Introduction}

Mortality modeling is crucial
in economy, demography and in life and social insurance, because
mortality rates determine insurance liabilities, prices of insurance products,
and social benefit schemes.
As such, many different stochastic mortality models
to estimate and forecast these rates
have been proposed,
starting from the Lee-Carter~\cite{LC} model.
For a broad overview and comparison
of existing models we refer to~\cite{Cairns}.
In this article we revisit two of these models
and study their calibration to Swiss mortality data.
We illustrate how machine learning techniques allow us to study the adequacy
of the estimated mortality rates.
By applying a regression tree boosting machine we analyze how the modeling
should be improved based on feature components of an individual, such as
its age or its birth cohort.
This (non-parametric) regression approach then
allows us to detect the weaknesses of different mortality models.

In a second part of this work we investigate cause-of-death mortality.
Given a death of an individual with a specific feature, we study the conditional probability
of its cause.
Based on Swiss mortality data we illustrate how regression tree algorithms can be applied to
estimate these conditional probabilities in a Poisson model framework.
The presented technique provides a simple way to detect patterns in these
probabilities over time.
For a parametric modeling approach and existing literature on
cause-of-death mortality we refer to~\cite{Arnold2015, Pavel} and references therein.

~

{\bf Organization of the paper.}
In the next section we introduce the notation and state the model assumptions.
In Section~\ref{Section: Boosting mortality rates} we analyze two standard models,
the Lee-Carter~\cite{LC} model and the Renshaw-Haberman~\cite{cohort} model,
and we explain how machine learning techniques can be applied
to investigate the adequacy of these models for a given dataset.
Note that we do not consider the most sophisticated models here,
but our goal is to study well-known standard models and to indicate how
their strengths and weaknesses can be detected with machine learning.
In Section~\ref{Section: cause} we refine these models to the analysis of cause-of-death mortality.

\section{Model assumptions}\label{Section: model}

In mortality modeling each individual person is identified
by its gender, its age, and the calendar year (also called period) in which the person is considered.
As such, each individual person is assigned
to a feature $\x=(g,a,t) \in \X= \G\times\A\times\T$,
with feature components
\begin{equation}\label{Equation: feature components}
\G=\{\text{female, male}\},\quad
\A=\{0,\ldots,\omega\}, \quad
\text{ and } \quad \T=\{t_\text{min},\ldots,t_\text{max}\}.
\end{equation}
Here, $a\in\A$ represents the age in years of the person,
$\omega\in\N$ denotes the maximal possible age the person can reach.
The component $\T \subset \N_0$ describes the calendar years considered.
This feature space $\X= \G\times\A\times\T$ could be extended by further feature components
such as the income or marital status of a person, however,
for our study we do not have additional individual information.

For a given feature $\x=(g,a,t) \in \X$ we denote the (deterministic) exposure by
$E_{\x}\ge1$ and the corresponding (random) number of deaths by $D_{\x}\ge0$.
That is, for a given feature $\x=(g,a,t)\in\X$ we have $E_\x$ people
with feature $\x$ alive at the beginning of the period
$(t,t+1]$, and during this period $D_\x$ of these people die.
Two assumptions commonly made in the literature are that
$D_\x$ are independent, for different $\x\in\X$,
and each $D_\x$ has a Poisson distribution with a parameter proportional to $E_\x$.
This also assumes that the force of mortality stays constant over each period $(t,t+1]$.
In this spirit, we consider the following model assumptions.

~

\begin{model ass}\label{Model}
Assume that the mortality rates are given by the regression function $q: \X \to [0,1]$, $\x \mapsto q(\x)$.
The numbers of deaths satisfy the following properties:
\begin{itemize}

\item
$(D_\x)_{\x\in\X}$ are independent in $\x\in\X$.

\item
$D_\x ~\sim~ \text{Pois}(q(\x) E_\x  )$ for all $\x\in\X$;

\end{itemize}
\end{model ass}

~

The main difficulty is to appropriately estimate the
mortality rates $q: \X \to [0,1]$ from historical data of a given population.
In particular, we would like to infer $q(\cdot)$ from observations.
Several models have been developed in the literature  to address this problem.
In the next section we consider two standard models
and we explain how machine learning techniques can be applied
to back-test these two models.

\section{Boosting mortality rates}\label{Section: Boosting mortality rates}

We investigate two different classical models for estimating mortality
rates: the Lee-Carter~\cite{LC} model and the Renshaw-Haberman~\cite{cohort} model.
By applying machine learning techniques we analyze the weaknesses of these models.
Note that this illustration has mainly pedagogical value.
First, the Lee-Carter model is a comparably simple model
that has been improved  in many directions in
various research studies.
Second, one of these improvements is the Renshaw-Haberman model
that we will compare to the Lee-Carter model.

The following analysis is based on historical Swiss mortality data
provided by the Human Mortality Database, see \url{www.mortality.org}.
The data we consider includes the exposures
$(E_\x)_{\x\in\X}$ and the numbers of deaths $(D_\x)_{\x\in\X}$ for
the feature space $\X= \G\times\A\times\T$
with feature components
\begin{equation*}
\G=\{\text{female, male}\}, \quad
\A=\{0,1,\ldots,97\}, \quad \text{and} \quad
\T=\{1876,1877,\ldots,2014\}.
\end{equation*}
Here, maximal age $a=97$ corresponds to ages of at least $97$, and the set
$\T$ consists of $139$ years of observations.
This results in $2\cdot98\cdot139=\,$27,244 data points
corresponding to  7,867,978 deaths within those $139$ years of observations.

\subsection{Back-testing the Lee-Carter model}\label{Section: LC}

\begin{figure}
\centering
\includegraphics[width=.49\linewidth]{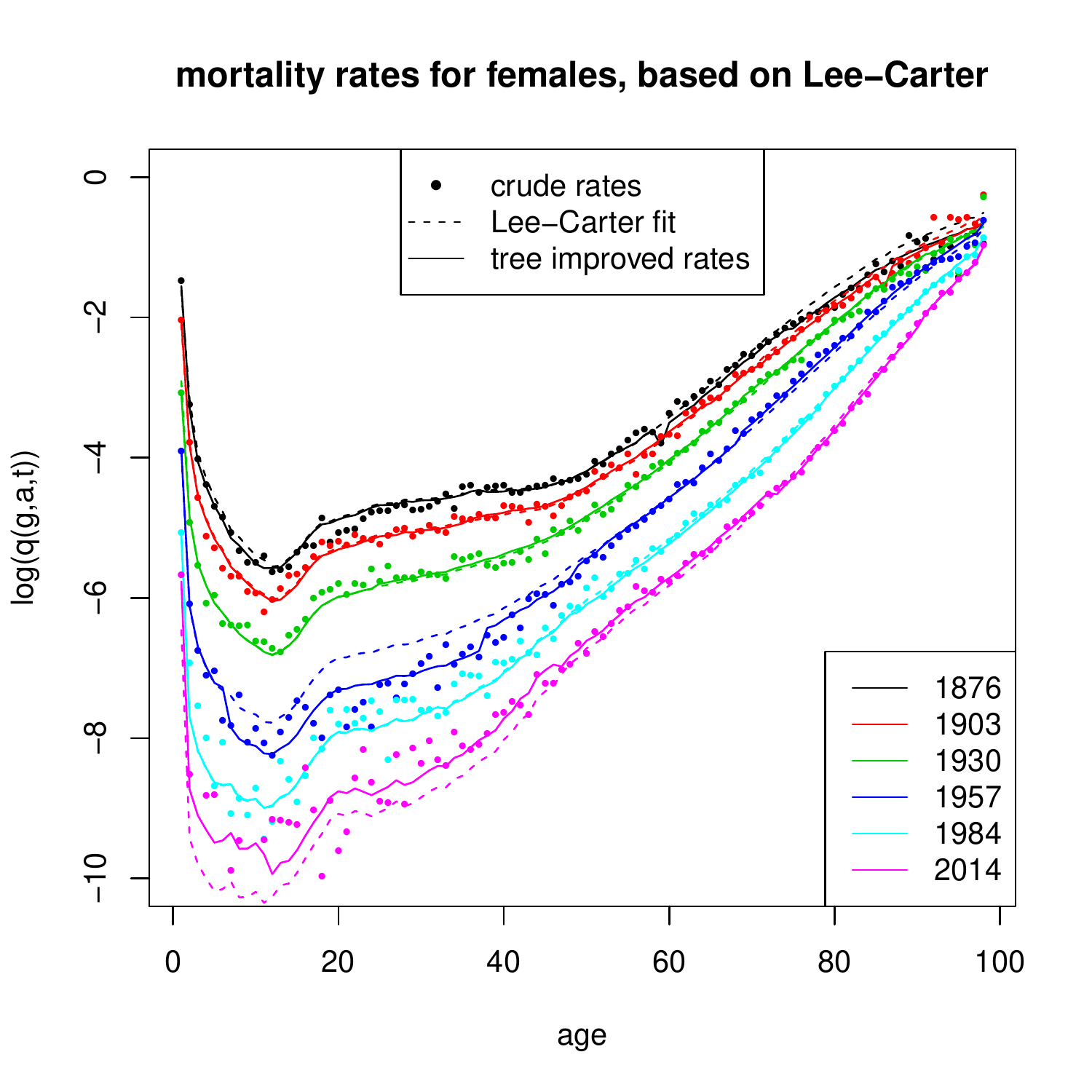}
\hfill
\includegraphics[width=.49\linewidth]{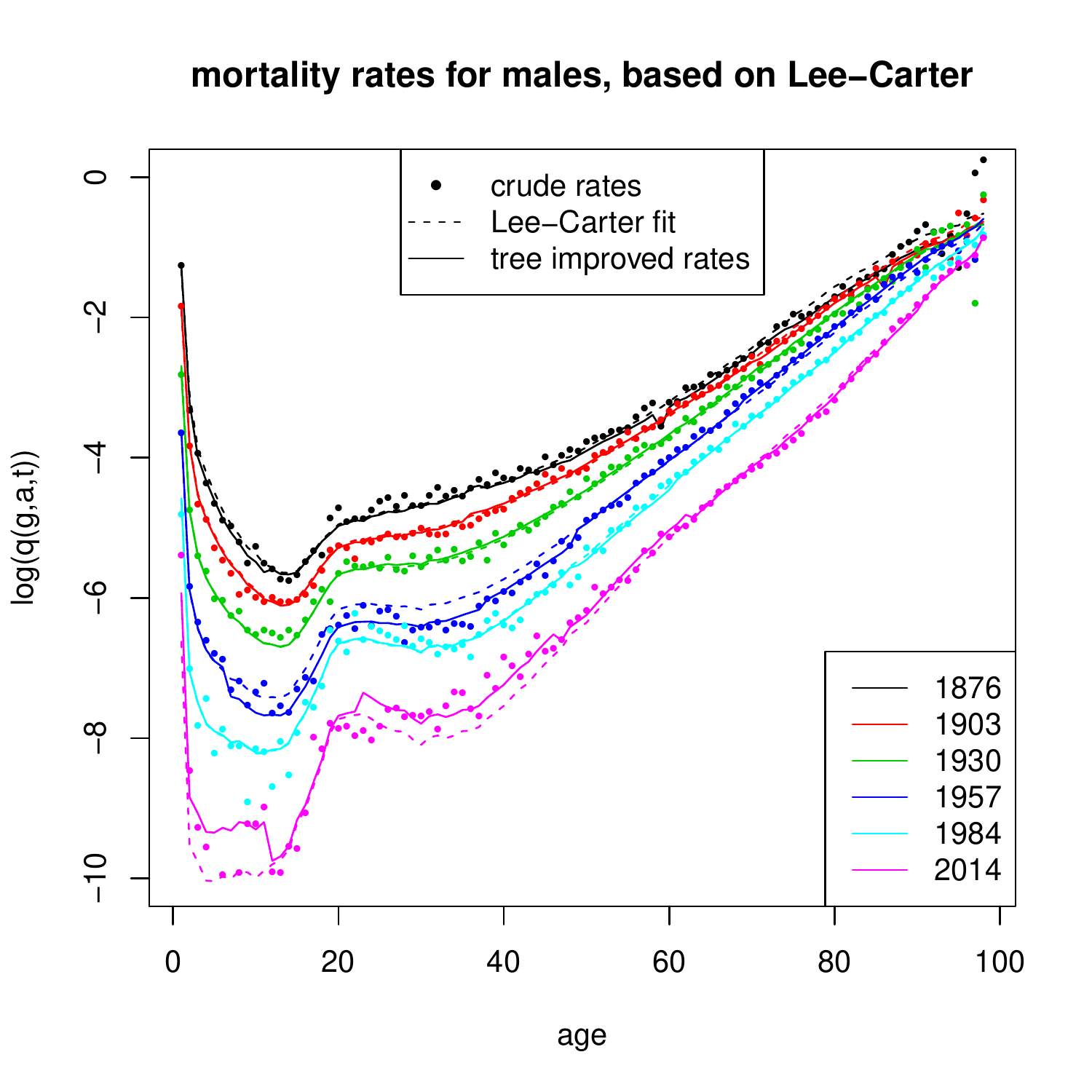}
\caption{\footnotesize
Logarithms of mortality rates for (lhs) females and (rhs) males, for different ages and calendar years.
The dots illustrate the crude mortality rates of the Swiss mortality data,
the dashed lines the Lee-Carter fit,
and the solid lines illustrate the tree improved estimates given by~\eqref{Equation: q.LC}.
}
\label{Figure: mortality rates LC}
\end{figure}

We first consider the classical Lee-Carter~\cite{LC} model
and we apply regression tree boosting to detect its weaknesses.
The Lee-Carter model is a comparably simple model
in which
the mortality rates are assumed to be of the form
\begin{equation*}
q^\text{LC}(\x)
~=~
\exp\left\{ \beta^{0,g}_a + \beta^{1,g}_a \kappa^g_t \right\},
\qquad \text{for $\x=(g,a,t)\in\X$,}
\end{equation*}
with the identifiability constraints $\sum_a \beta^{1,g}_a=1$ and $\sum_t \kappa^g_t=0$ for each $g\in\G$.
For fixed gender $g\in\G$ we fit the parameters $\beta^{0,g}_a$, $\beta^{1,g}_a$, and $\kappa^g_t$
to the Swiss mortality data using the {\tt `StMoMo'} {\sf R}-package, see~\cite{stmomo},
as follows
\begin{flushleft}
 \tt > \quad
LC <- fit(lc(link="log"),\, Dxt = deaths,\, Ext = exposures)  \\
 \tt > \quad
q.LC <- fitted(LC, type = "rates") \hfill \eqnum\label{Equation: LC code}
\end{flushleft}
%
The {\tt exposures} $(E_{\x})_{\x\in\X}$ and the numbers of {\tt deaths} $(D_{\x})_{\x\in\X}$
are the input data, and we apply the above command  to each gender $g\in\G$ separately.
{\tt q.LC} then provides the Lee-Carter
mortality rates $(q^\text{LC}(\x))_{\x\in\X}$ fitted to the Swiss data.
These rates are presented in Figure~\ref{Figure: mortality rates LC}
by dashed lines, and compared to the
crude (observed) mortality rates $D_\x/E_\x$ illustrated by dots in Figure~\ref{Figure: mortality rates LC}.
We aim at back-testing these fitted mortality rates by using machine learning techniques.
For this, we initialize Model Assumptions~\ref{Model}
with the rates $q(\x)=q^\text{LC}(\x)$ obtained from the Lee-Carter fit~\eqref{Equation: LC code},
i.e., we consider for $\x\in\X$,
\begin{equation}\label{LC model}
D_\x ~\sim~ \text{Pois}(\mu(\x)d_\x  ),
\qquad \text{with $\mu(\x)\equiv1$ and $d_\x =q^\text{LC}(\x) E_\x$}.
\end{equation}
Observe that $d_\x$ describes the expected number of deaths according to the
Lee-Carter fit.
To back-test this model we analyze whether the constant factor
$\mu(\x)\equiv1$ is an appropriate choice for the Swiss data considered.
For instance, for given feature $\x\in\X$ this factor $\mu(\x)$
should be increased if the Lee-Carter
mortality rate $q^\text{LC}(\x)$ underestimates
the crude rate $D_\x/E_\x$.
As such, our aim is to calibrate the factor $\mu(\x)$ in~\eqref{LC model}
based on the chosen features $\x\in\X$.

To do so,  we apply
one step of the {\it Poisson Regression Tree Boosting Machine}
to the working data $( D_\x,\, \x,\,  d_\x )_{\x \in \X}$,
see~\cite{Breiman} and  Section~6.4 in~\cite{data}.
This standardized binary split (SBS) tree growing algorithm
selects at each iteration step a feature component
(gender, age or calendar year) and splits the feature space $\X$
in a rectangular way with respect to this chosen feature component.
The explicit choice of each split is based on an optimal improvement of
a given loss function that results from that split.
The algorithm then provides an SBS
Poisson regression tree estimator $\hat \mu(\x)$, $\x\in\X$,
that is calibrated on each rectangular subset of $\X$ obtained from these splits.
That is, we obtain $\hat \mu(\cdot)$ as a calibration of $\mu(\cdot)$
using this non-parametric
regression approach; we refer to~\cite{data} for further details
on this regression tree boosting.

Observe that the SBS tree growing algorithm calibrates $\mu(\cdot)$  by
generating {\it only}
rectangular splits of the feature space $\X$.
However, we would like to calibrate the factor $\mu(\cdot)$ also with respect to birth cohorts,
which requires diagonal splits of the feature space.
For this reason we extend the feature space $\X$ to the feature space
$\bar\X = \{(g,a,t,c=t-a) \mid g\in\G,\, a\in\A, \, t\in\T  \}$,
where $c=t-a$ provides the cohort of feature $\x=(g,a,t)\in\X$.
Observe that there is a one-to-one correspondence between $\X$ and $\bar\X$,
but this extension is necessary to allow for sufficient
degrees of freedom in SBSs.
From this we see that a smart model design includes a thoughtful choice of the feature
space, allowing for the desired interactions and dependencies.

\begin{figure}[t]
\centering
\includegraphics[width=.49\linewidth]{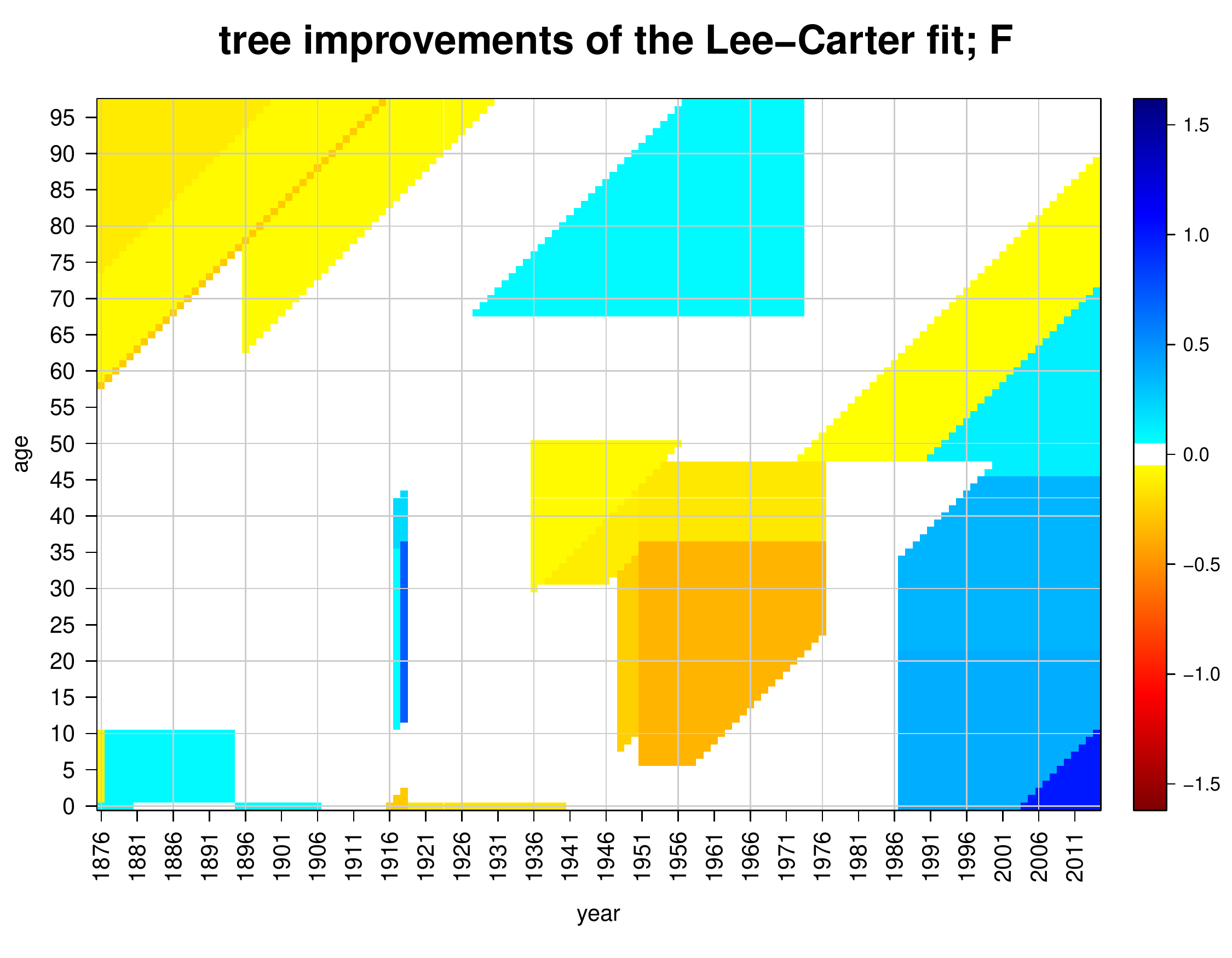}
\hfill
\includegraphics[width=.49\linewidth]{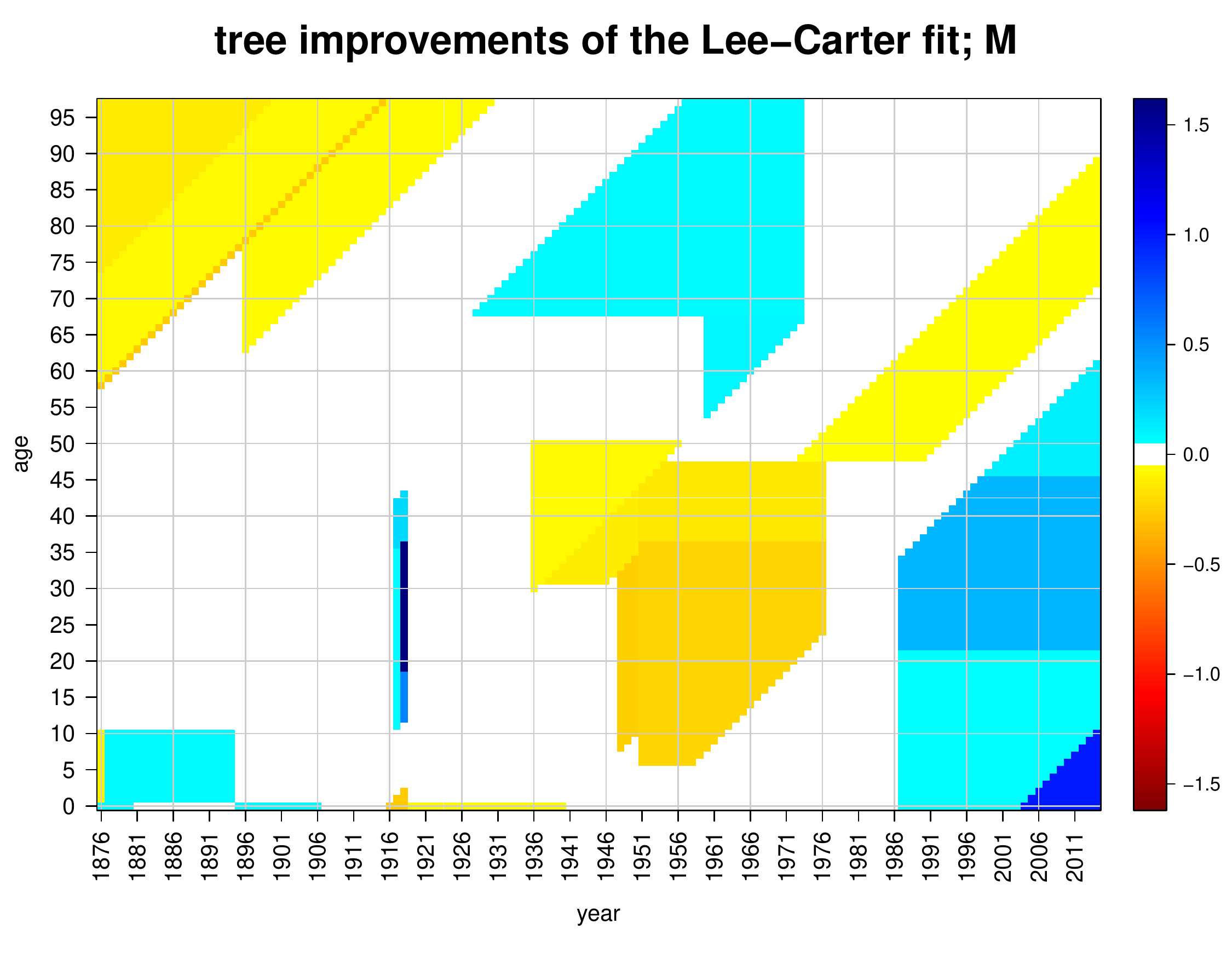}
\\
\includegraphics[width=.49\linewidth]{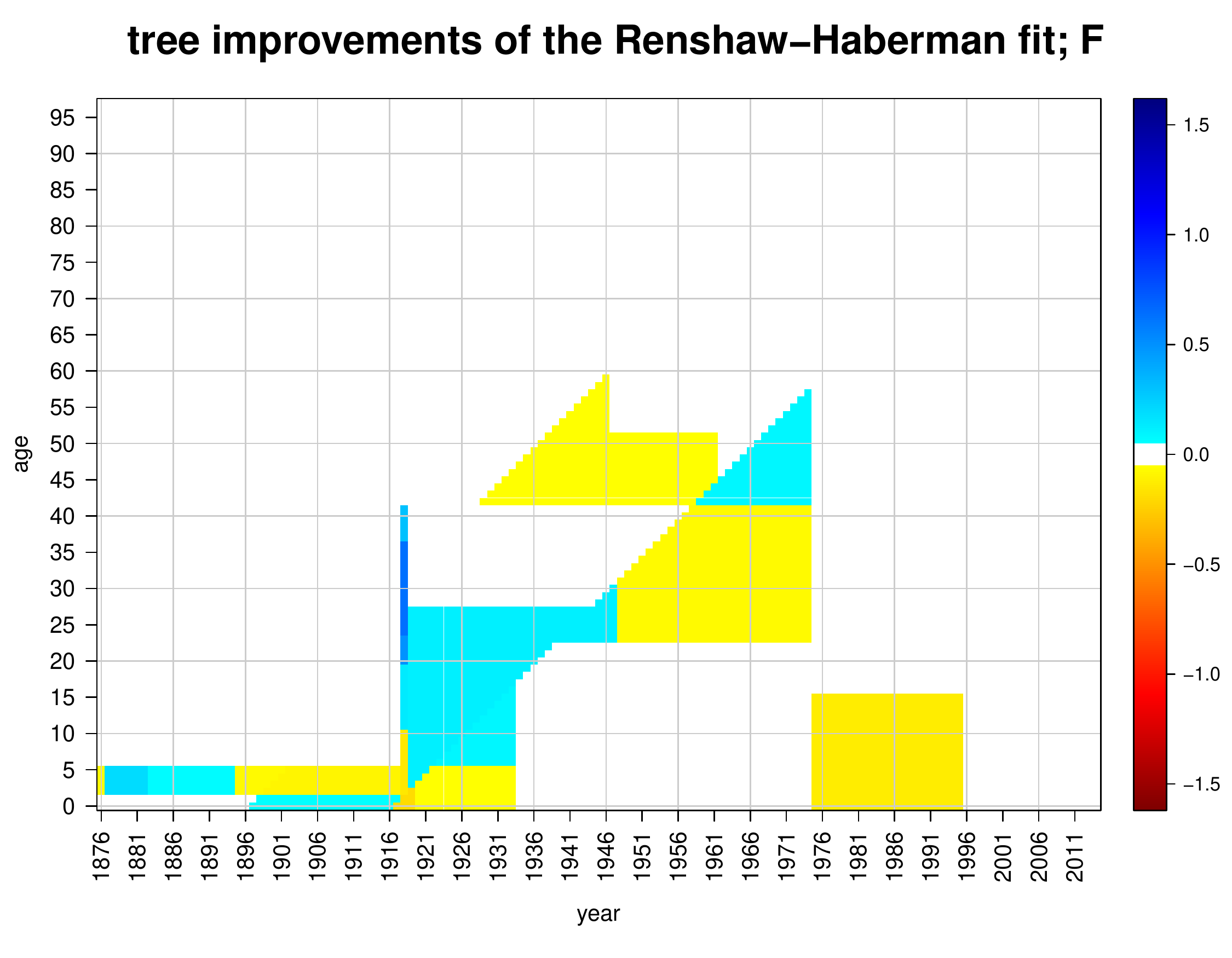}
\hfill
\includegraphics[width=.49\linewidth]{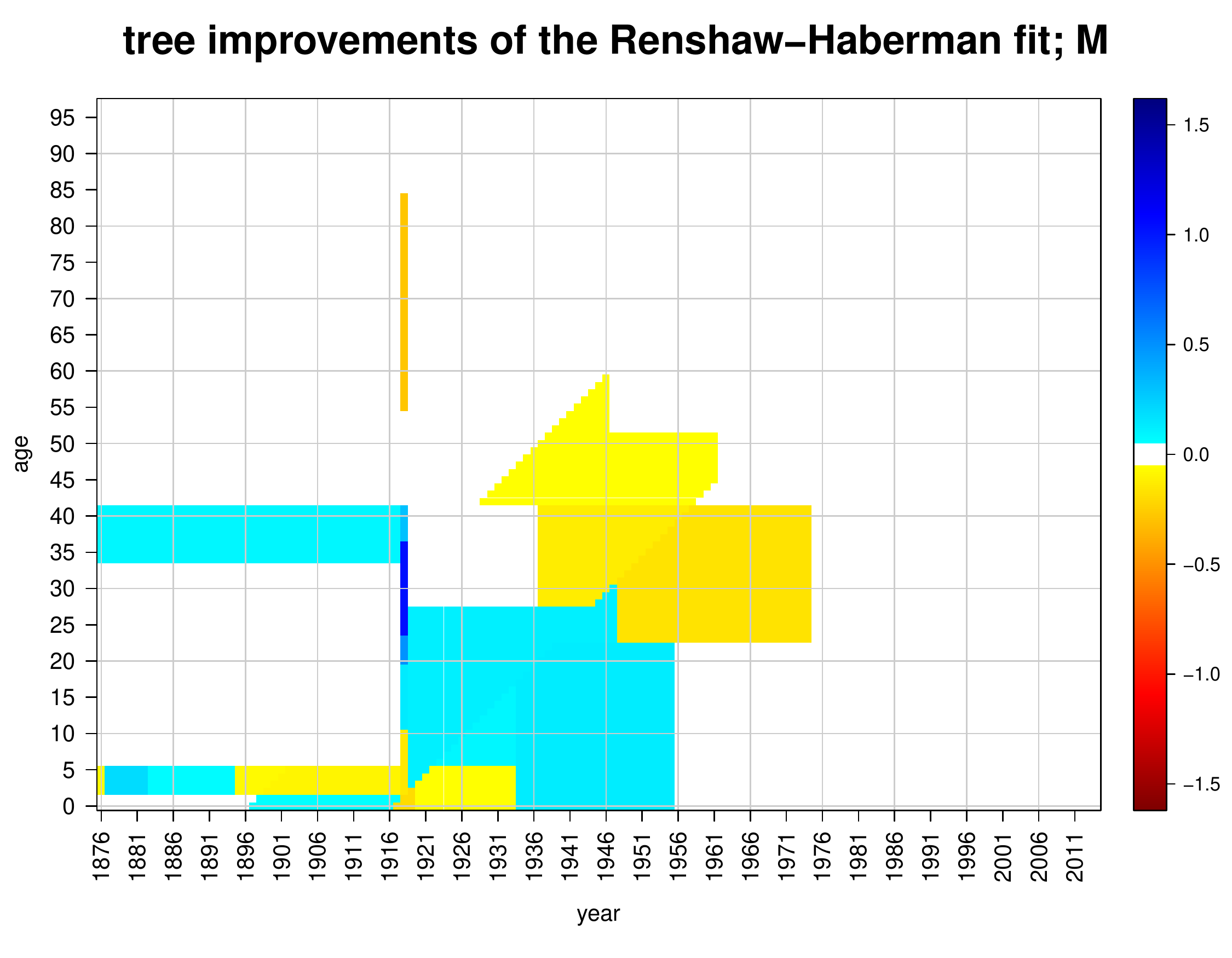}
\caption{\footnotesize
Changes in mortality rates $\Delta q^\text{LC}(\x)$ (first row) and
$\Delta q^\text{RH}(\x)$ (second row),
respectively, obtained by the boosting step of the tree growing algorithm.
The first column shows the changes for females, the second one for males.
Small changes in the range $[-5\%,5\%]$ are illustrated in white color.
}
\label{Figure: correction}
\end{figure}

We apply the SBS tree growing algorithm to calibrate $\mu(\cdot)$ in~\eqref{LC model}
with respect to the (extended) feature space $\bar\X$.
This is obtained in {\sf R} using the {\tt `rpart'} package, see~\cite{rpart},
and the input
\begin{flushleft}
 \tt > \quad tree <- rpart(cbind(volume,deaths)  $\sim$ gender + age + year + cohort, \\
\hspace{8cm}						 data = data, method = "poisson",\\
\hspace{8cm}						cp = 2e-3)
						
 \tt > \quad mu <- predict(tree) \hfill \eqnum\label{Equation: tree code}
\end{flushleft}
Here, the {\tt data} contains for each feature $\x=(g,a,t)\in\X$
the {\tt volume} $d_\x$ and the number of {\tt deaths} $D_\x$, and we optimize subject to
{\tt gender} $g$, {\tt age} $a$, {\tt year} $t$,
and {\tt cohort} $c=t-a$.
The cost-complexity parameter  {\tt cp} allows us to control the number of iteration steps
performed by the regression tree algorithm,
for more details on this control parameter we refer to Section~5.2 of~\cite{data}.
The regression tree estimator $\hat\mu(\cdot)$ is then given by  {\tt mu}.
This estimator allows us to define the regression tree improved mortality rates
\begin{equation}\label{Equation: q.LC}
q^\text{tree}(\x)
~=~
\hat\mu(\x) q^\text{LC}(\x),
\qquad \text{ for $\x\in\X$.}
\end{equation}
These mortality rates are illustrated in Figure~\ref{Figure: mortality rates LC} by lines
and compared to the Lee-Carter fit $q^\text{LC}(\x)$ presented by dashed lines.
%
%
%
In order to analyze the improvements in the initialized mortality rates $q^\text{LC}(\x)$
obtained by the tree growing algorithm, we consider the relative changes
\begin{equation*}
\Delta q^\text{LC}(\x)
~=~
\frac{q^\text{tree}(\x)-q^\text{LC}(\x)}{q^\text{LC}(\x)}
~=~
\hat\mu( \x) -1,
\qquad \text{ for $\x\in\X$.}
\end{equation*}
These relative changes are presented in the first row of Figure~\ref{Figure: correction}.
We observe substantial changes in the calibrated mortality rates
for many parts of the feature space $\X$.
For instance, we see that the algorithm remarkably increases the Lee-Carter mortality
rates for the calendar year $t=1918$, when the Spanish flu had its spike.
This is not surprising, since the Lee-Carter model does not appropriately capture
the impacts of such special events, see also the discussions in~\cite{LC, RH2003}.
A main observation is that the algorithm performs many splits with respect to birth cohorts
(diagonals in Figure~\ref{Figure: correction}),
many of them are performed in the very first iteration steps.
This means that these splits
particularly reduce the value of the loss function considered by the algorithm
and lead to a substantial improvement in the fit.
As an example, the algorithm improves the initialized mortality
rates $q^{LC}(\x)$ for features $\x$ with birth cohort $c=t-a=1816$,
which is called ``the year without a summer'';
we refer to~\cite{1816} for historical information.
Again, this is not surprising, since the Lee-Carter framework
does not account for such cohort effects, see also~\cite{RH2003}.
We conclude that these cohort effects motivate us to perform
a similar analysis on the Renshaw-Haberman model which aims at
addressing this drawback of the Lee-Carter model.

\subsection{Back-testing the Renshaw-Haberman model}\label{Section: RH}

In this section we consider the Renshaw-Haberman~\cite{cohort} model,
which is a cohort-based extension to the classical Lee-Carter model
discussed in the previous section.
In the Renshaw-Haberman model
the mortality rates are assumed to be of the form
\begin{equation*}
q^\text{RH}(\x)
~=~
\exp\left\{ \beta^{0,g}_a + \beta^{1,g}_a \kappa^g_t+ \beta^{2,g}_a\gamma^g_{t-a}  \right\},
\qquad \text{for $\x=(g,a,t)\in\X$,}
\end{equation*}
with the identifiability constraints
\begin{equation*}
\sum_a\beta^{1,g}_a = \sum_a\beta^{2,g}_a=1
\qquad \text{and} \qquad
\sum_{a,t}\gamma^g_{t-a}=\sum_t \kappa^g_t=0,
\qquad \text{for each $g\in\G$. }
\end{equation*}
Observe that, compared to the Lee-Carter framework, we have additional
terms  $\beta^{2,g}_a\gamma^g_{t-a}$ that allow the model to capture
cohort effects.
We fit the Renshaw-Haberman model to the Swiss mortality data.
This is achieved in {\sf R} similarly to the Lee-Carter model
using the command {\tt rh()} instead of {\tt lc()} in input~\eqref{Equation: LC code}.
From this we obtain the fitted mortality rates $q^\text{RH}(\x)$, $\x\in\X$,
and we initialize Model Assumptions~\ref{Model} with these rates
$q(\x)  = q^\text{RH}(\x)$.
Then, we apply the same regression tree boosting machine as
explained in Section~\ref{Section: LC}
to obtain regression tree improved mortality rates $q^\text{tree}(\x)$
as well as the corresponding relative
changes $\Delta q^\text{RH}(\x)=q^\text{tree}(\x)/ q^\text{RH}(\x)-1$,  $\x\in\X$.

The second row of Figure~\ref{Figure: correction} illustrates
these relative changes $\Delta q^\text{RH}(\x)$ in the mortality rates.
We first compare these changes to the changes $\Delta q^\text{LH}(\x)$
obtained in Section~\ref{Section: LC} based on the Lee-Carter model,
see first row of Figure~\ref{Figure: correction}.
We observe that for the Renshaw-Haberman fit the regression tree algorithm proposes
less adjustments  than for the Lee-Carter fit.
In particular, for ages above $60$ there are no changes $\Delta q^\text{RH}(\x)$
outside the range $[-5\%,5\%]$,
except for the calendar year $t=1918$.
On the one hand, this illustrates that for these ages the Renshaw-Haberman fit
is quite appropriate and better than the Lee-Carter fit.
On the other hand, year $t=1918$
indicates that the Renshaw-Haberman model may not fully capture
special events such as epidemics, see also the split for birth cohort $c=t-a=1917$
in Figure~\ref{Figure: correction}.
This is mainly explained by the fact that special events are too heavy-tailed
for this model choice.
Moreover, not surprisingly, the algorithm performs only a few splits
with respect to birth cohorts in the Renshaw-Haberman model; indeed
it captures cohort effects
more appropriately compared to the Lee-Carter model.
Finally, compared to the Lee-Carter model, we observe that the Renshaw-Haberman framework
provides better mortality fits for recent calendar years between $t=1986$ and $t=2014$.

\section{Boosting cause-of-death mortality}\label{Section: cause}

In this section we discuss cause-of-death mortality under the Poisson framework of
Model Assumptions~\ref{Model}.
Given a death with specific feature $\x\in\X$, we
investigate the conditional probability of its cause.
We illustrate how these probabilities
can be estimated from real data by applying Poisson regression tree boosting
similarly as introduced in Section~\ref{Section: Boosting mortality rates}.
This allows us to detect patterns in these probabilities over time.
We first introduce the setup and then  apply the boosting machine  to Swiss mortality data.

\subsection{Cause-of-death mortality framework}\label{Section: framework}

Consider the feature space $\X=\G\times\A\times\T$ with
components given by~\eqref{Equation: feature components}.
Again, one could consider further feature components such as
the socio-economic status of a person.
We fix Model Assumptions~\ref{Model} with given  (estimated) mortality rates $q(\x)$, $\x\in\X$,
and we consider the set $\K=\{1,\ldots,K\}$ that describes
$K\in\N$ different possible causes of death.
Conditionally given a death with feature $\x\in\X$,
we denote by $\theta(k| \x)\in[0,1]$ the conditional
probability that the corresponding cause is $k\in\K$,
and we denote by $D_{\x,k}\ge0$ the number of such deaths.
Since $\K$ provides a partition, we get
$\sum_k D_{\x,k} = D_{\x}$ and $\sum_k \theta(k| \x) = 1$ for all $\x\in\X$.
Furthermore, under Model Assumptions~\ref{Model} we obtain
\begin{equation}\label{Equation: death cause assumption}
D_{\x,k} ~\sim~ \text{Pois}\left(  \theta(k| \x) q(\x)E_{\x}  \right),
\end{equation}
and $(D_{\x,k})_{\x,k}$ are independent, see Theorem~2.14 in~\cite{W}.
Note that~\eqref{Equation: death cause assumption}  is subject to
$\theta(k| \x)>0$, otherwise we set $D_{\x,k}=0$, $\p$-a.s.
As an initial model (prior choice) we assume that the probabilities $\theta(k| \x)$
are independent of any features $\x\in\X$ and of any cause $k\in\K$, and we simply set
\begin{equation*}
\theta(k| \x)~=~\theta_K ~=~ \frac{1}{K},
\qquad \text{for $k\in\K$ and $\x\in\X$.}
\end{equation*}
Alternative initializations, such as observed relative frequencies, could be chosen as well.
This provides the input (starting value) for the subsequent boosting machine to calibrate
the conditional probabilities $\theta(\cdot| \x)$.
Using~\eqref{Equation: death cause assumption} we write
\begin{equation*}
D_{\x,k} ~\sim~ \text{Pois}(\mu(\x,k)d_{\x,k}  )
\qquad \text{with $\mu(\x,k)\equiv1$ and $d_{\x,k} = \theta(k| \x) q(\x)E_{\x}$},
\end{equation*}
for $\x\in\X$ and $k\in\K$.
According to our initial model assumptions,
$d_{\x,k}$ denotes the expected number of deaths with feature $\x\in\X$
and with cause of death being $k\in\K$.
Our aim is to calibrate the factor $\mu(\cdot):\X\times\K \to \R_+$ by a
non-parametric regression approach in complete analogy to Section~\ref{Section: LC}.

For this we apply the SBS tree growing algorithm
to the working data $( D_{\x,k},\, (\x,k),\,  d_{\x.k} )_{\x,k}$
based on the extended feature space $\{(g,a,t,c=t-a,k) \mid g\in\G,\, a\in\A,\, t\in\T,\, k\in\K\}$.
That is, we calibrate the factor $\mu(\cdot)$ with respect to the feature components
gender, age, calendar year, birth cohort and cause of death by using the
{\sf R}-command  similar to the one in \eqref{Equation: tree code}.
This provides us
a tree based estimator $\hat \mu(\x,k)$
of $\mu(\x,k)$,
and allows us to define the regression tree estimated probabilities
\begin{equation}\label{Equation: theta tree}
\theta^\text{tree}(k| \x)~=~\hat\mu(\x,k)\theta(k| \x),
\qquad \text{for $k\in\K$ and $\x\in\X$.}
\end{equation}
Here, we interpret $\hat\mu(\x,k)$ as a refinement of the
initial conditional probability $\theta(k| \x)=\theta_K=1/K$.
Instead, $\hat\mu(\x,k)$ could also be interpreted as an improvement of the unconditional
probability $\theta(k| \x)q(\x)$.
However, in our context, we consider the initialized mortality rates $q(\x)$ as fixed.

\subsection{Boosting Swiss cause-of-death mortality data}\label{Section: Swiss}

We apply regression tree boosting as explained in the previous section
to Swiss cause-of-death mortality data provided
by the Swiss Federal Statistical Office and by
the Human Mortality Database, see also Section~\ref{Section: Boosting mortality rates}.
This data includes the exposures
$(E_{\x})_{\x}$ and the numbers of deaths $(D_{\x,k})_{\x,k}$ for
the feature space $\X= \G\times\I\times\T$
with feature components
\begin{equation*}
\G=\{\text{female, male}\}, \quad
\I=\{1,2,\ldots,6\}, \quad \text{and} \quad
\T=\{1990,1991,\ldots,2014\},
\end{equation*}
and for $\K=\{1,\ldots,12\}$ describing $12$ different causes of death, see below.
Because our data on cause-of-death mortality only contains limited information about the
age of a person,
we have replaced the feature component $\A$ by the feature component $\I$ that
represents six disjoint age buckets.
Age group $i=1$ corresponds to age $0$,
$i=2$ to ages $1-14$,
$i=3$ to ages $15-44$,
$i=4$ to ages $45-64$,
$i=5$ to ages $65-84$,
and $i=6$ corresponds to ages of at least $85$.
The set $\K$ represents the following $K=12$ different possible causes of death:
\begin{table}[h!]\setlength{\tabcolsep}{12pt}
  \centering
  \begin{tabular}{lll}
   1) infectious diseases & 5) circulatory system & \hspace{2mm}9) congenital malformation \\
    2) malignant tumors & 6) respiratory organs & 10) perinatal causes \\
    3) diabetes mellitus & 7) alcoholic liver cirrhosis & 11) accidents and violent impacts \\
    4) dementia & 8) urinary organs & 12) others/unknown
  \end{tabular}
\end{table}
%


The data $D_{\x,k}$ for cause $k=4$ (dementia) is {\it not} available for the first five
years of observations $t\in\{1990,\ldots,1994\}$.
However, the {\tt rpart()} command of the {\tt `rpart'} package
is able to cope with such missing data, see~\cite{rpart} for details.

\begin{figure}
\centering
\includegraphics[width=0.5\linewidth]{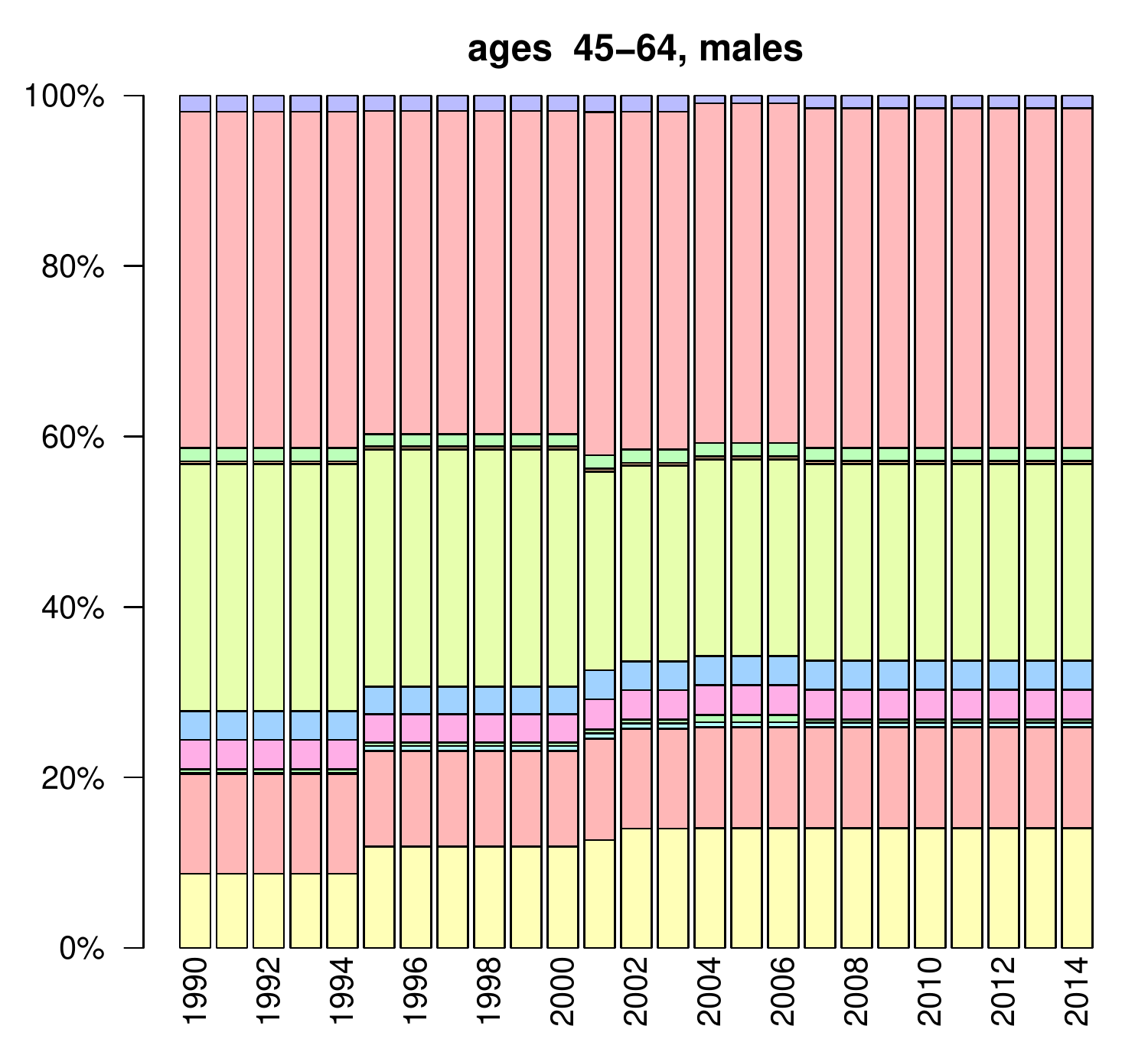}
\includegraphics[width=.36\linewidth]{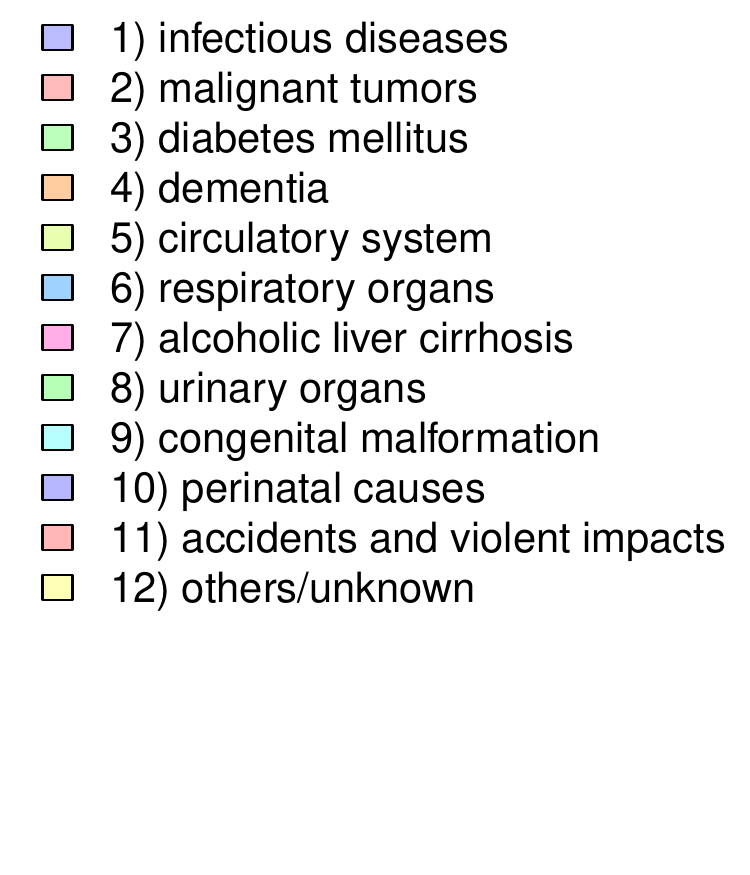}
\caption{\footnotesize
Regression tree estimated probabilities $\theta^\text{tree}(k| \x)$
for the $12$ different causes of death considered,
conditionally for males aged between $45$ and $64$.
}
\label{Figure: death causes time}
\end{figure}

We initialize Model Assumptions~\ref{Model} with the regression tree improved
mortality rates $q(\x)= q^\text{tree}(\x)$, $\x\in\X$, obtained in Section~\ref{Section: RH}
based on the Renshaw-Haberman fit.
Observe that this needs some care because we work here on a feature space that is
different to the one in Section~\ref{Section: Boosting mortality rates},
see Remark~\ref{Remark: q bucket}, below.
Then, we apply the SBS tree growing algorithm based to the feature space $\X\times\K$;
note that here we do not calibrate $\hat\mu$ with respect to birth cohorts.
This provides us with the regression tree estimated conditional probabilities $\theta^\text{tree}(k| \x)$
for $k\in\K$ and $\x\in\X$, according to~\eqref{Equation: theta tree}.
The resulting probabilities for females and
the causes of death $k=2$ (malignant tumors) and $k=5$ (circulatory system), respectively,
are presented in the first row of Figure~\ref{Figure: death causes}; the
remaining probabilities are summarized in Appendix~\ref{Appendix: Figures},
in Figure~\ref{Appendix, Figure: death causes, females} for females
and in Figure~\ref{Appendix, Figure: death causes, males} for males.
We have applied a polynomial smoothing model to these plots
in order to present the tree estimated probabilities
in a more accessible way.
The second row of Figure~\ref{Figure: death causes} shows the
corresponding Pearson's residuals given by
\begin{equation}\label{Equation: residuals}
\delta^\text{tree}_{\x|k}
~=~
\frac{D_{\x,k} - \theta^\text{tree}(k| \x) D_\x}{ \sqrt{ \theta^\text{tree}(k| \x) D_\x }},
\qquad \text{for $\x\in\X$ and $k\in\K$,}
\end{equation}
where we set $\delta^\text{tree}_{\x|k}=0$ if the denominator equals $0$
or if the data on $D_{\x,k}$ is missing.
We observe that the regression tree boosting machine has
suitably estimated the probabilities $\theta(k| \x)$,
there are no structures visible in the residual plots.
However, note that the interpretation of these estimates $\theta^\text{tree}(k| \x)$
needs some care.
For instance, the classification of certain causes of death may have changed over time,
see also the discussion in~\cite{Richards}.
\begin{figure}
\centering
\includegraphics[width=.45\linewidth]{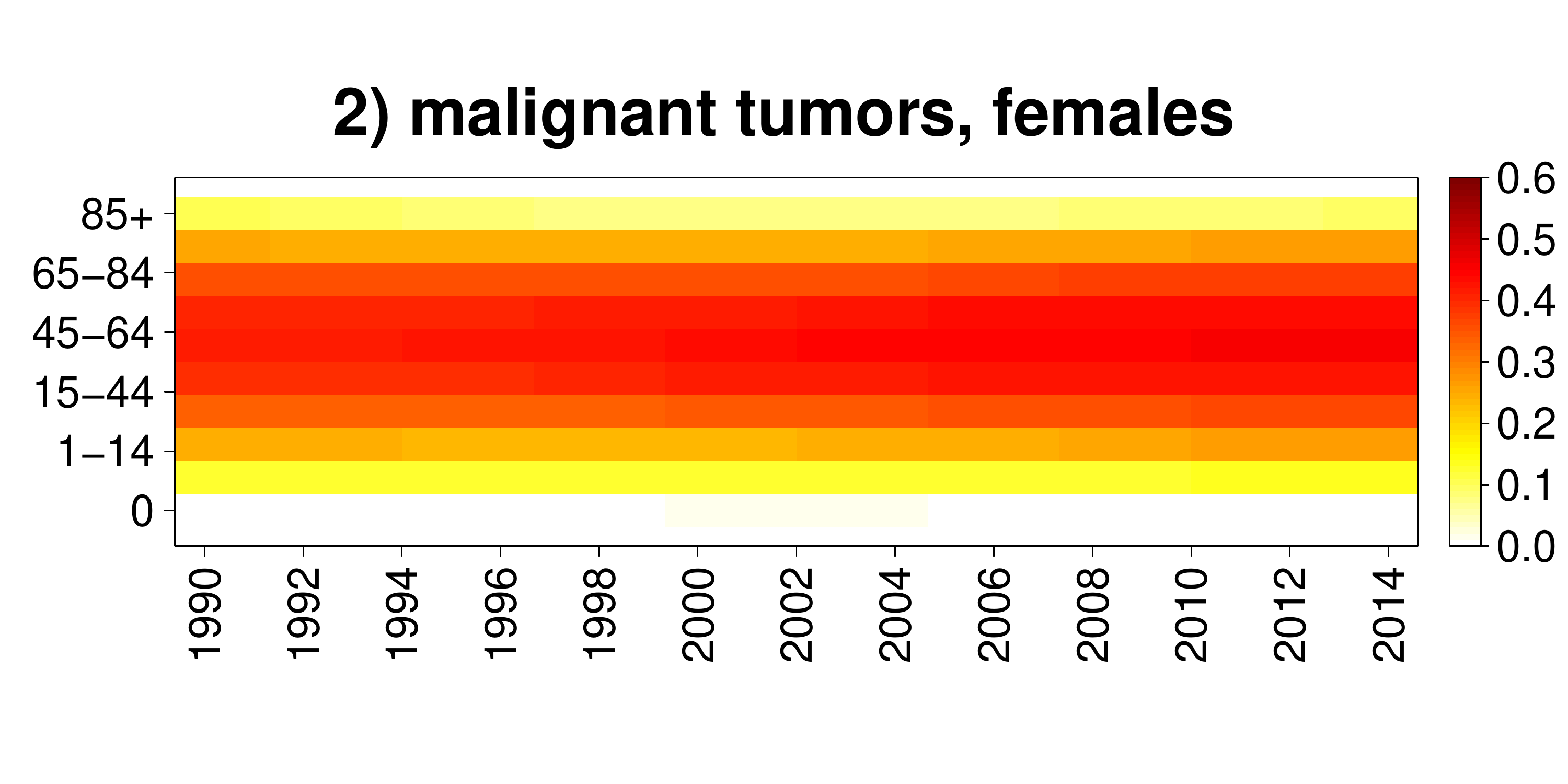}
\hfill
\includegraphics[width=.45\linewidth]{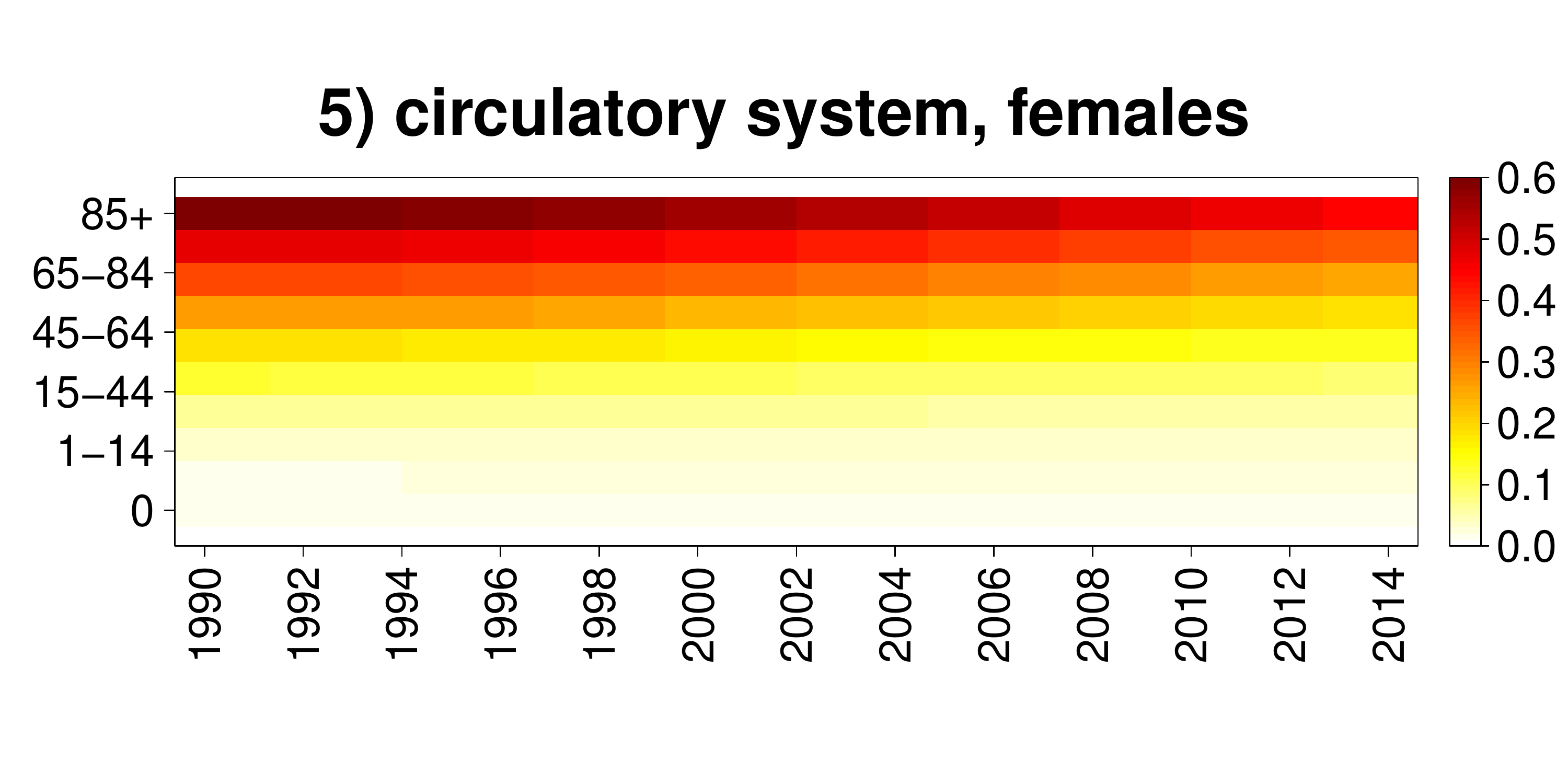}
\\[-0.5cm]
\includegraphics[width=.45\linewidth]{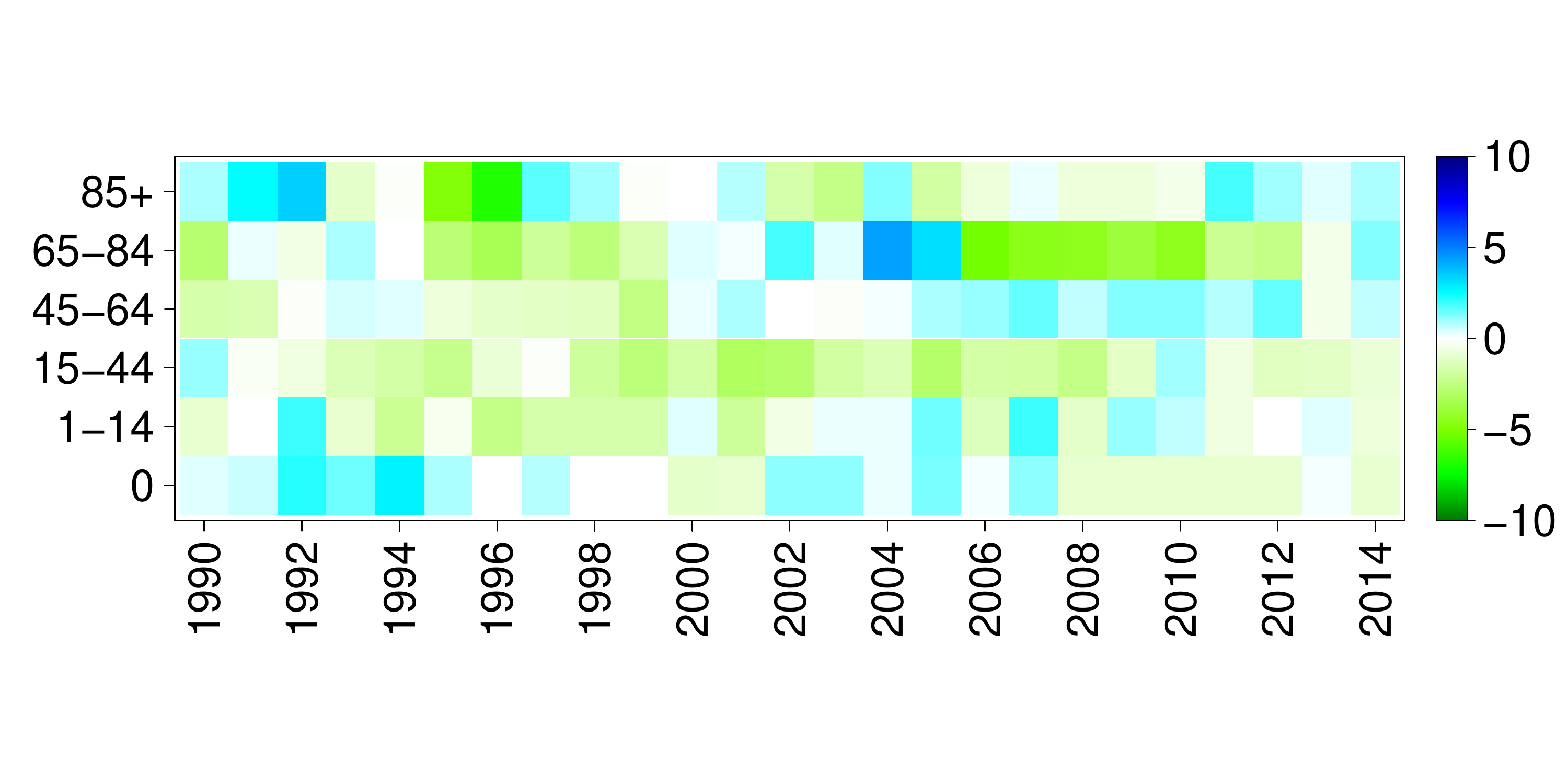}
\hfill
\includegraphics[width=.45\linewidth]{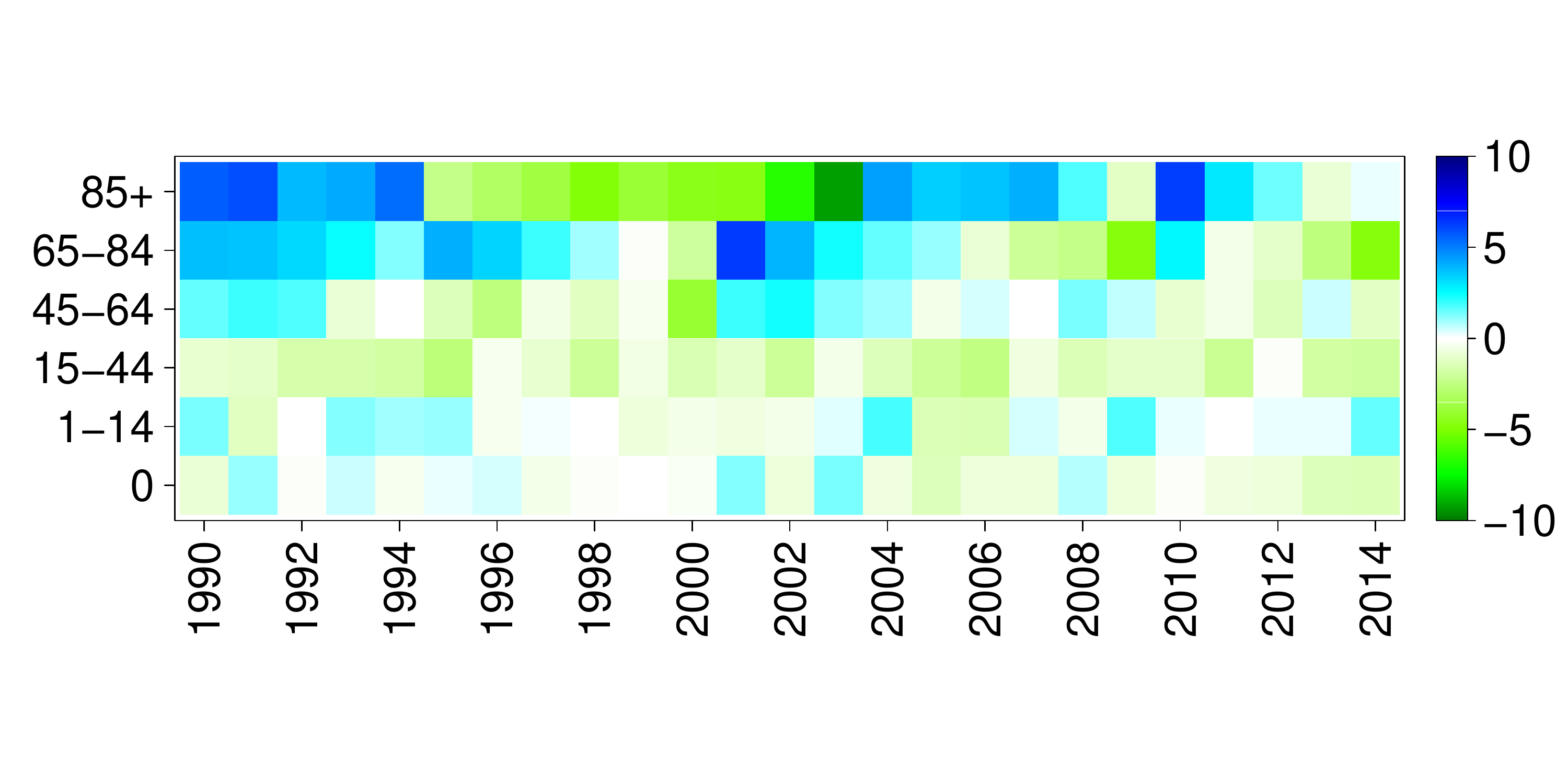}
\caption{\footnotesize
The first row illustrates the regression tree estimated probabilities $\theta^\text{tree}(k| \x)$
for females and for the causes of death $k\in\{2,5\}$.
The second row shows the corresponding Pearson's residuals
defined in~\eqref{Equation: residuals}.
}
\label{Figure: death causes}
\end{figure}
Figure~\ref{Figure: death causes time} shows
the evolution of the different cause-of-death probabilities over time
for males aged between $45$ and $64$;
the probabilities for the
remaining age classes are presented in Appendix~\ref{Appendix: Figures},
in Figure~\ref{Appendix, Figure: death causes time, females} for females
and in Figure~\ref{Appendix, Figure: death causes time, males} for males.

~

\begin{remark}\label{Remark: q bucket}

Consider the feature space $\X=\G\times\A\times\T$ with feature components
as in~\eqref{Equation: feature components},
and consider Model Assumptions~\ref{Model} with given mortality rates $q(\x)$, $\x\in\X$.
Additionally, consider the condensed feature space $\tilde\X=\G\times\I\times\T$,
with $\I=\{1,\ldots,I\}$ representing  $I\ge1$ disjoint and non-empty
age buckets $\A_1, \ldots, \A_I \subset \A$ that form a partition of $\A$.
In the following we briefly sketch how to construct mortality rates
$\tilde q(\tilde\x)$, $\tilde\x\in\tilde\X$, from $q(\cdot)$ in a way that
is consistent with Model Assumptions~\ref{Model} for the
condensed feature space $\tilde\X$.

For $\x=(g,a,t)\in\X$ and $\tilde\x=(\tilde g,i, \tilde t) \in \tilde\X$
we write $\x\sim\tilde\x$ if $g=\tilde g$, $a\in \A_i$ and $t=\tilde t$.
Consider the mortality rates given by the regression function $\tilde q:\tilde\X \to [0,1]$ with
\begin{equation}\label{Equation: tilde q}
\tilde q(\tilde \x)~=~\frac{1}{E_{\tilde \x}}\sum_{\x\sim\tilde \x} E_{\x} \,q(\x),
\qquad \tilde\x\in\tilde\X,
\end{equation}
with $E_{\tilde \x} = \sum_{\x\sim\tilde \x} E_{\x}$ denoting the total exposure
with condensed feature $\tilde\x\in\tilde\X$.
Denote by $D_{\tilde \x} = \sum_{\x\sim\tilde \x}D_{\x}$
the total number of deaths with condensed feature $\tilde \x\in\tilde \X$.
By Model Assumptions~\ref{Model} we  obtain
\begin{equation*}
D_{\tilde \x} ~\sim~ \text{Pois}\left( \sum_{\x\sim\tilde \x}q(\x) E_{\x}  \right)
~\overset{(\text{d})}{=}~\text{Pois}(\tilde q(\tilde \x) E_{\tilde \x}  ),
\end{equation*}
and these random variables are independent in $\tilde \x \in\tilde\X$,
see Theorem~2.12 in~\cite{W}.
That is, the mortality rates $\tilde q(\cdot)$ given by~\eqref{Equation: tilde q} are defined
in a way that is consistent with Model Assumptions~\ref{Model}
for the condensed feature space $\tilde\X$.
In particular, our model assumptions are closed towards aggregations
that give more coarse partitions.

\end{remark}

\section{Conclusions}

In a first analysis we have illustrated how machine learning
techniques, in particular the regression tree boosting machine,
can be used in order to back-test parametric mortality models.
These techniques allow us to detect the weaknesses of such models
based on real data.
Moreover, regression tree boosting can further be applied to improve
the fits of such models
with respect to feature components that are not
captured by these models.
Typical examples are education, income or marital status of a person.

In the second part we have investigated cause-of-death mortality
under a Poisson model framework.
We have presented how regression tree boosting can be applied to
estimate cause-of-death mortality rates from real data.
This technique provides a simple way to detect patterns in these
probabilities over time.

%
%
%
%

\bibliographystyle{abbrv}
\bibliography{./bibliography}

\newpage

\appendix

\section{Figures on Swiss cause-of-death mortality}\label{Appendix: Figures}
\begin{figure}[!ht]
\centering
\includegraphics[width=.32\linewidth]{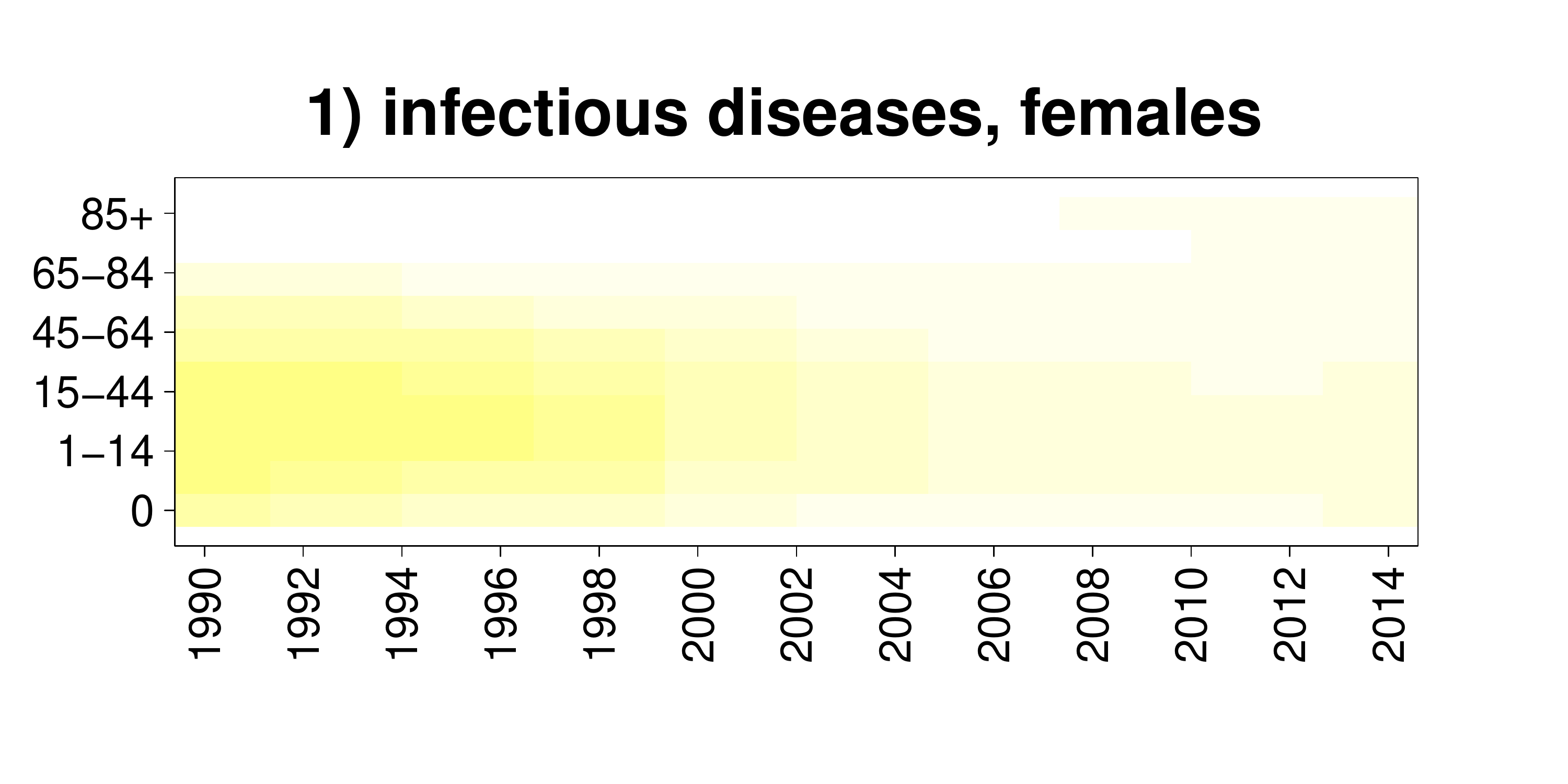}
\hfill
\includegraphics[width=.32\linewidth]{./Figures/C2_F.pdf}
\hfill
\includegraphics[width=.32\linewidth]{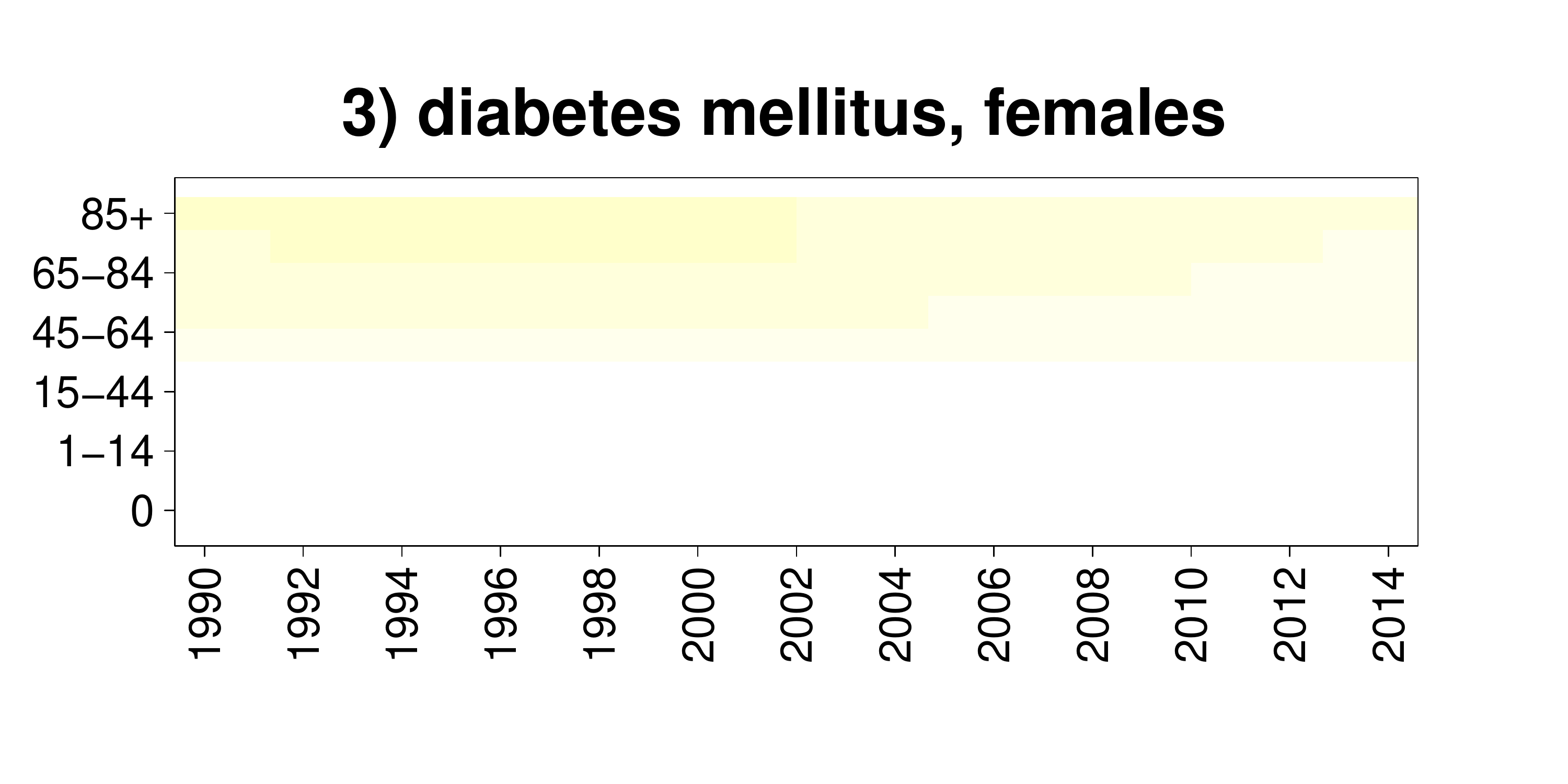}
\\[-0.5cm]
\includegraphics[width=.32\linewidth]{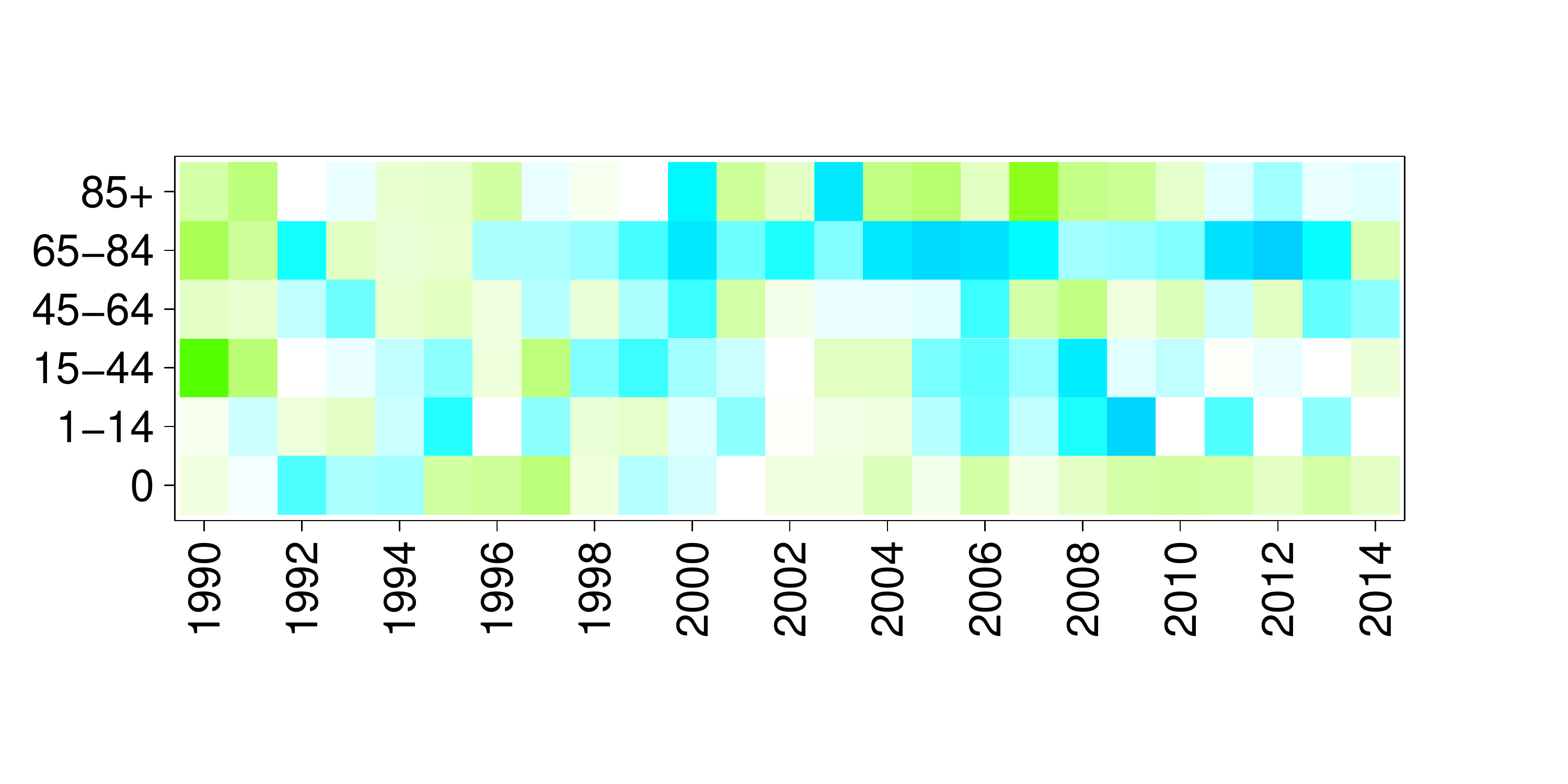}
\hfill
\includegraphics[width=.32\linewidth]{./Figures/R2_F.pdf}
\hfill
\includegraphics[width=.32\linewidth]{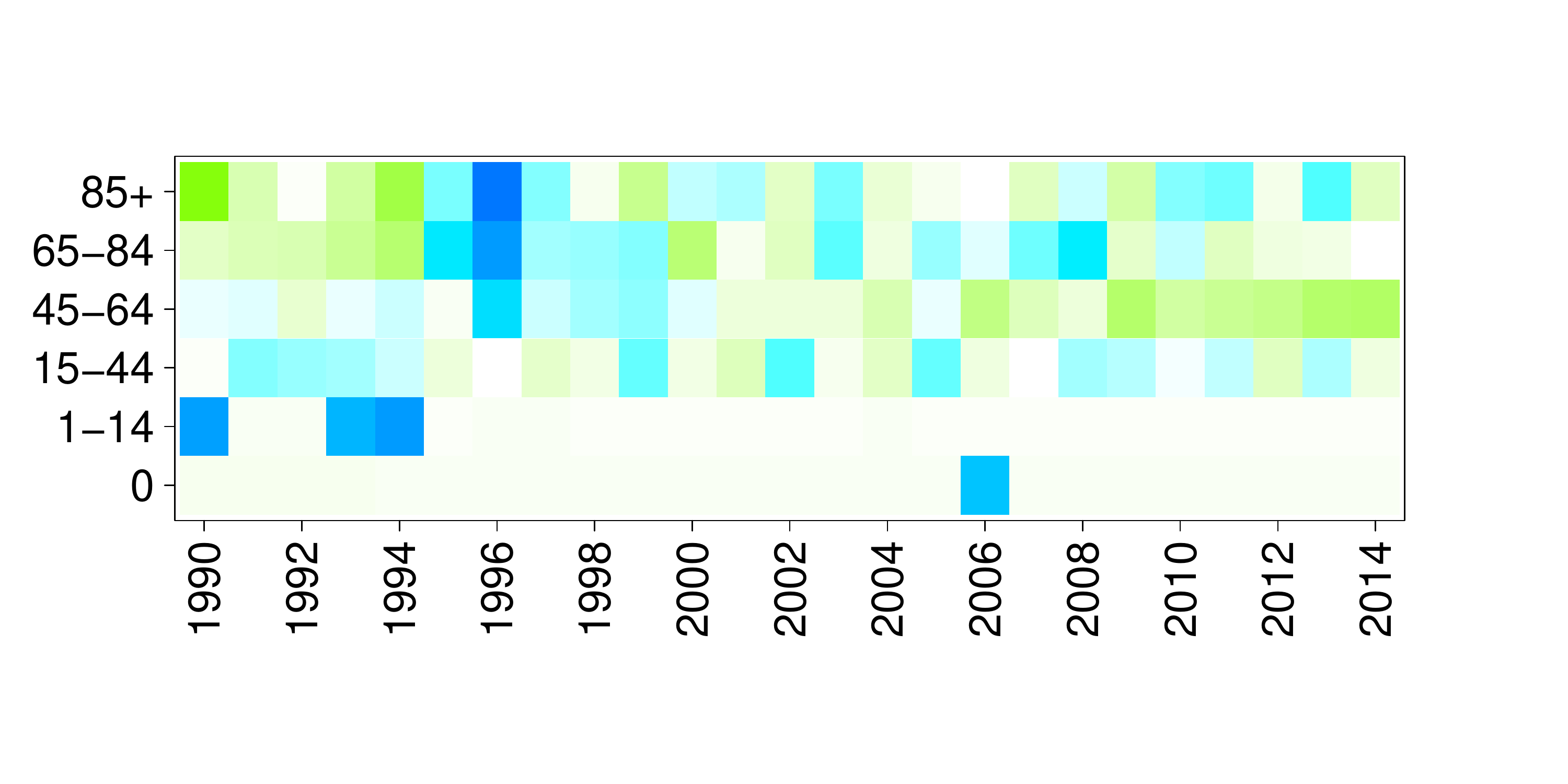}
\\
\hrule
\includegraphics[width=.32\linewidth]{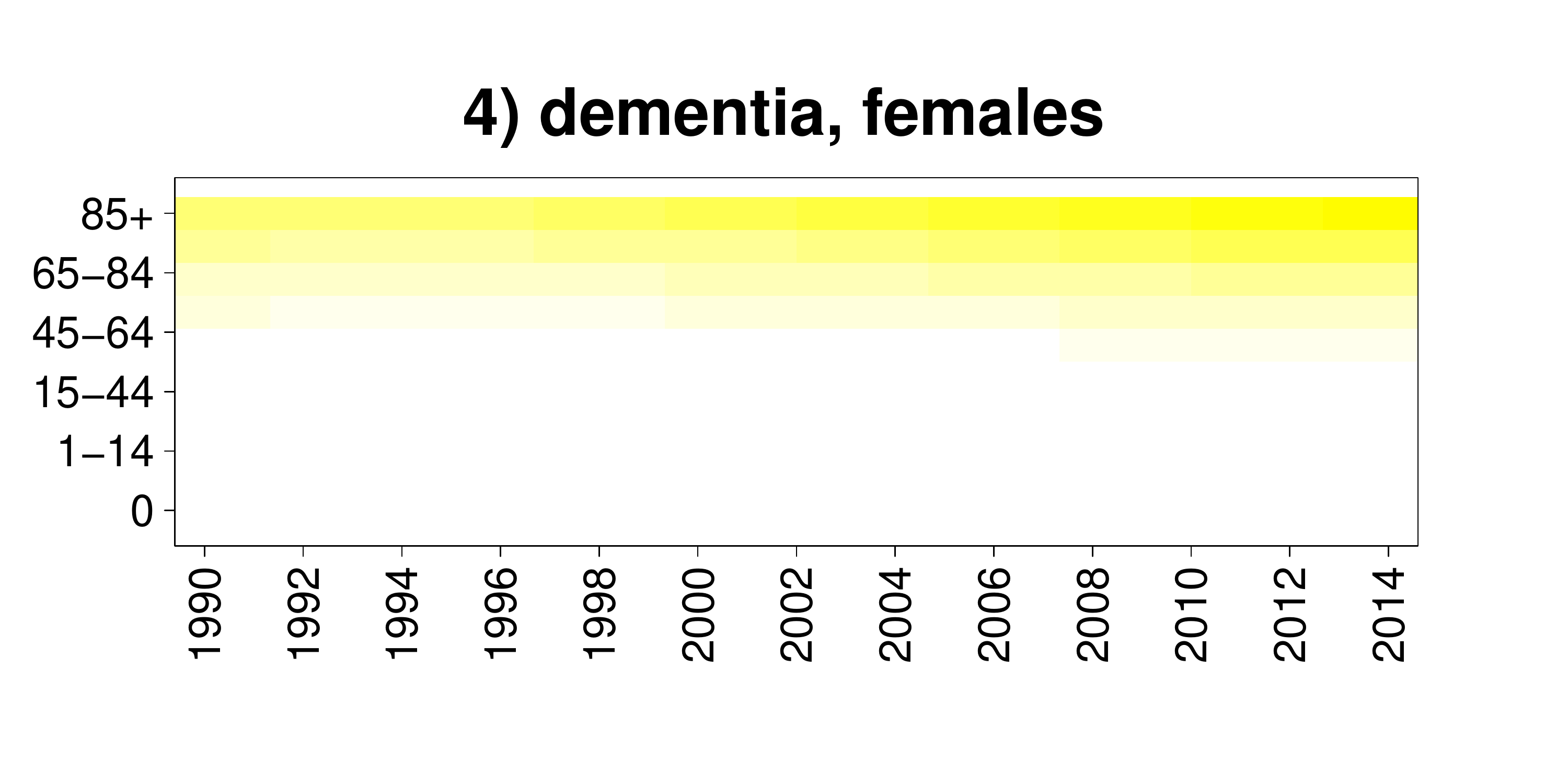}
\hfill
\includegraphics[width=.32\linewidth]{./Figures/C5_F.pdf}
\hfill
\includegraphics[width=.32\linewidth]{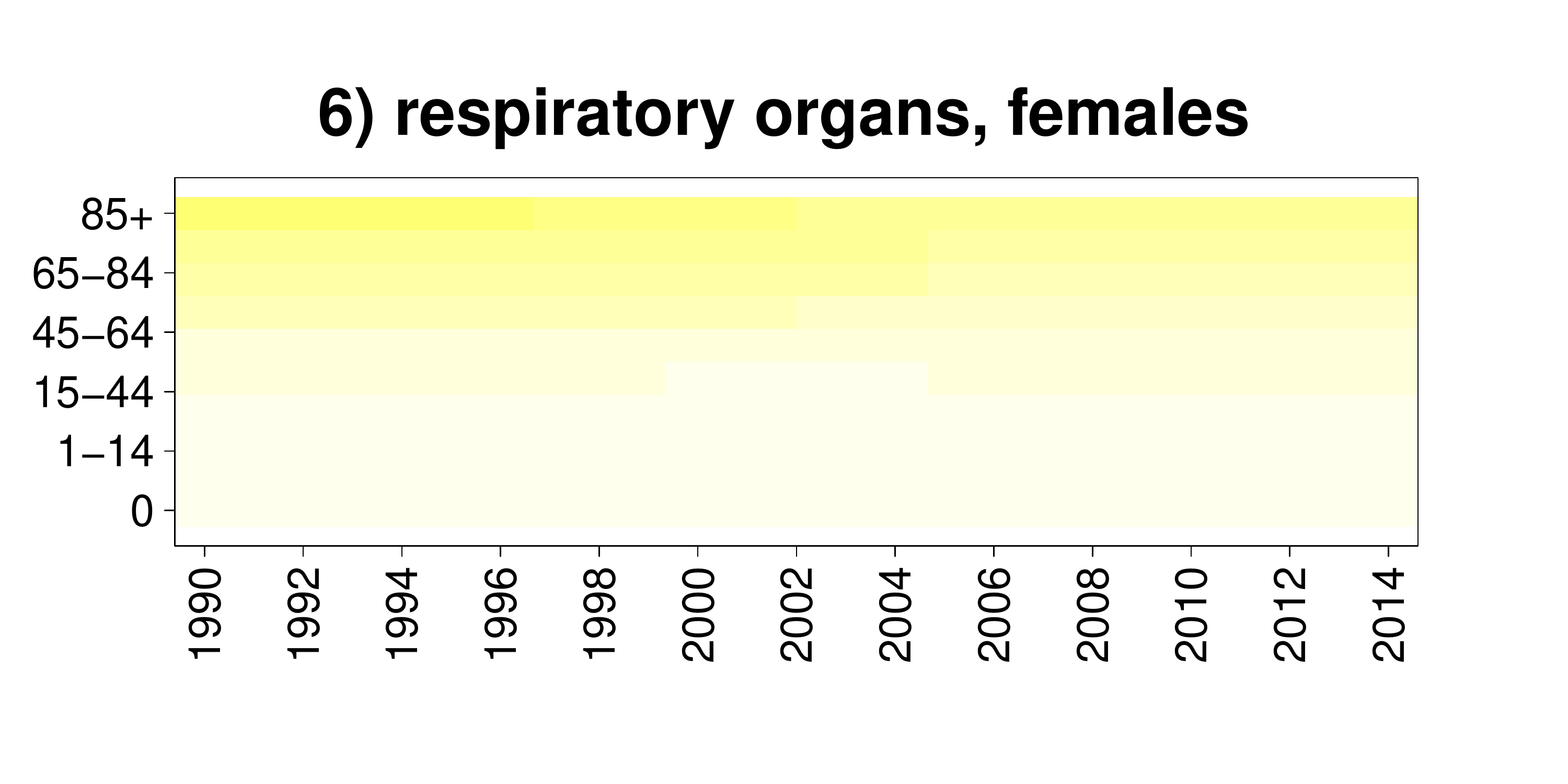}
\\[-0.5cm]
\includegraphics[width=.32\linewidth]{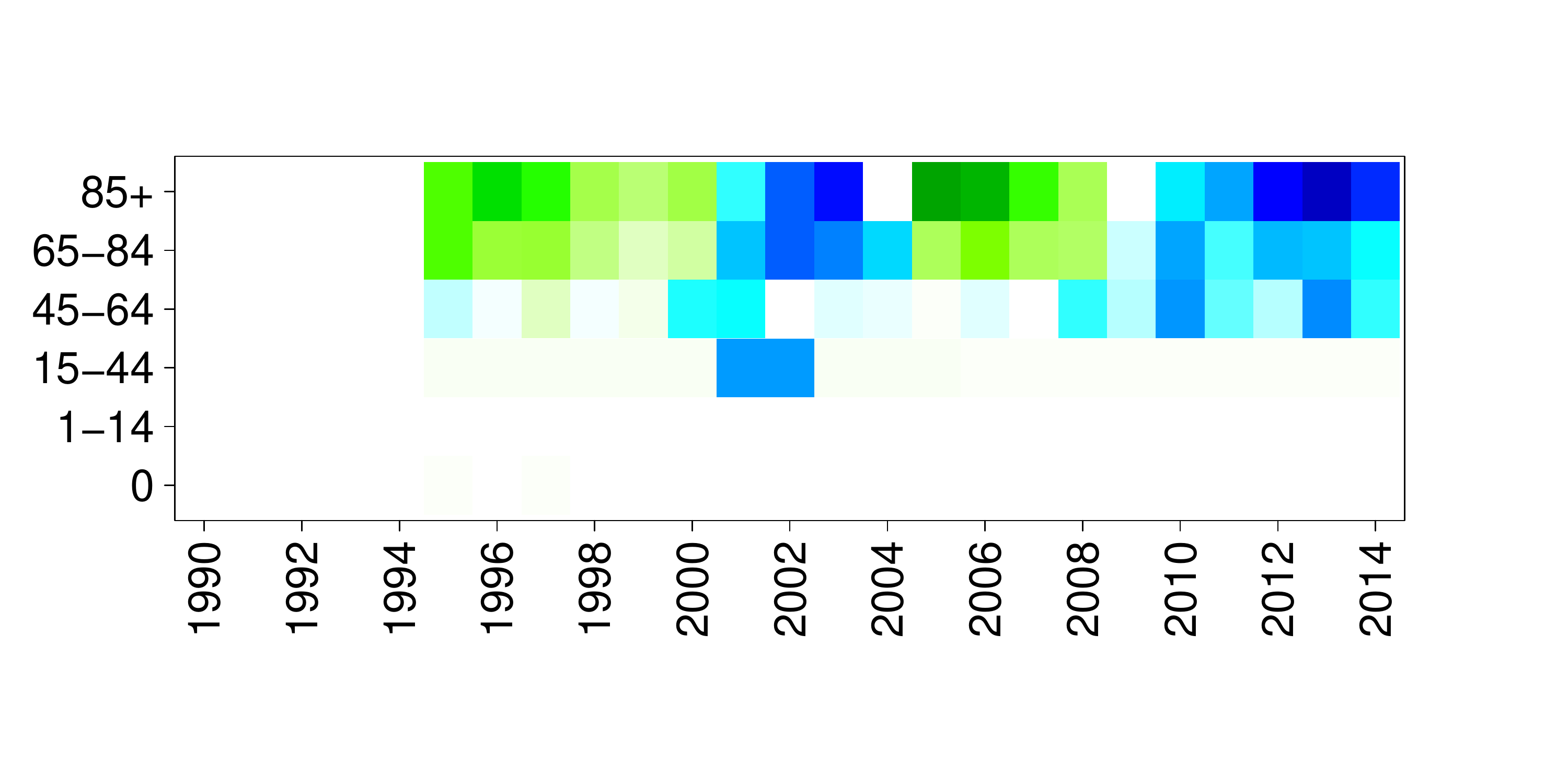}
\hfill
\includegraphics[width=.32\linewidth]{./Figures/R5_F.pdf}
\hfill
\includegraphics[width=.32\linewidth]{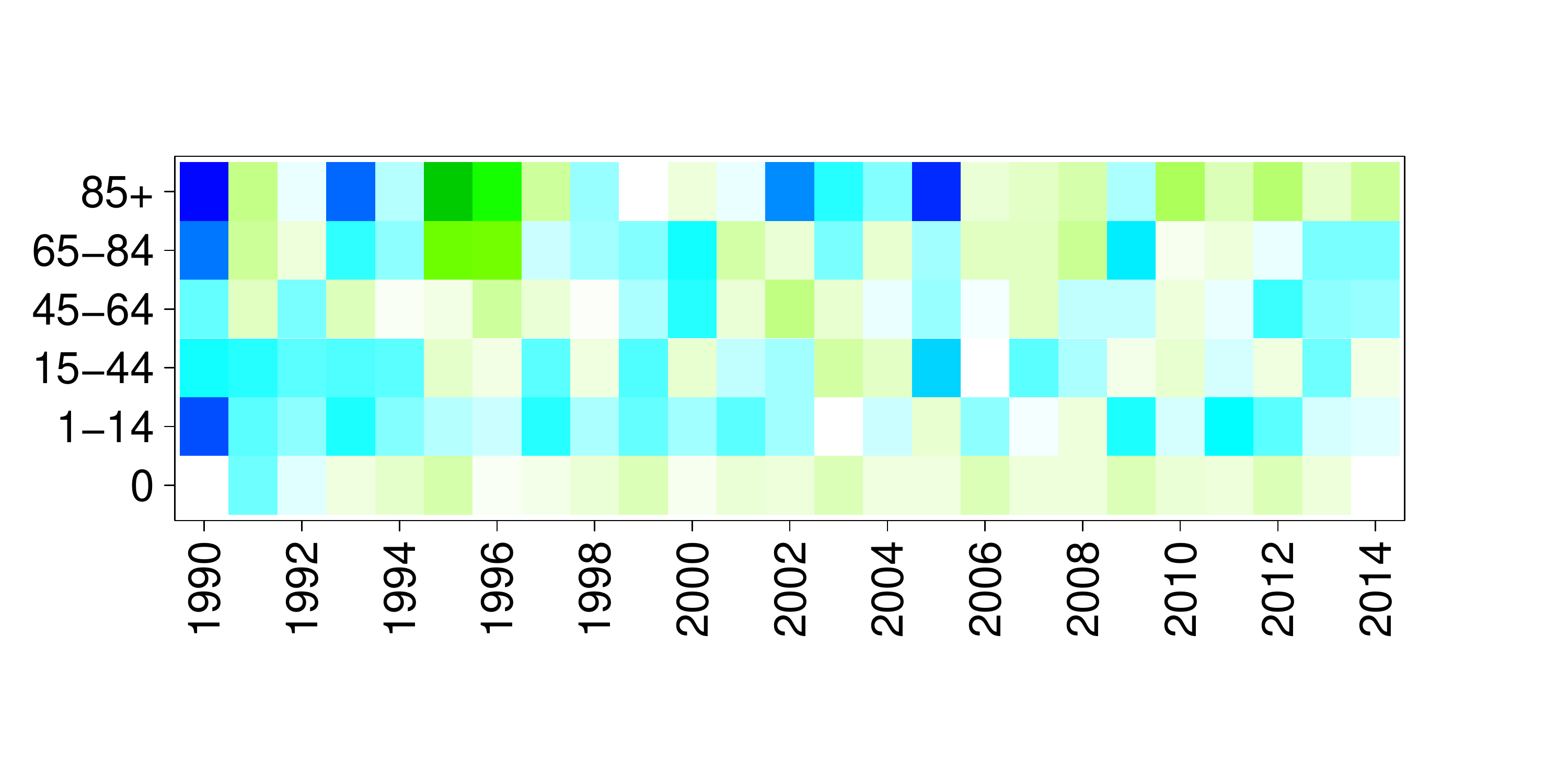}
\\
\hrule
\includegraphics[width=.32\linewidth]{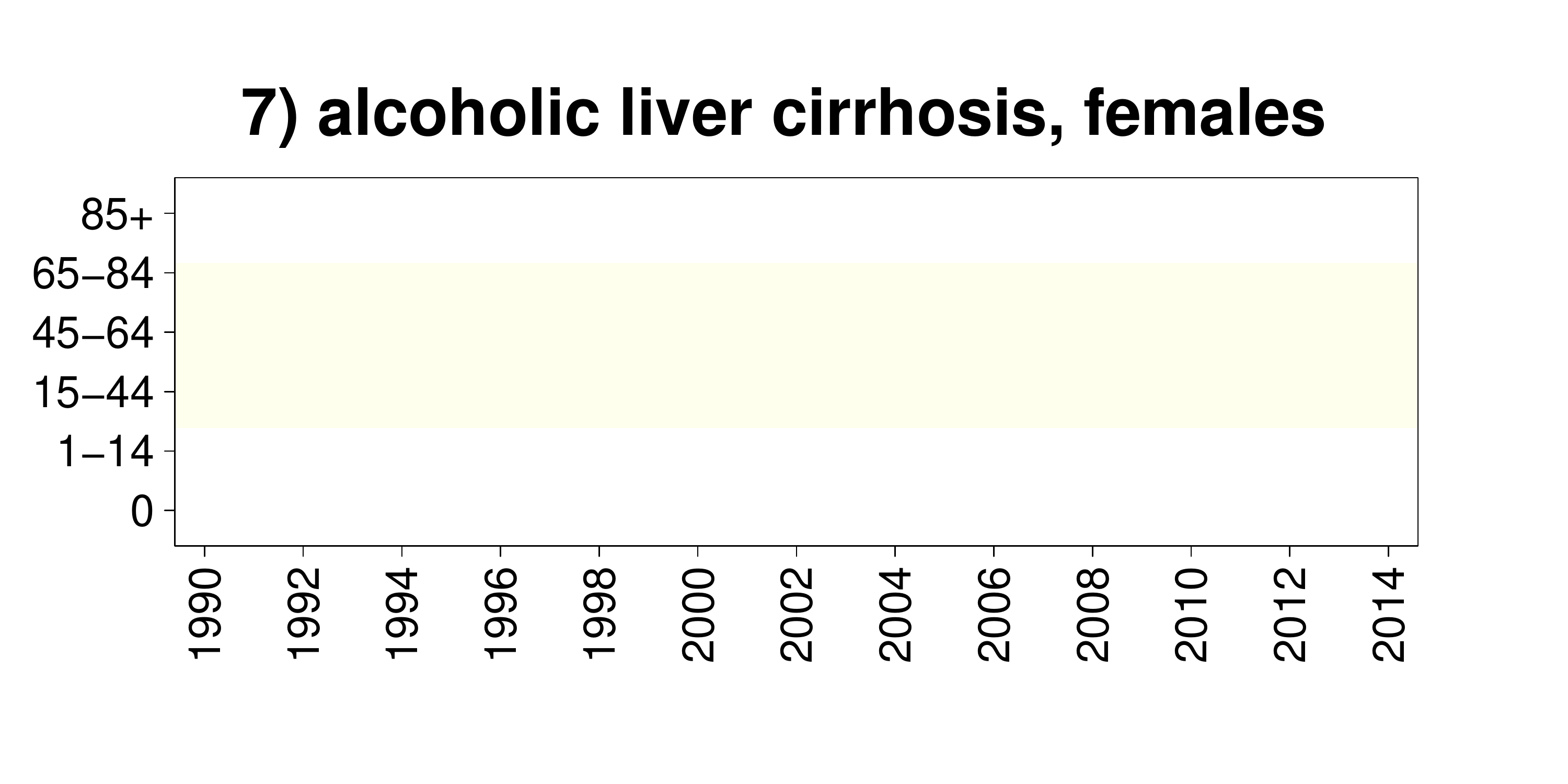}
\hfill
\includegraphics[width=.32\linewidth]{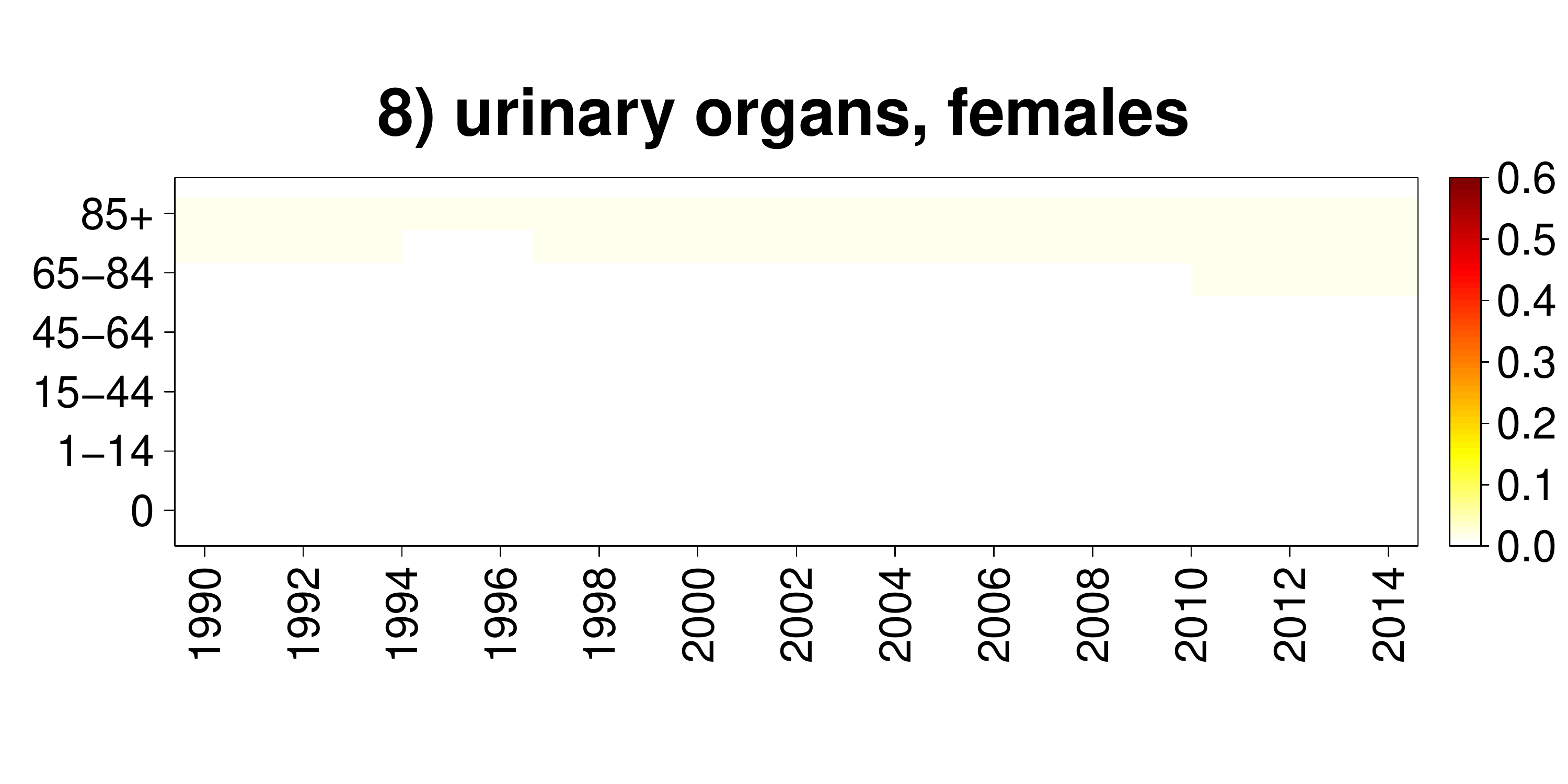}
\hfill
\includegraphics[width=.32\linewidth]{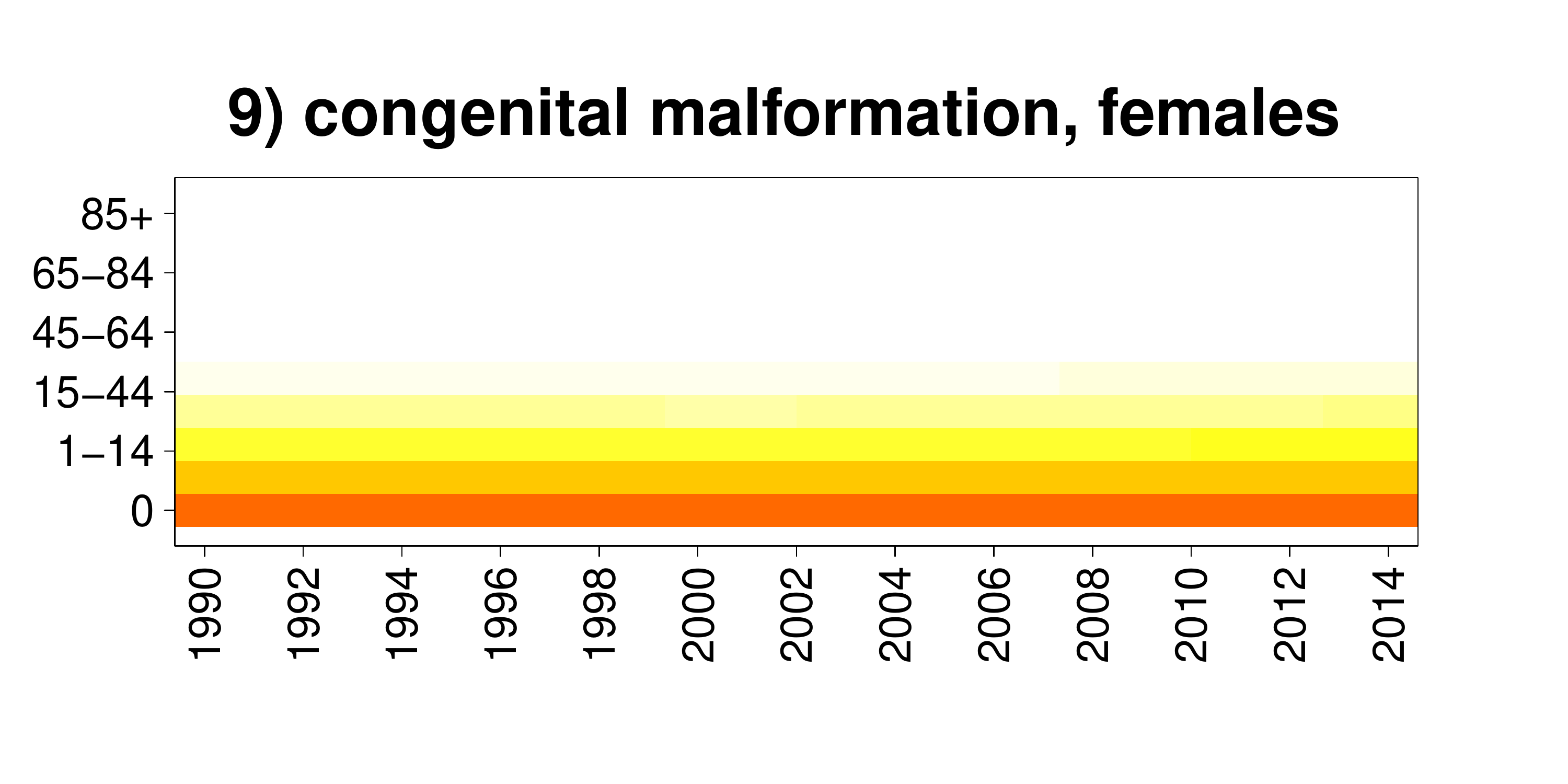}
\\[-0.5cm]
\includegraphics[width=.32\linewidth]{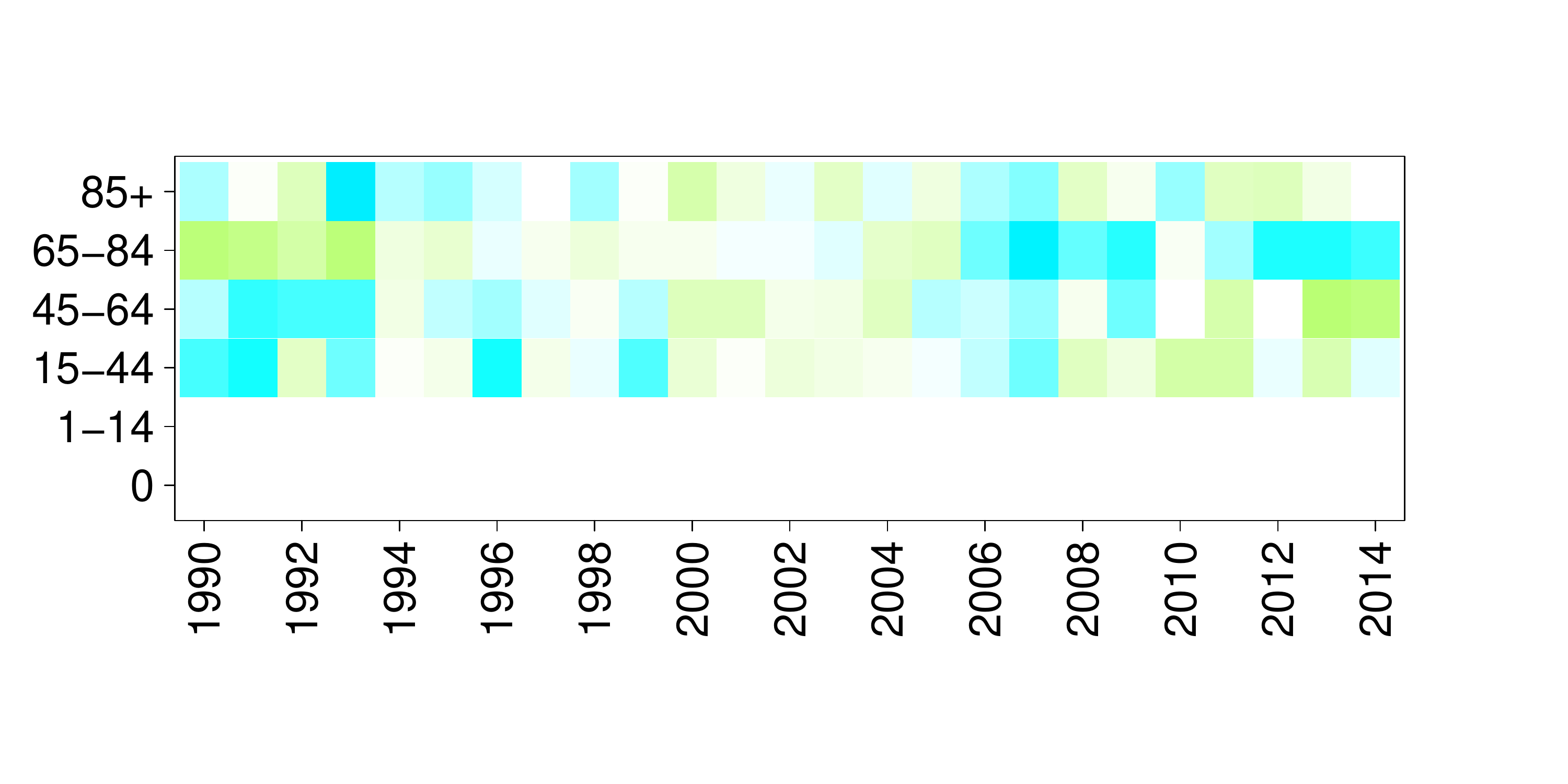}
\hfill
\includegraphics[width=.32\linewidth]{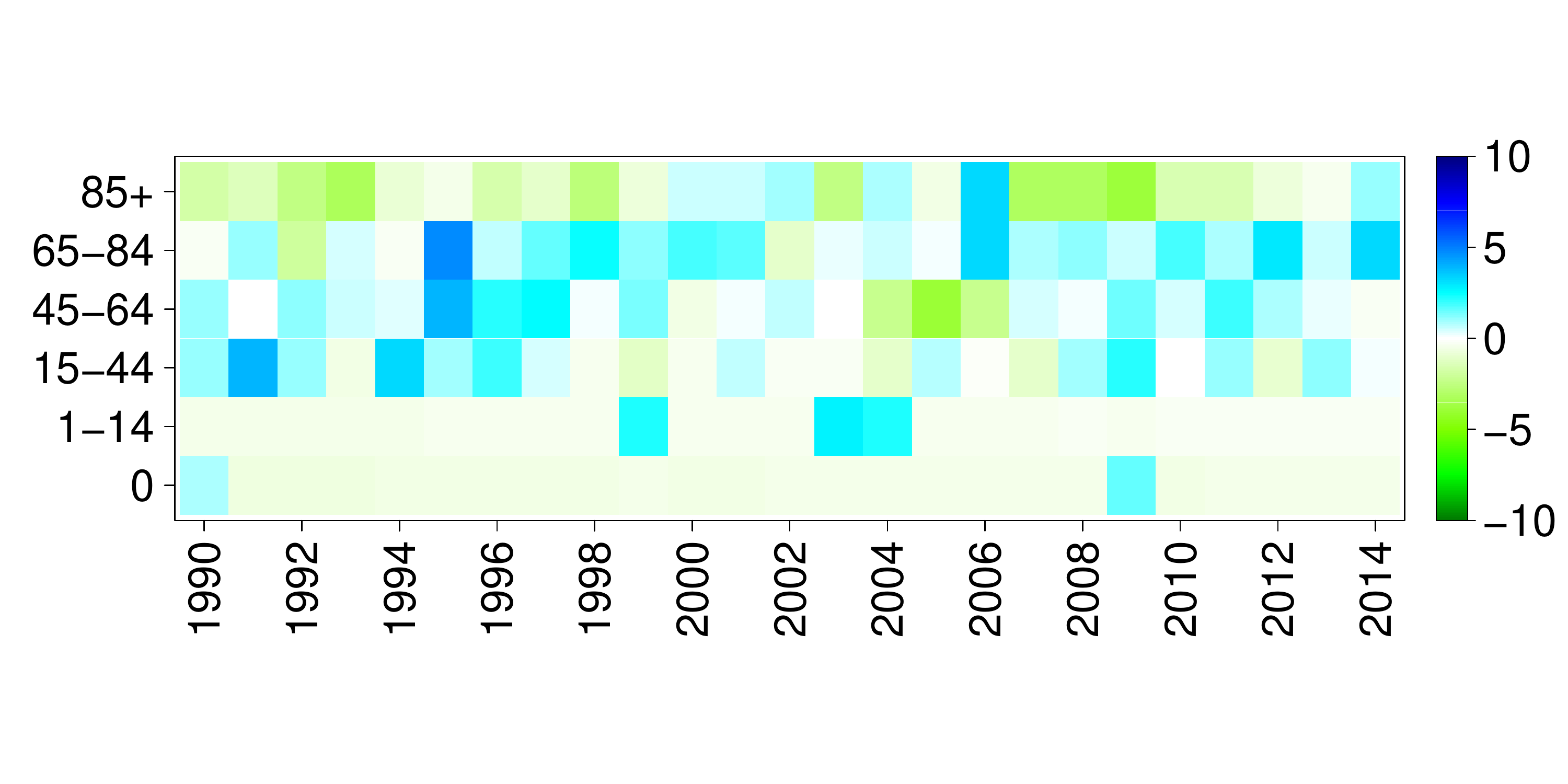}
\hfill
\includegraphics[width=.32\linewidth]{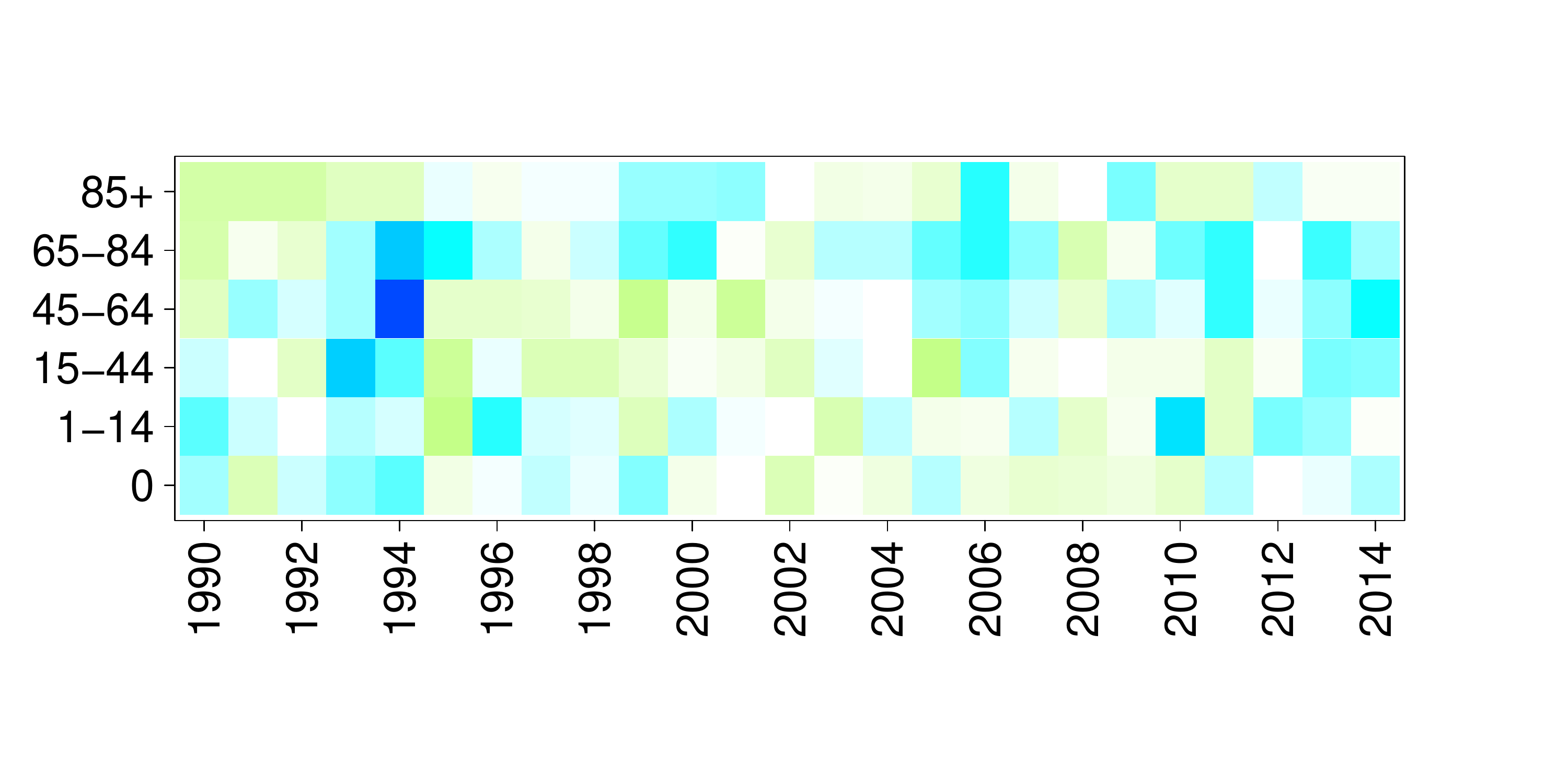}
\\
\hrule
\includegraphics[width=.32\linewidth]{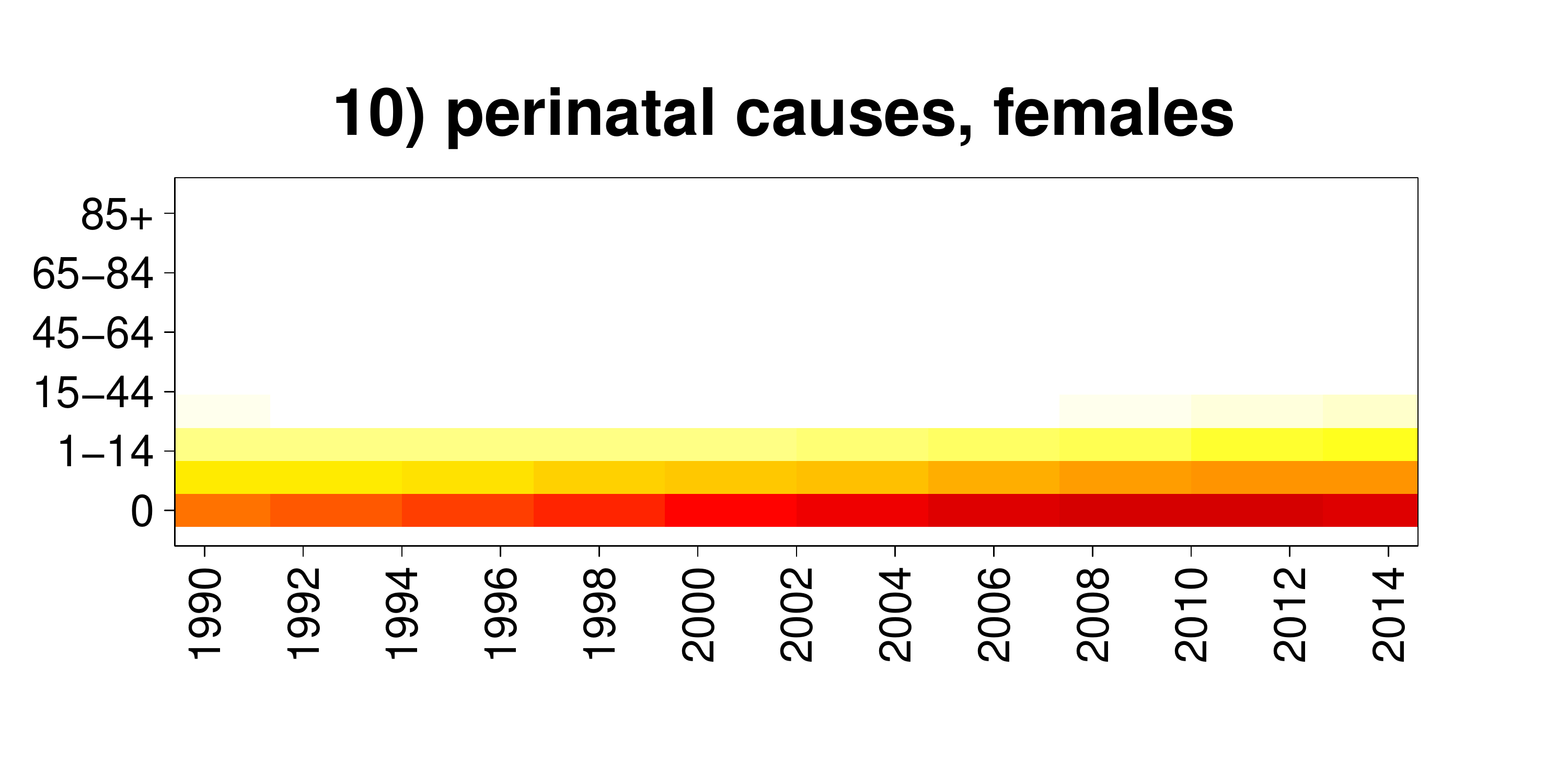}
\hfill
\includegraphics[width=.32\linewidth]{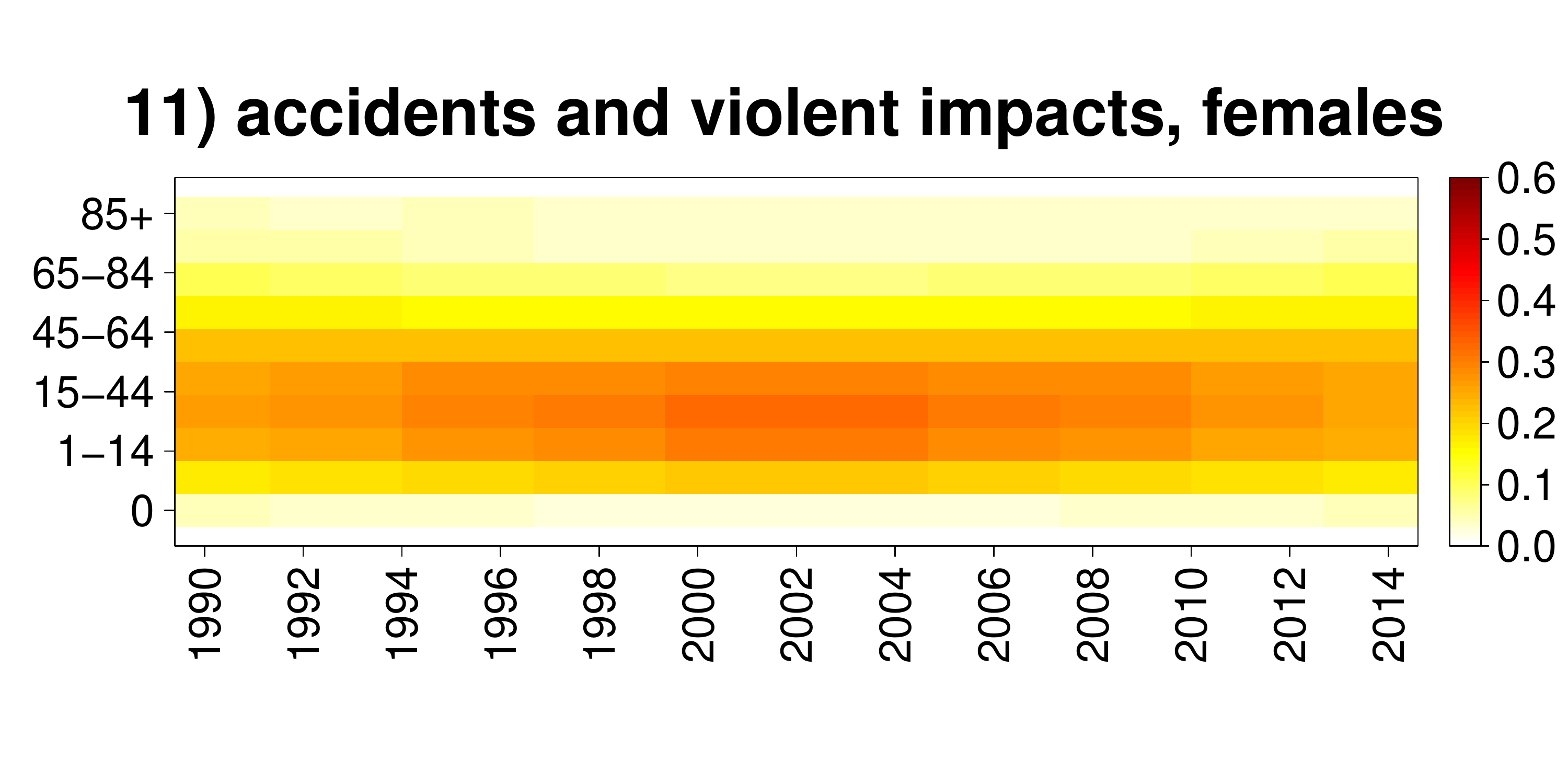}
\hfill
\includegraphics[width=.32\linewidth]{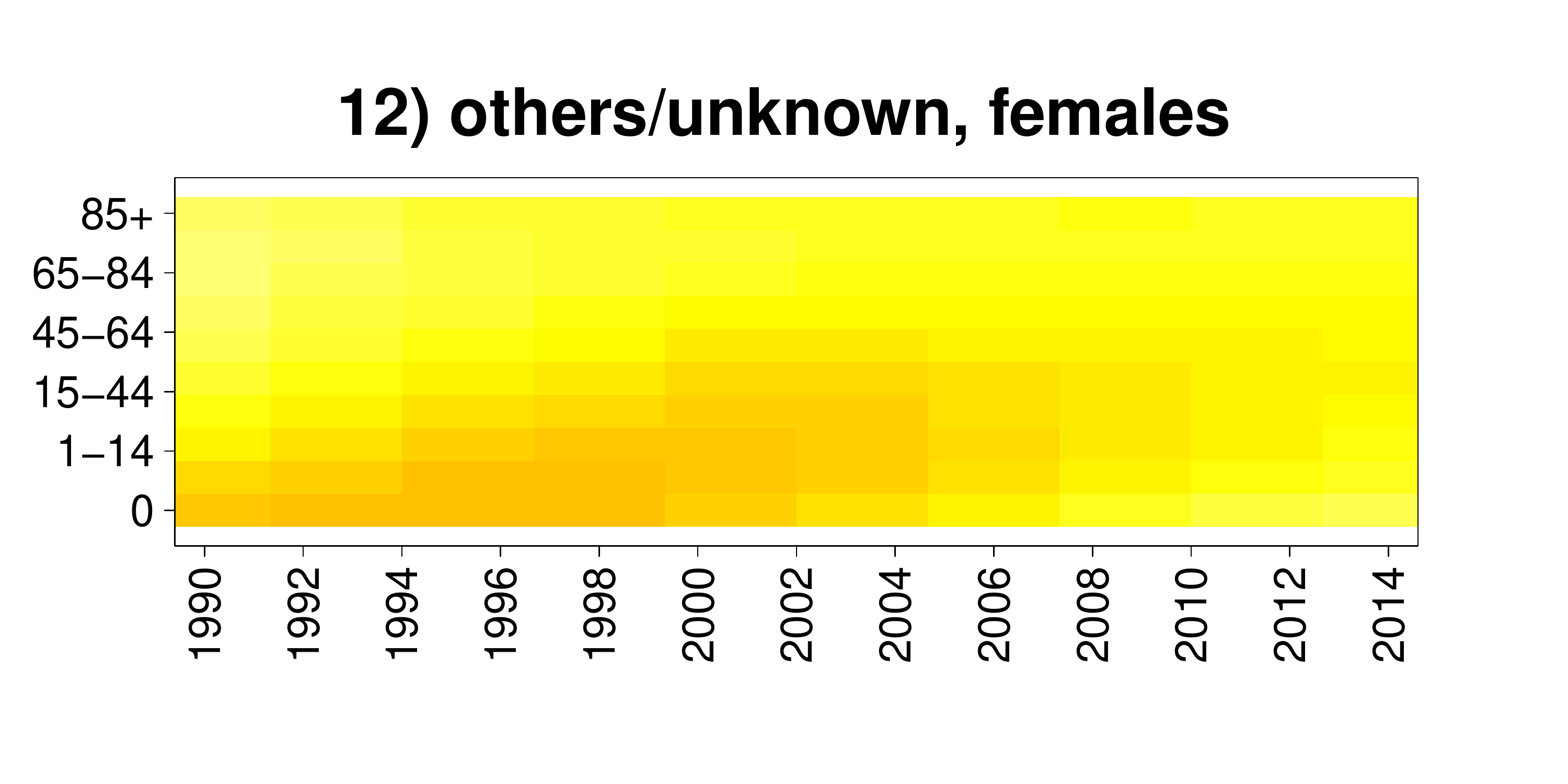}
\\[-0.5cm]
\includegraphics[width=.32\linewidth]{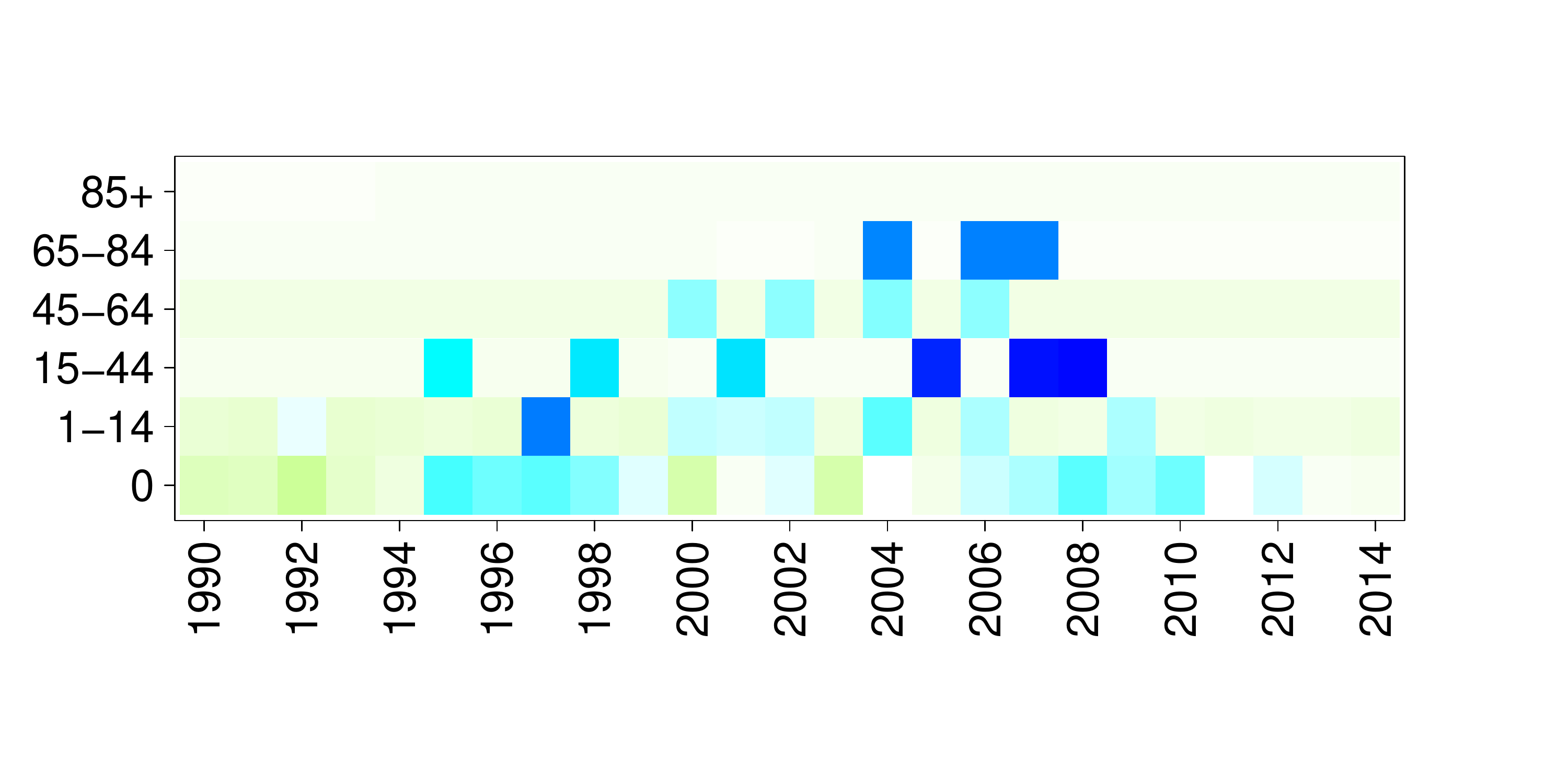}
\hfill
\includegraphics[width=.32\linewidth]{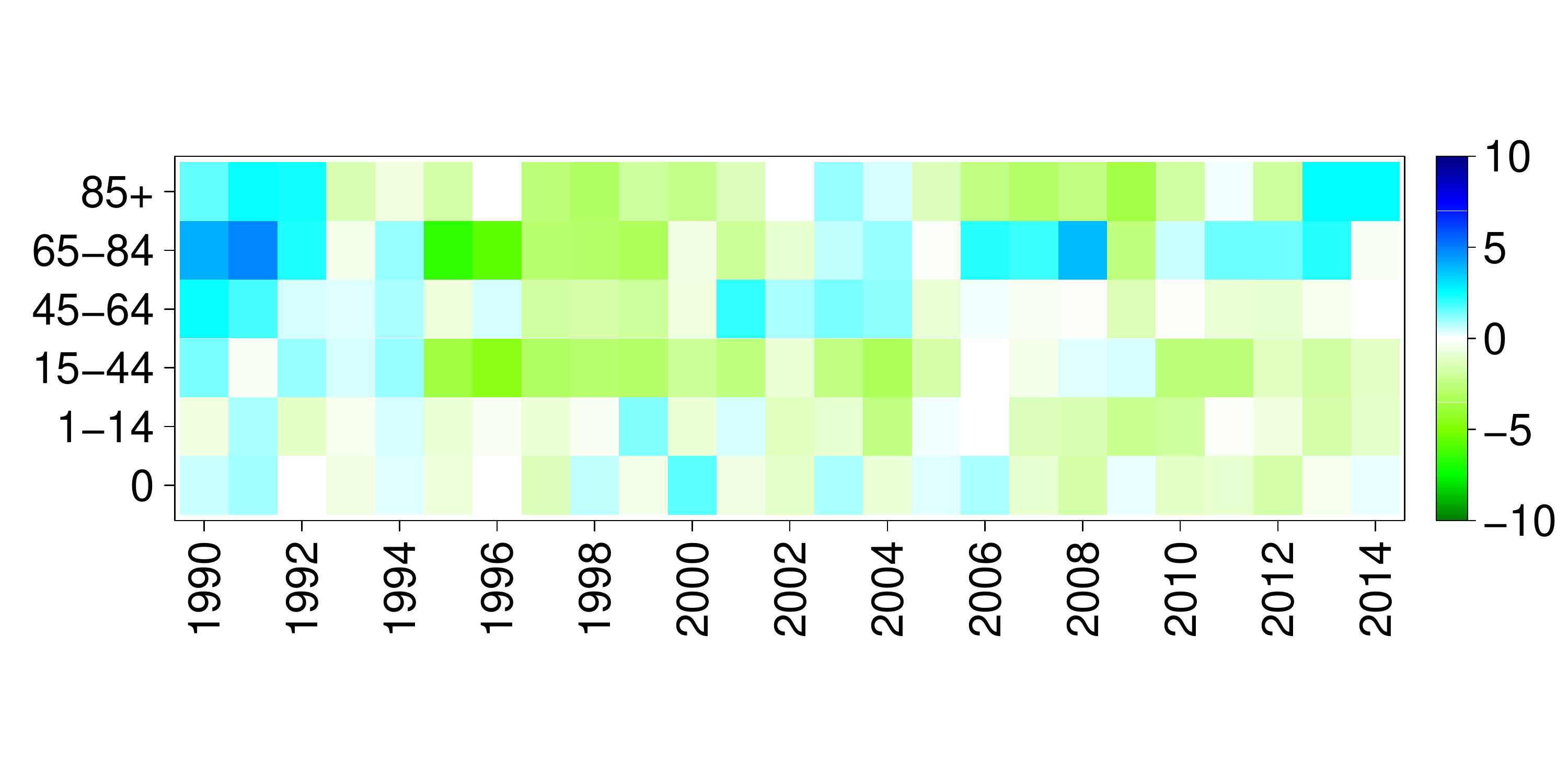}
\hfill
\includegraphics[width=.32\linewidth]{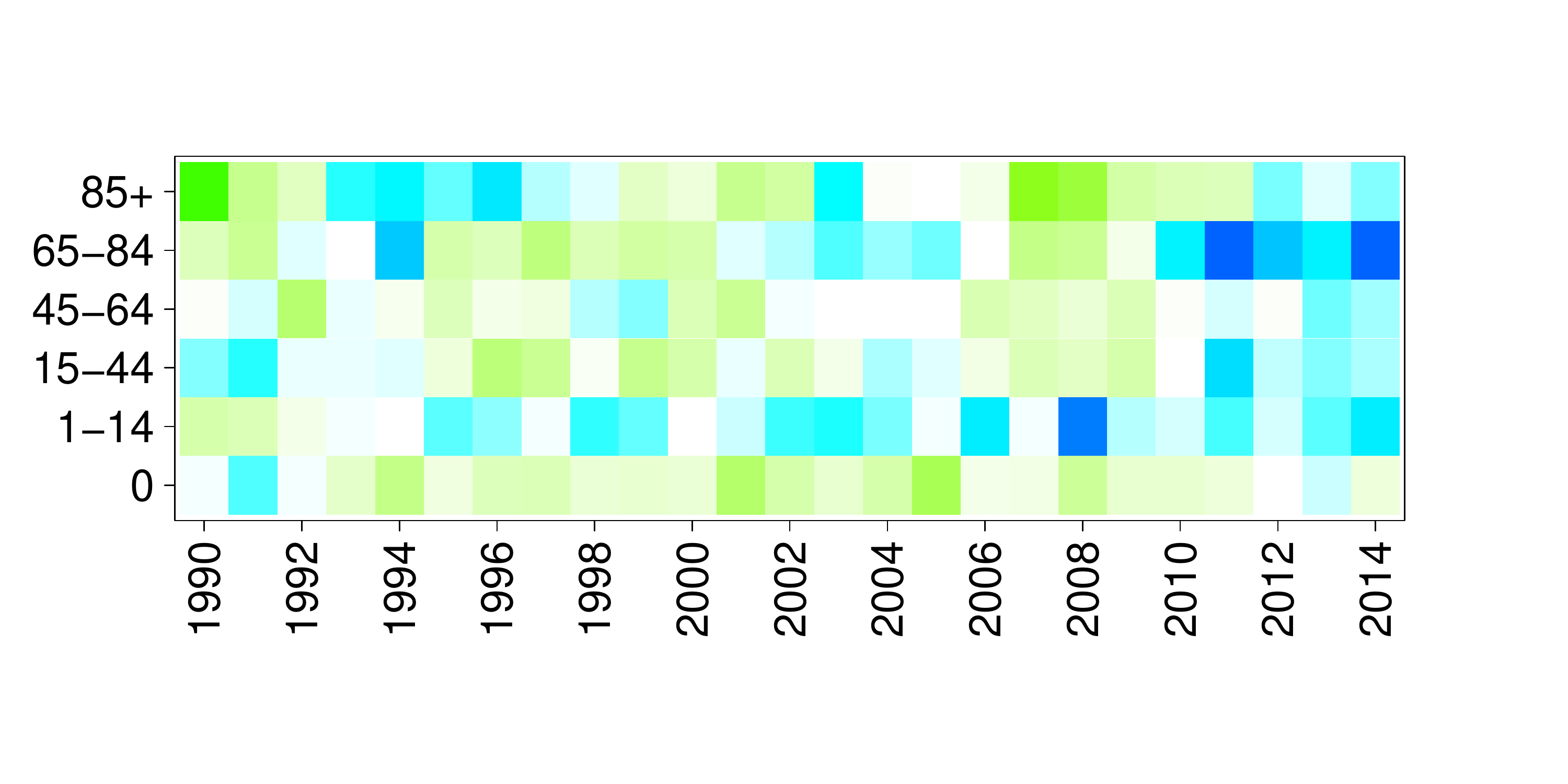}
\caption{\footnotesize
The odd rows illustrate the regression tree estimated probabilities $\theta^\text{tree}(k| \x)$
for females.
These plots all have the same scale given in the middle plot in each odd row.
The even rows show the corresponding Pearson's residuals
given by~\eqref{Equation: residuals}.
These plots all have the same scale given in the middle plot in each even row.
}
\label{Appendix, Figure: death causes, females}
\end{figure}
\begin{figure}[!ht]
\centering
\includegraphics[width=.32\linewidth]{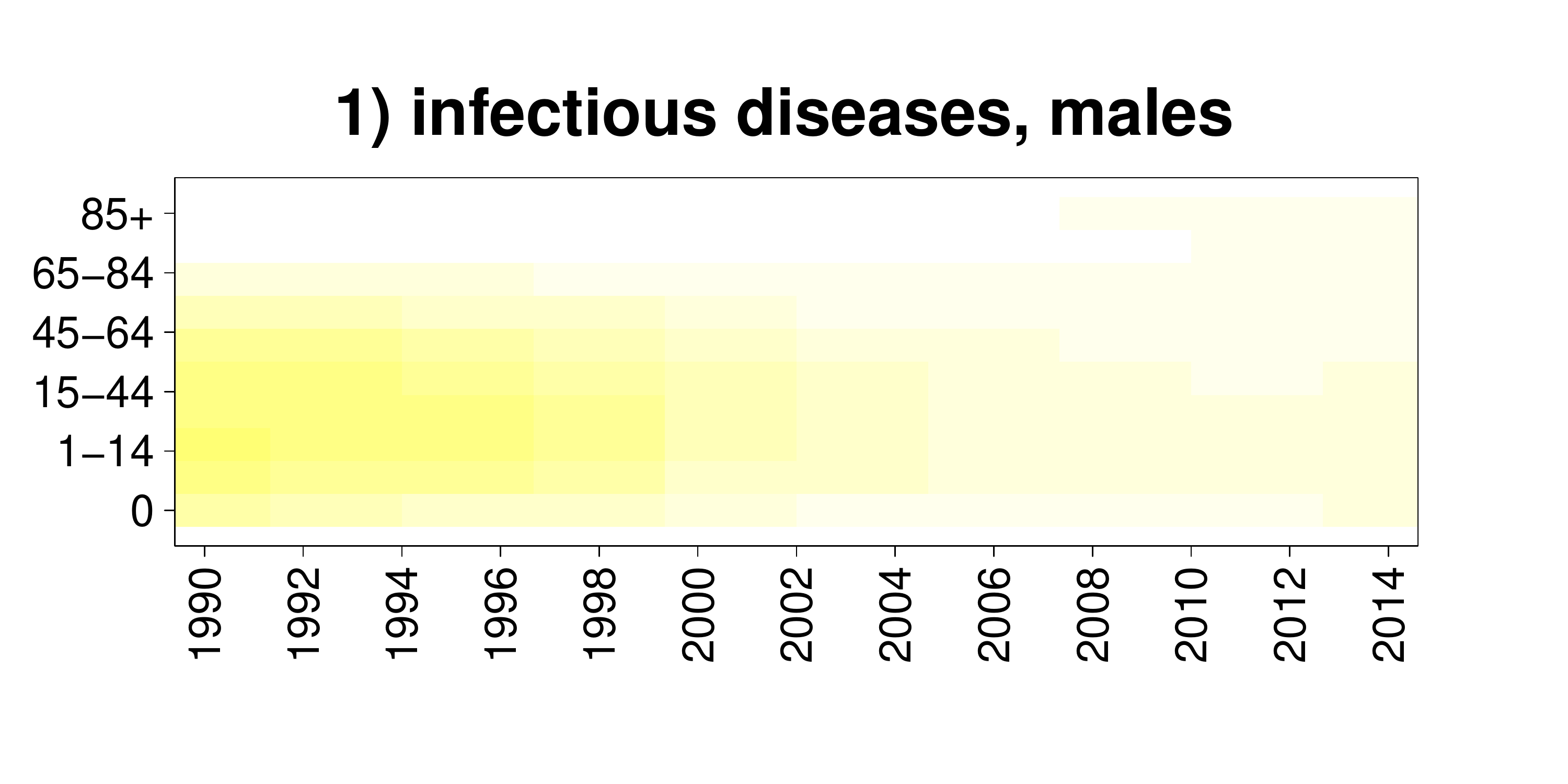}
\hfill
\includegraphics[width=.32\linewidth]{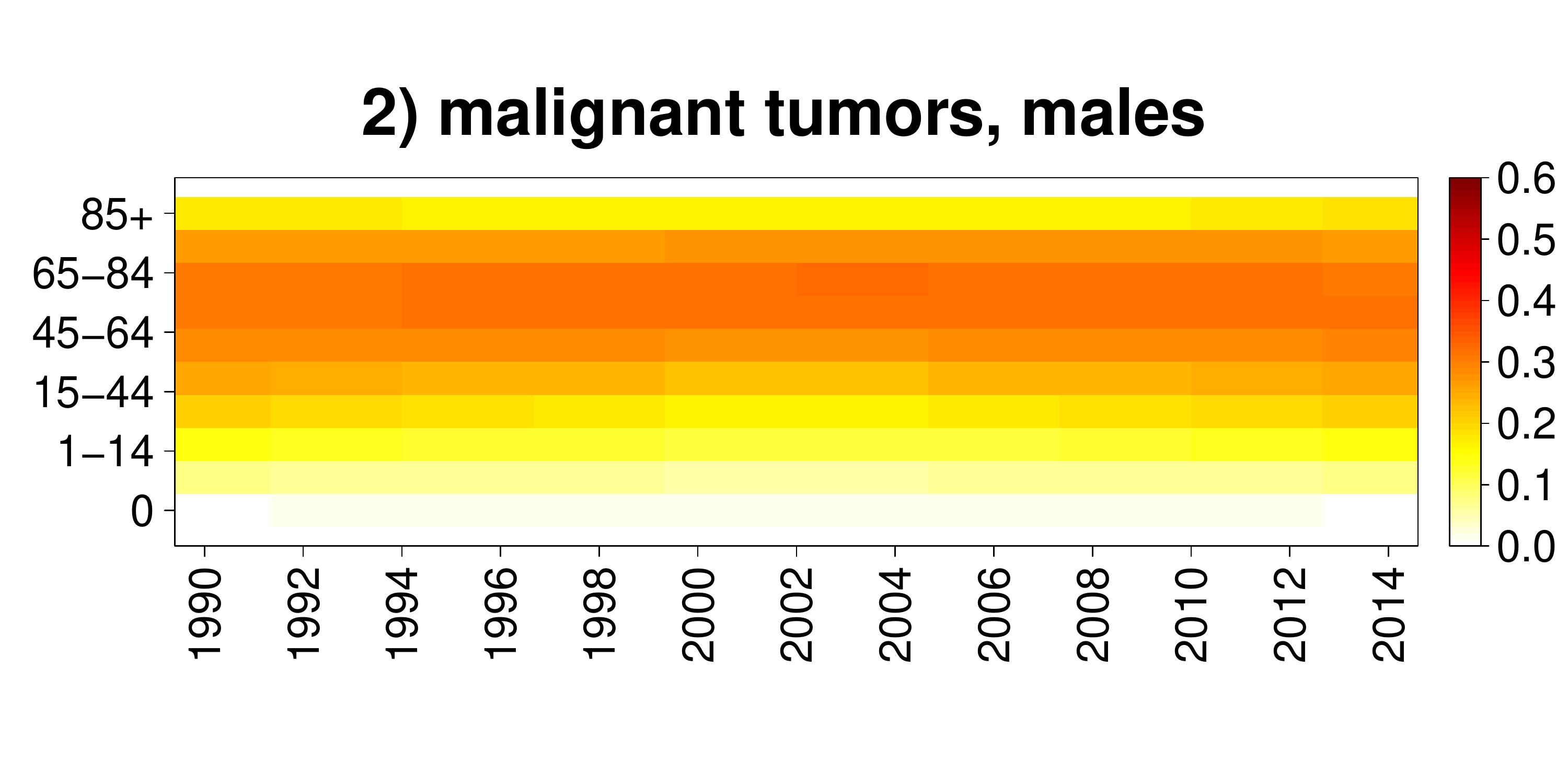}
\hfill
\includegraphics[width=.32\linewidth]{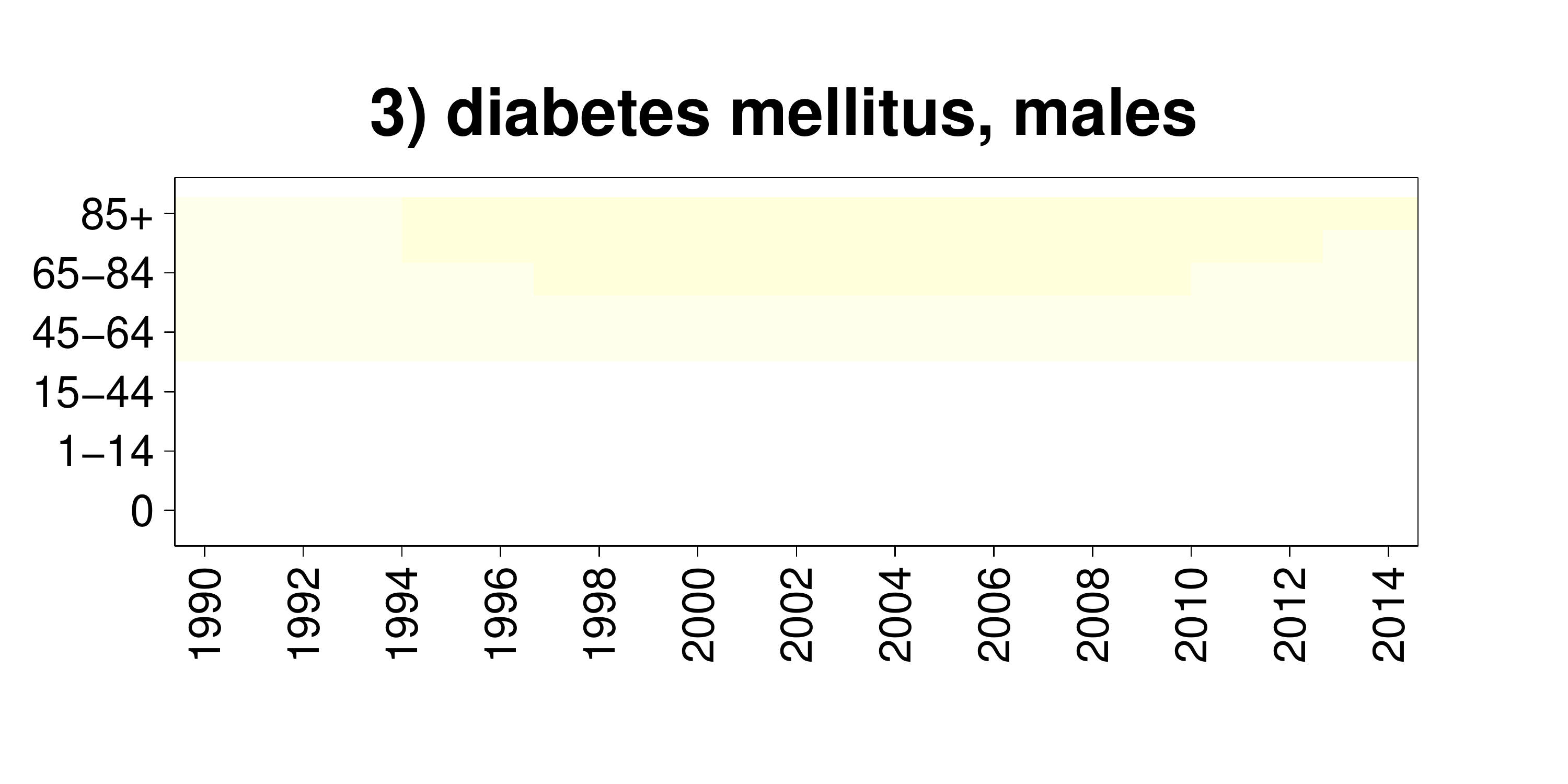}
\\[-0.5cm]
\includegraphics[width=.32\linewidth]{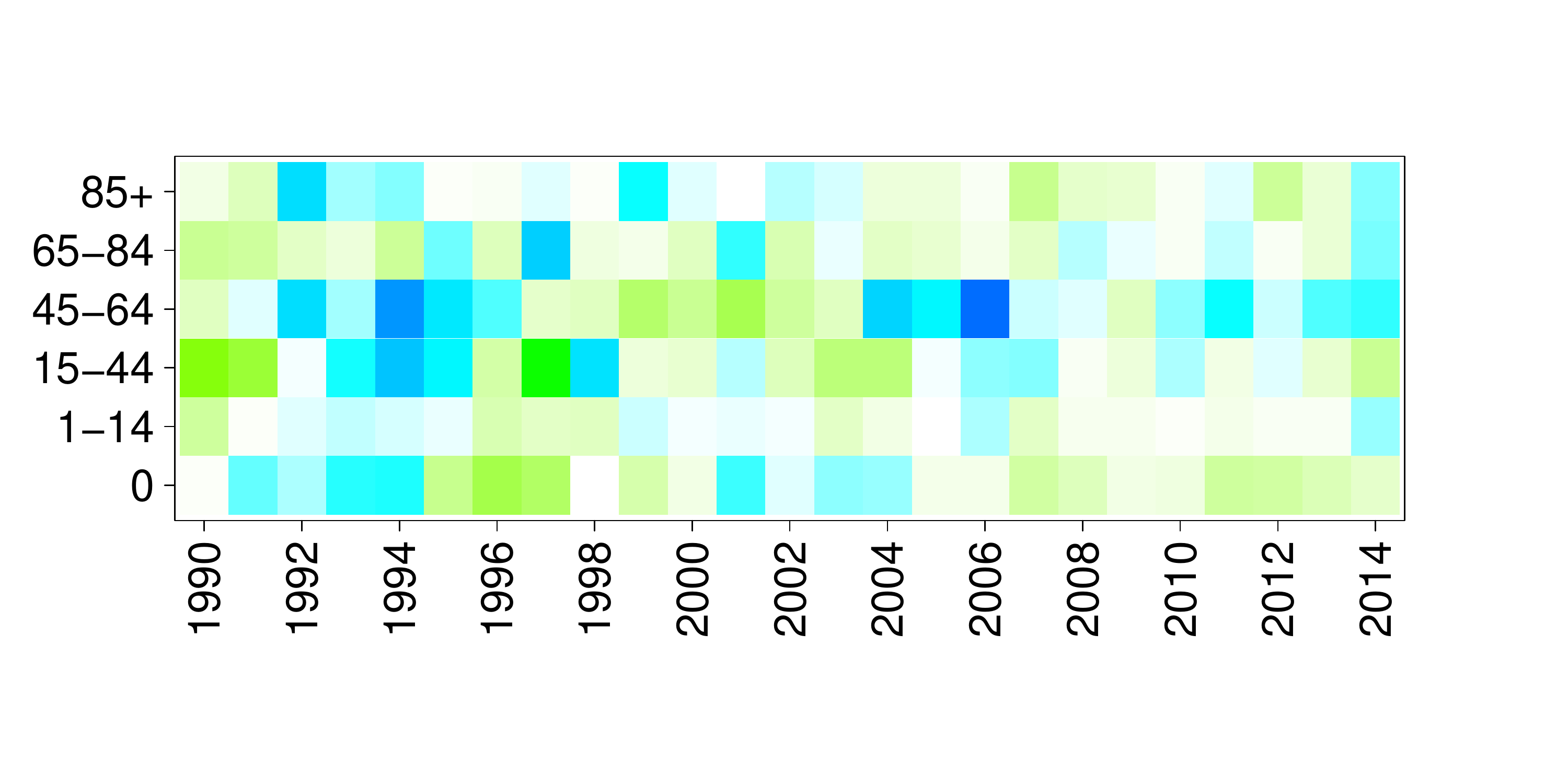}
\hfill
\includegraphics[width=.32\linewidth]{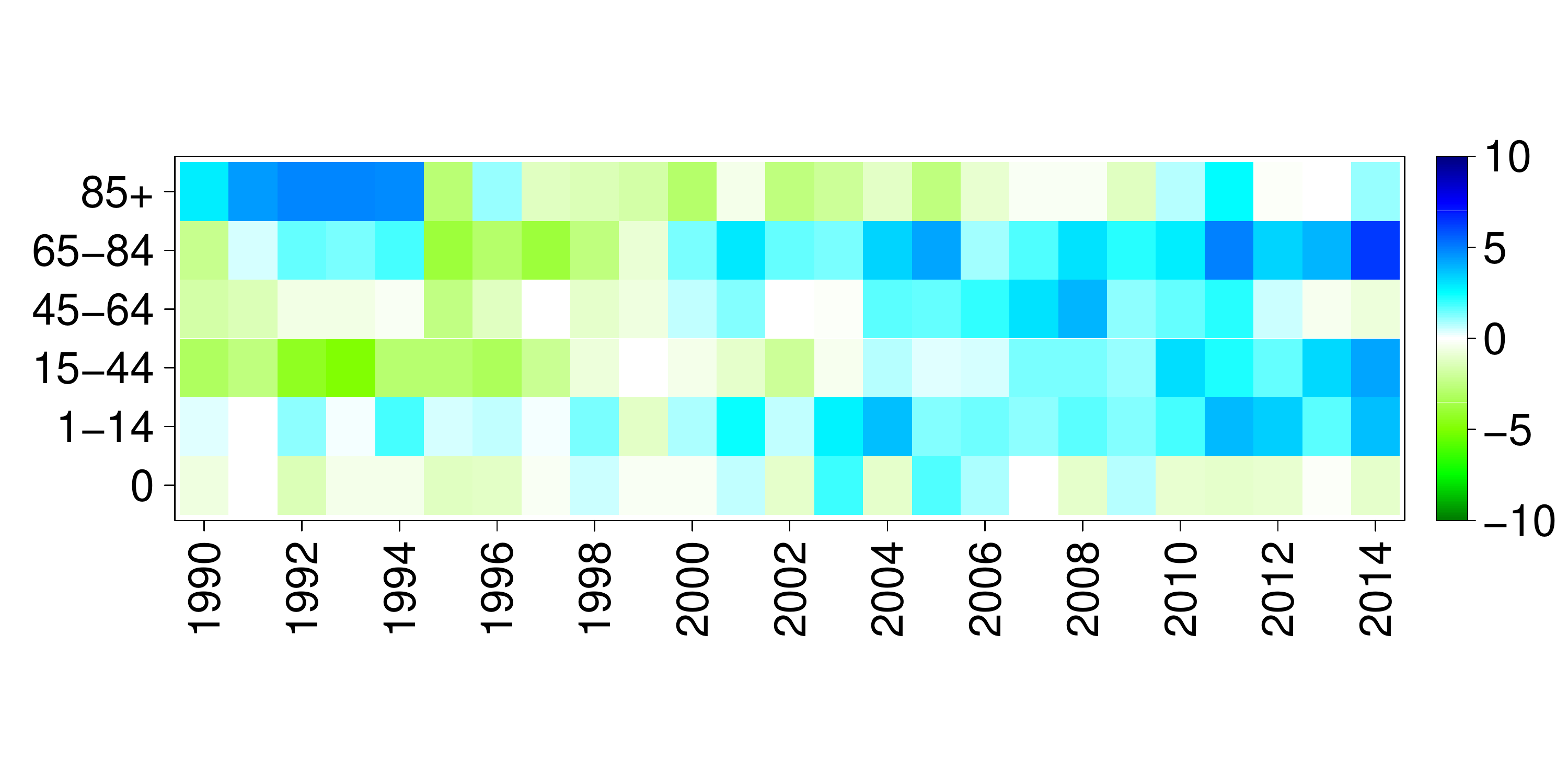}
\hfill
\includegraphics[width=.32\linewidth]{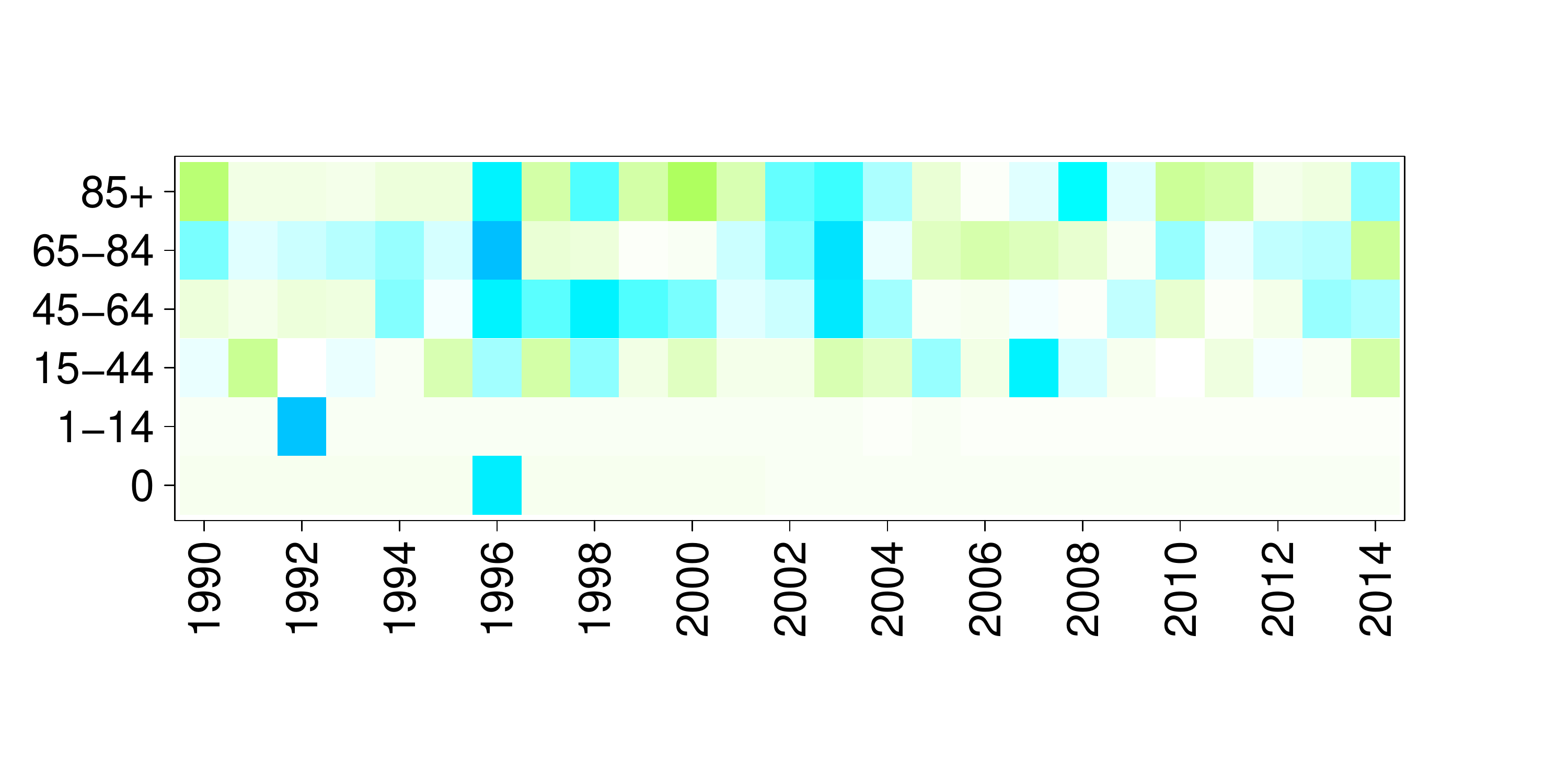}
\\
\hrule
\includegraphics[width=.32\linewidth]{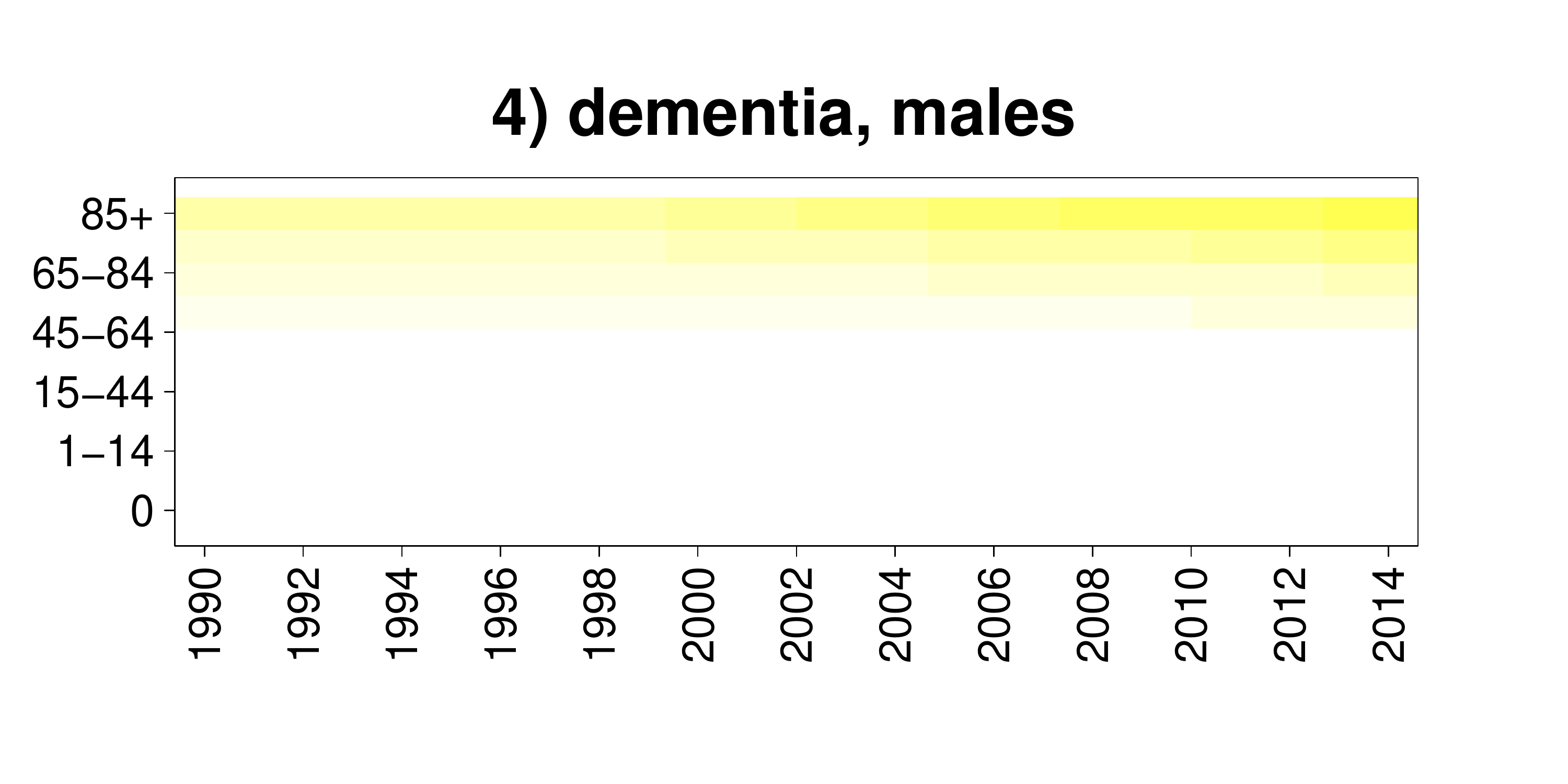}
\hfill
\includegraphics[width=.32\linewidth]{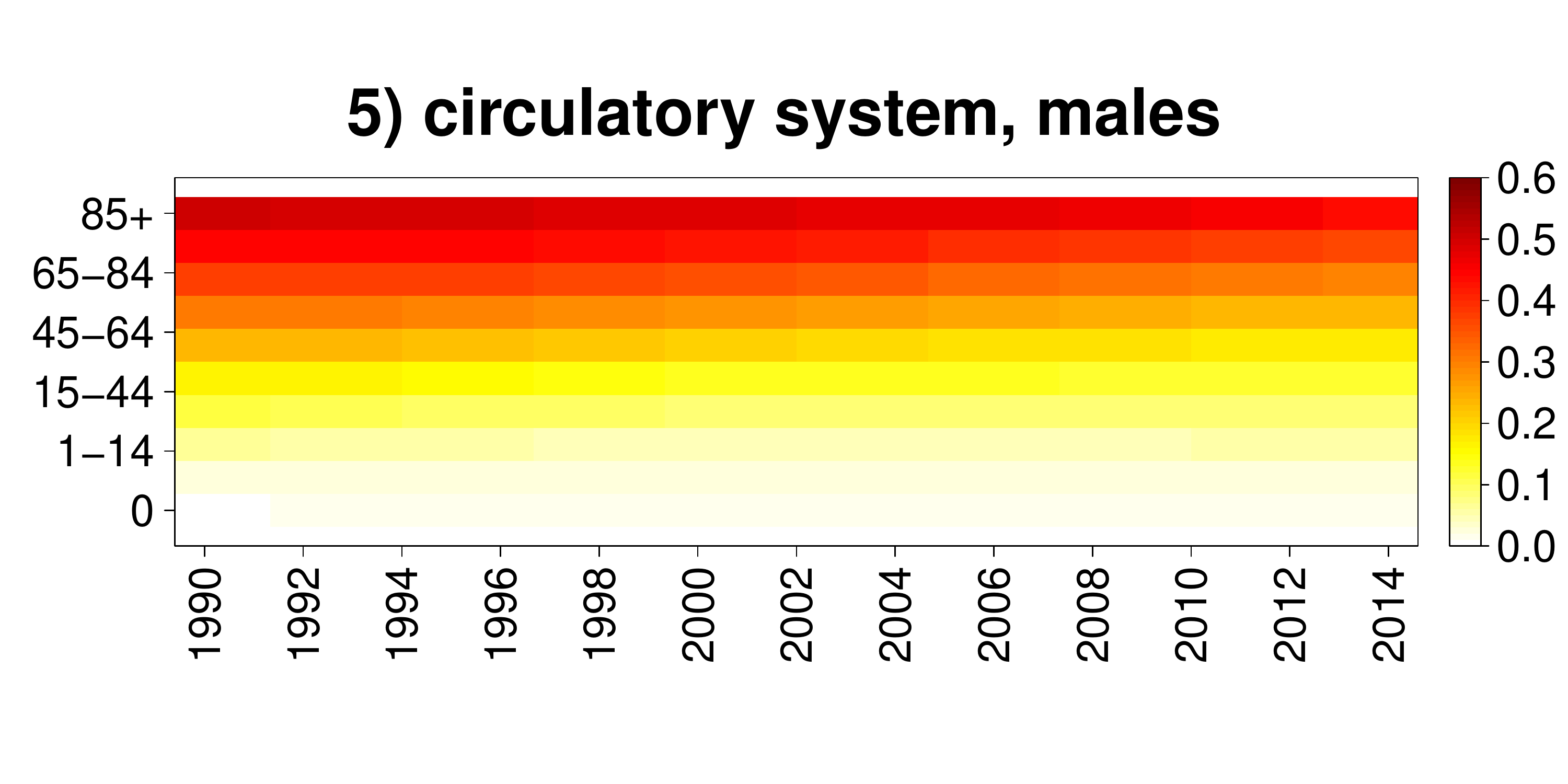}
\hfill
\includegraphics[width=.32\linewidth]{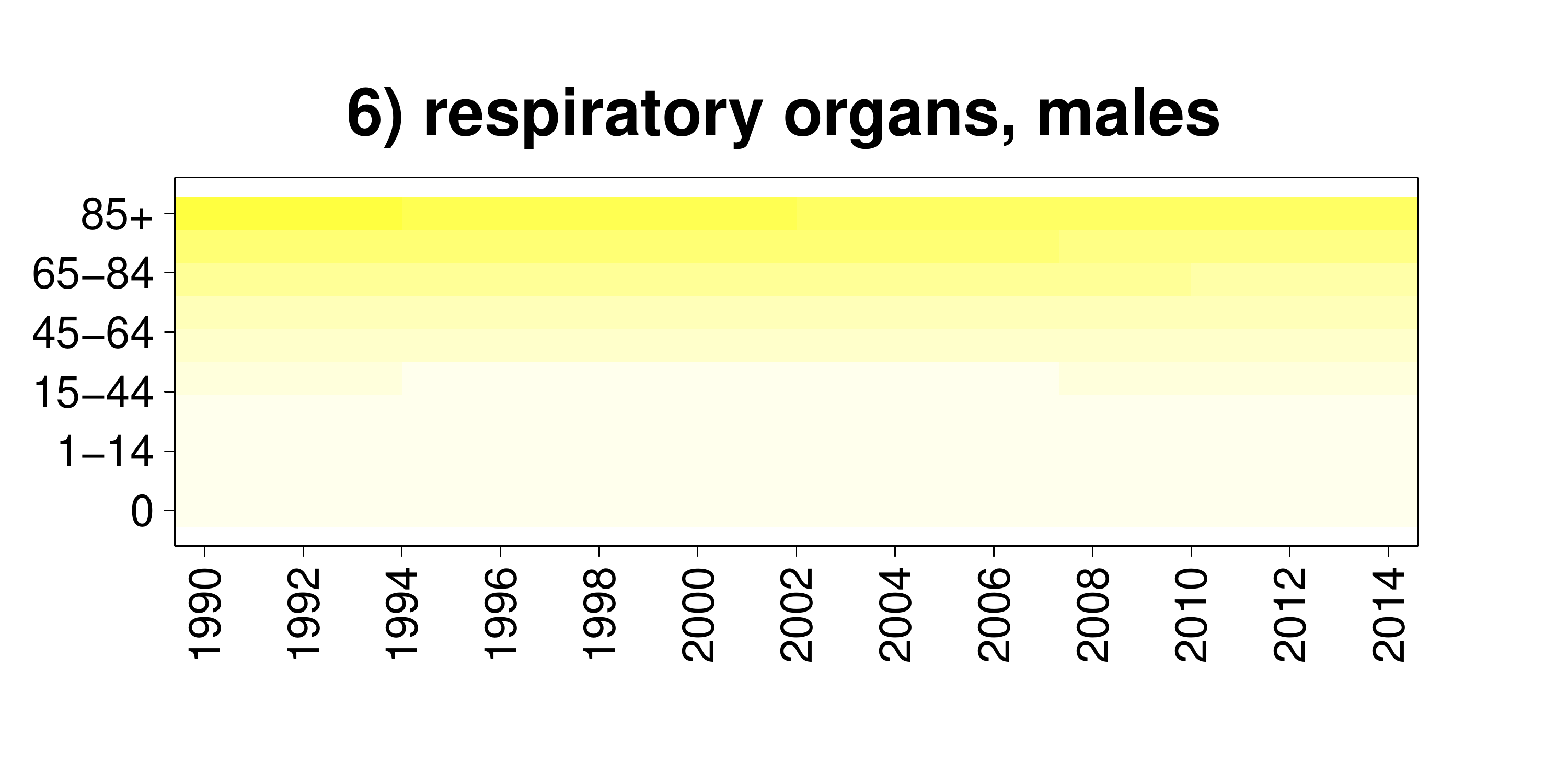}
\\[-0.5cm]
\includegraphics[width=.32\linewidth]{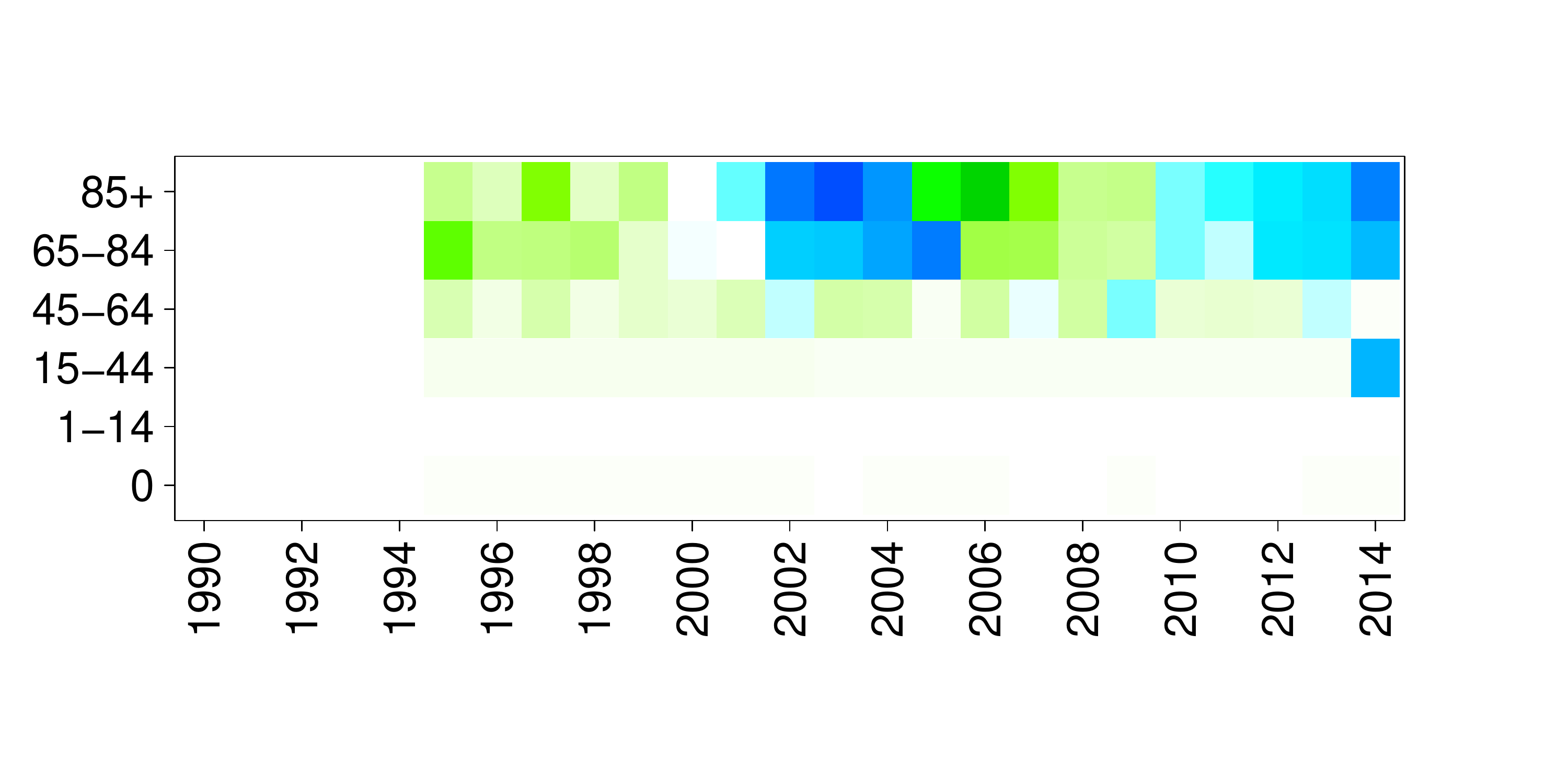}
\hfill
\includegraphics[width=.32\linewidth]{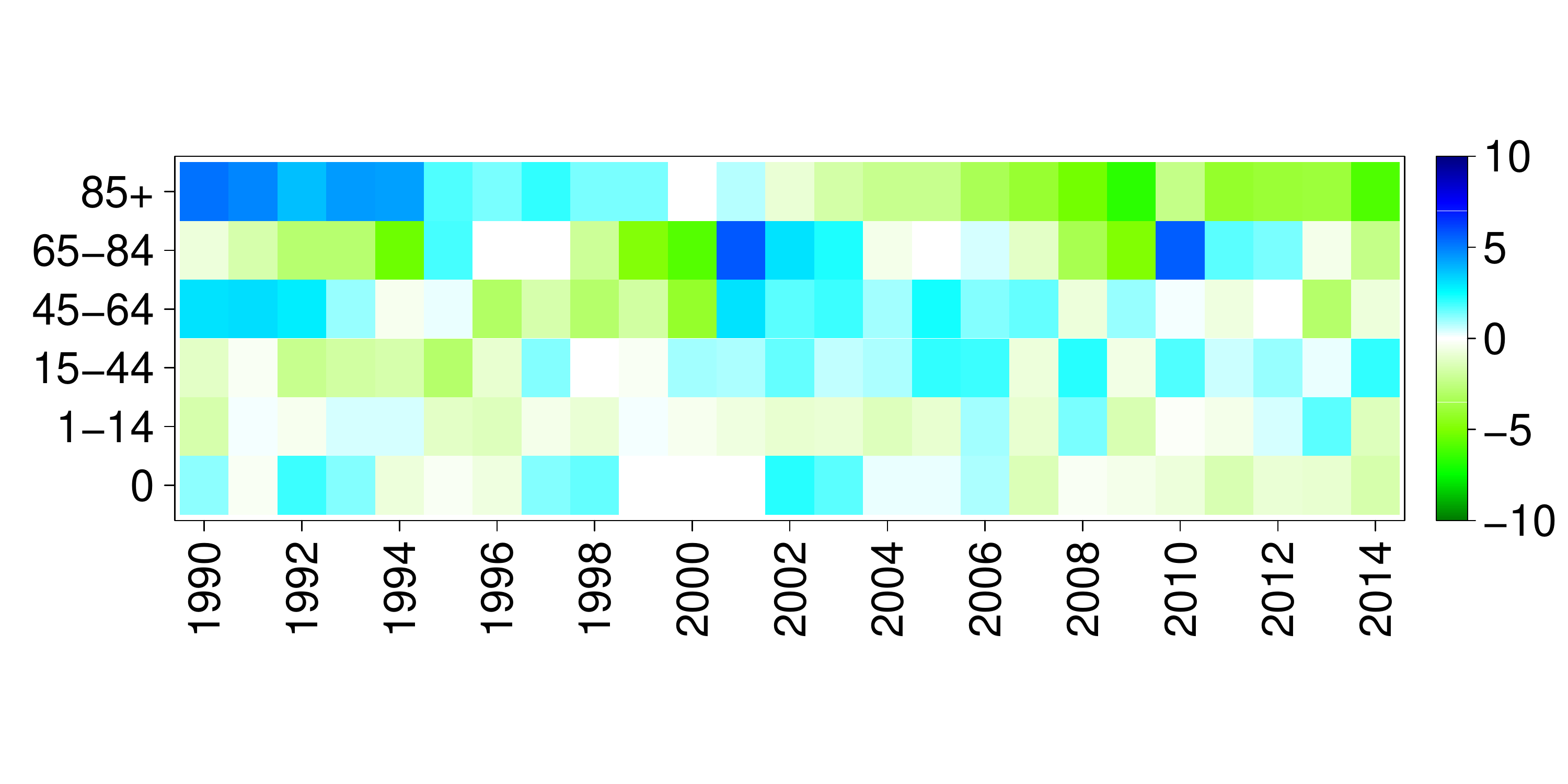}
\hfill
\includegraphics[width=.32\linewidth]{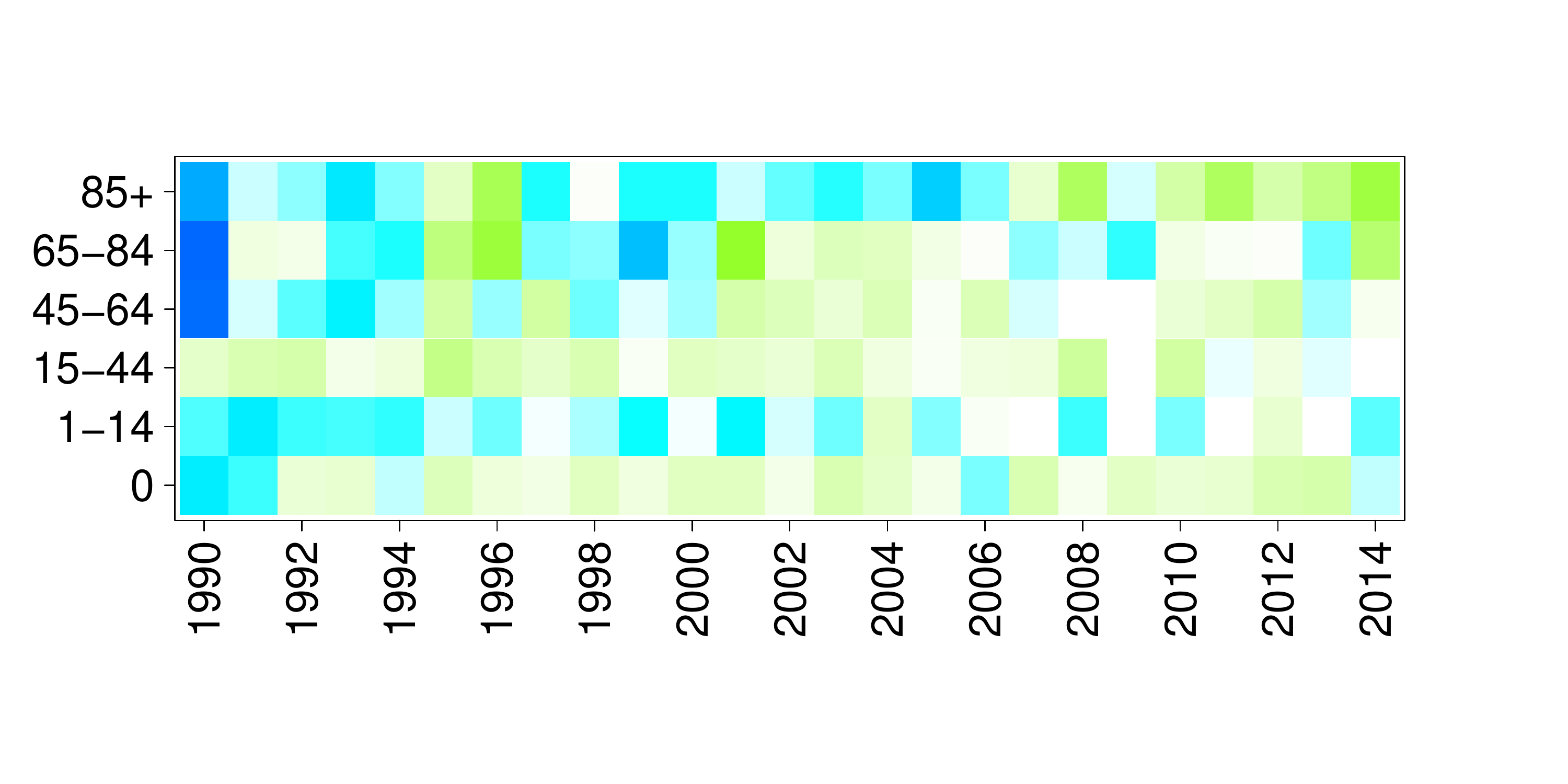}
\\
\hrule
\includegraphics[width=.32\linewidth]{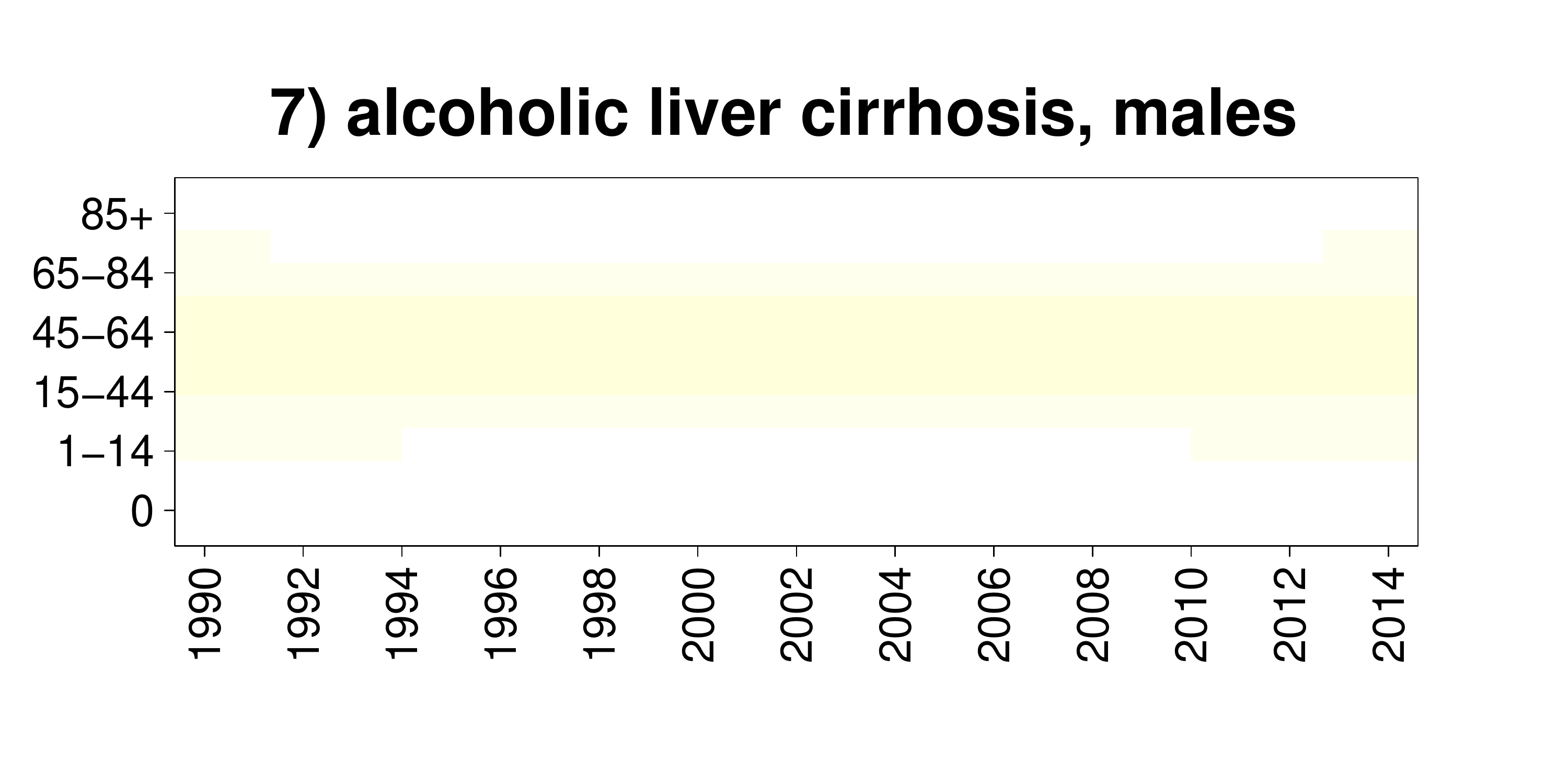}
\hfill
\includegraphics[width=.32\linewidth]{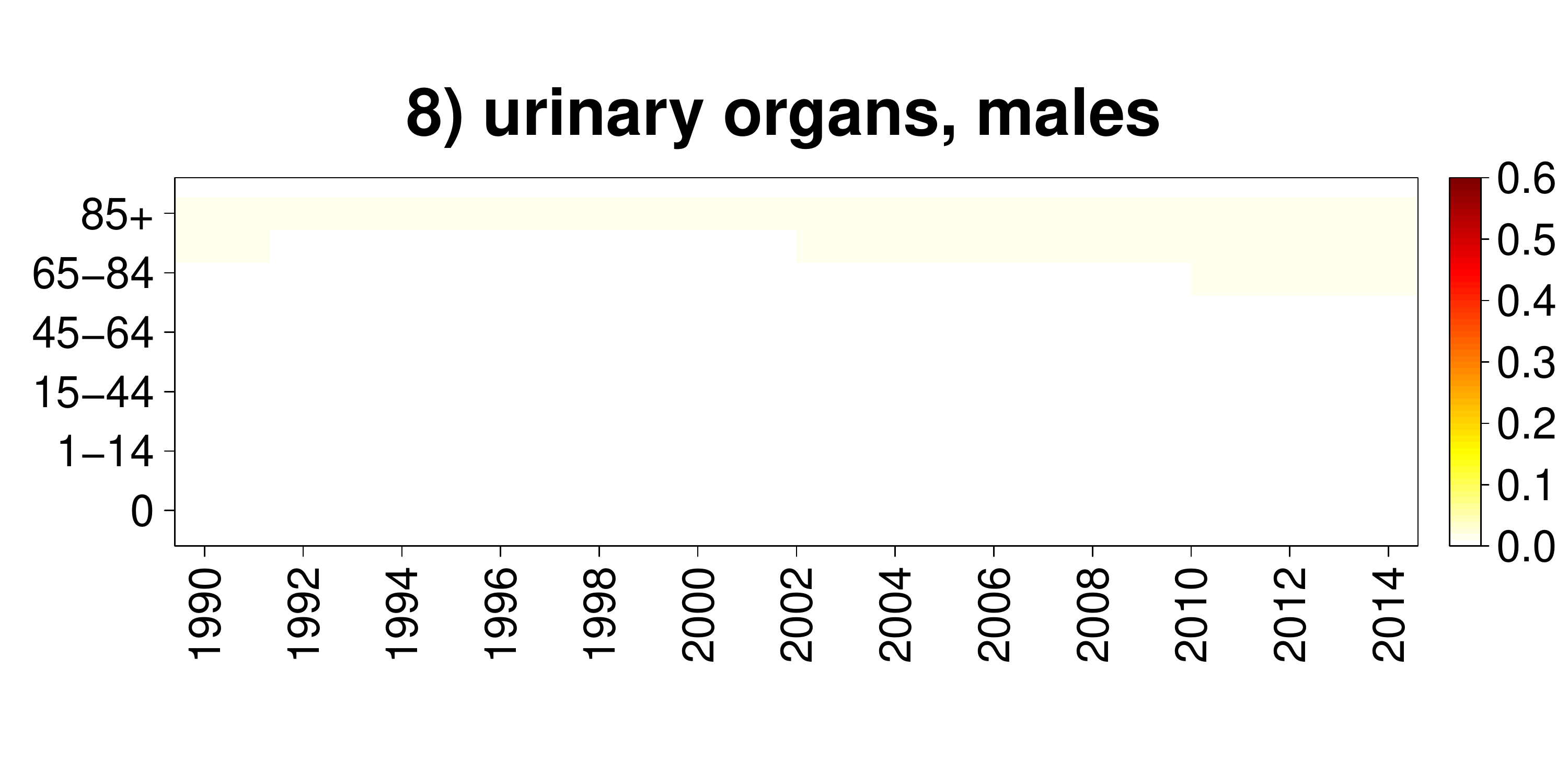}
\hfill
\includegraphics[width=.32\linewidth]{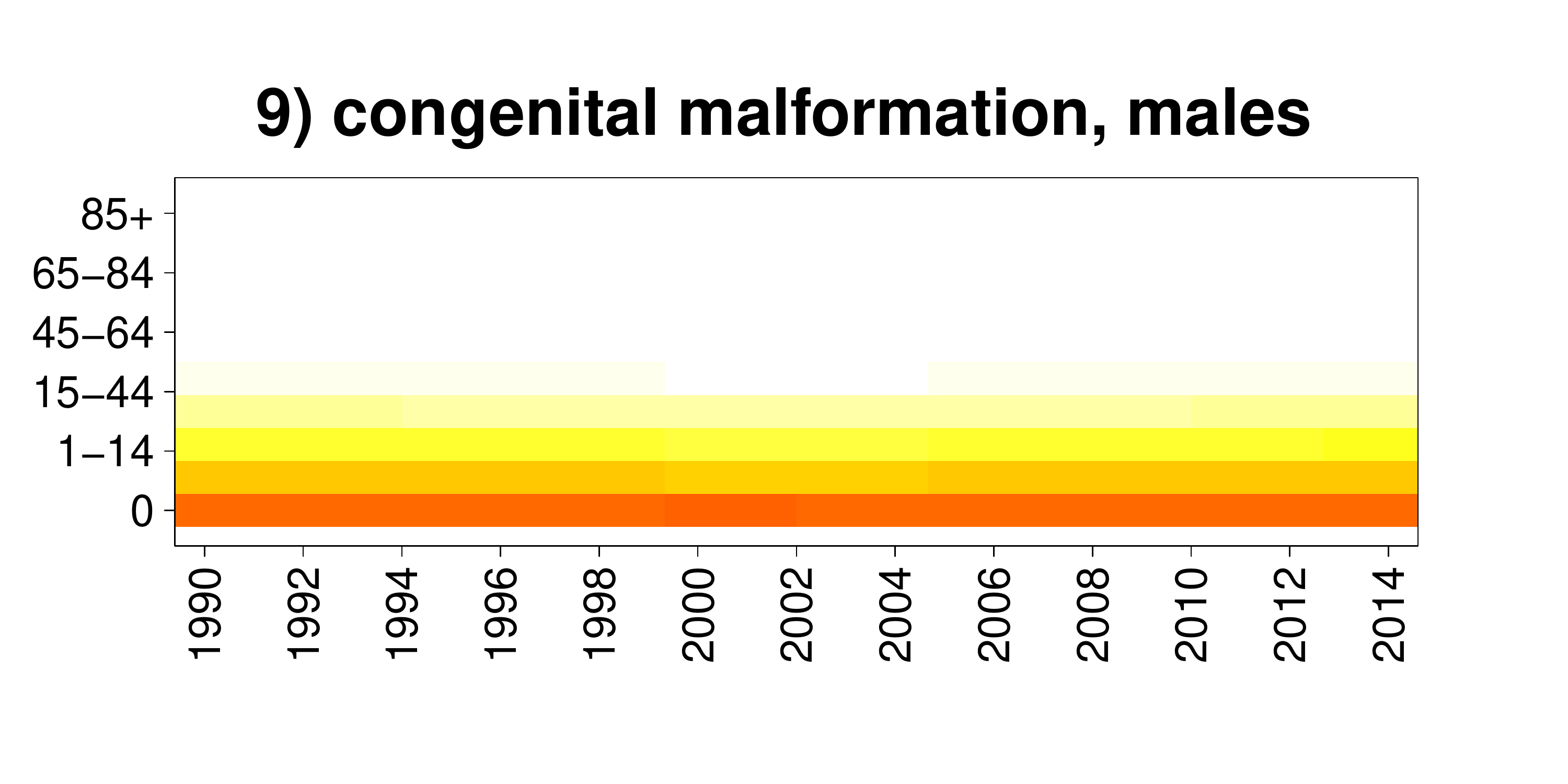}
\\[-0.5cm]
\includegraphics[width=.32\linewidth]{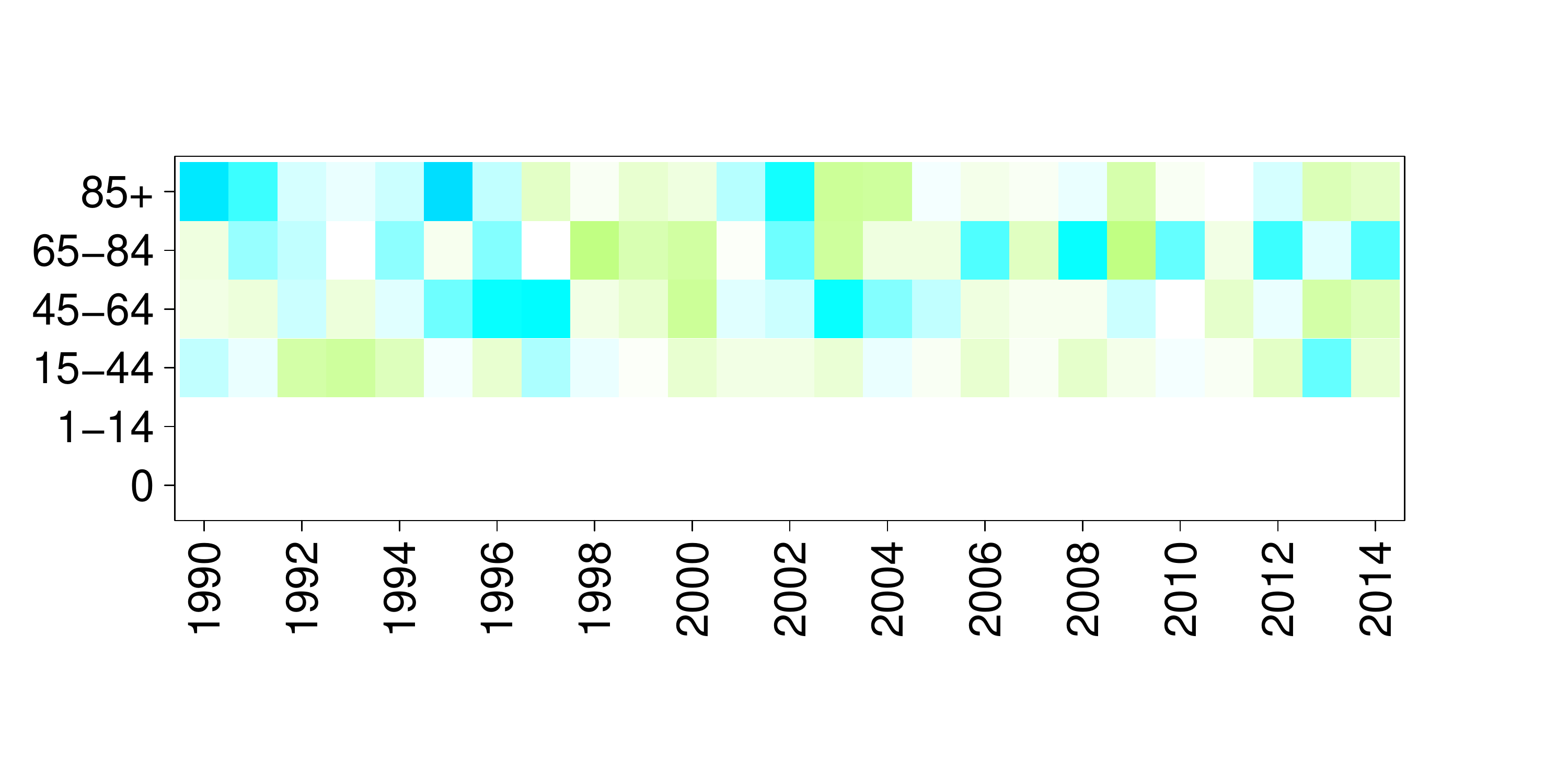}
\hfill
\includegraphics[width=.32\linewidth]{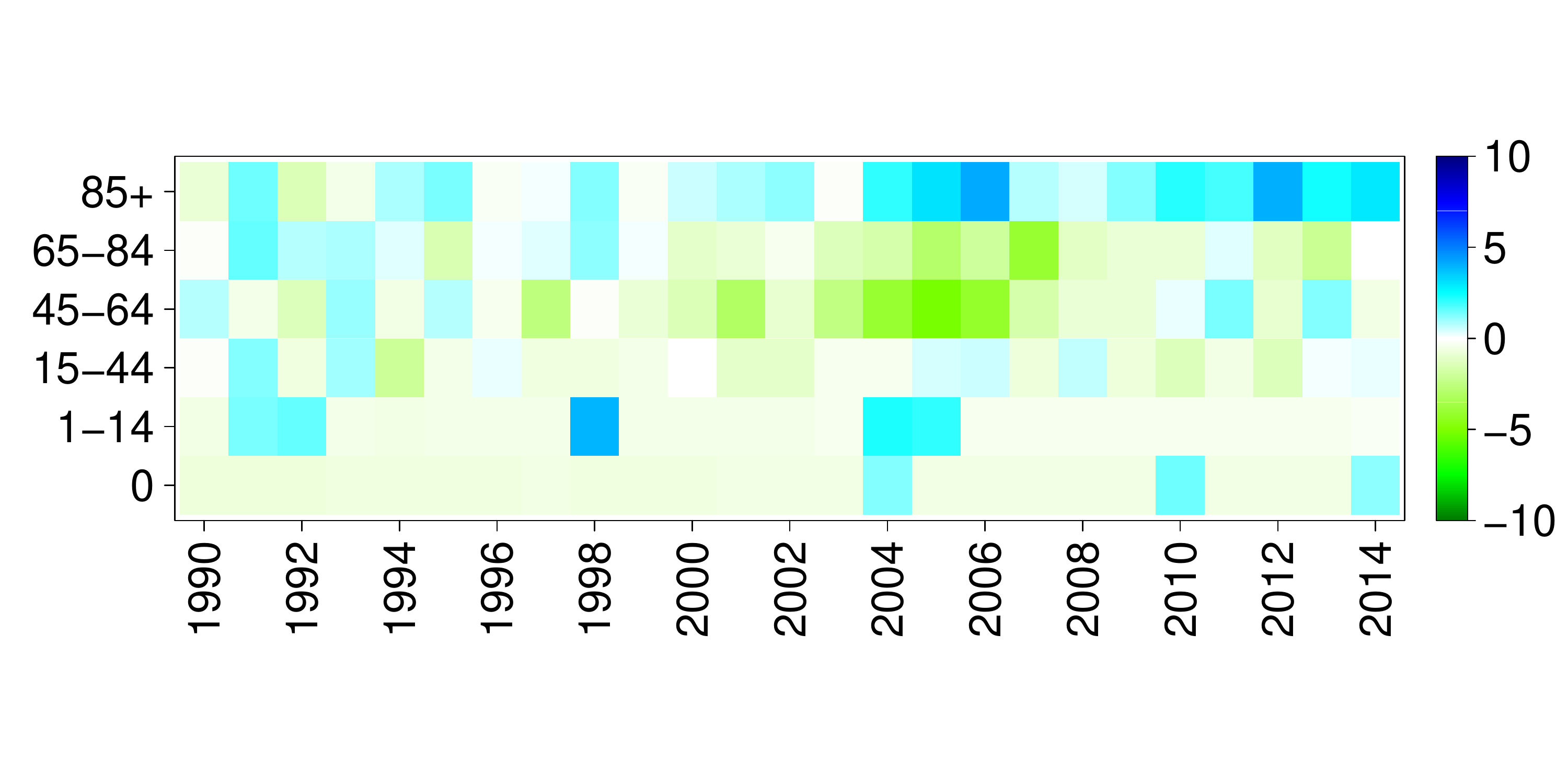}
\hfill
\includegraphics[width=.32\linewidth]{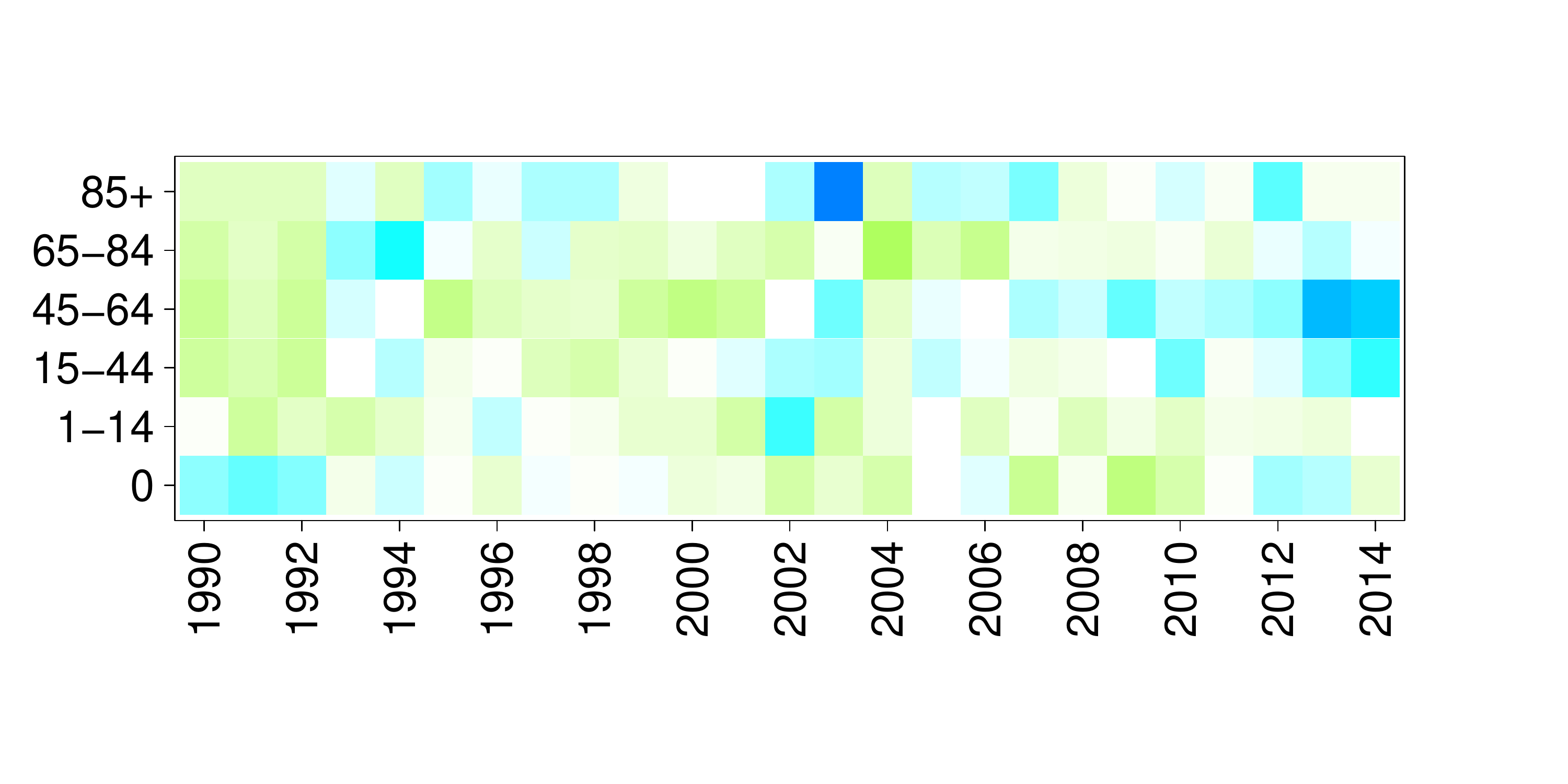}
\\
\hrule
\includegraphics[width=.32\linewidth]{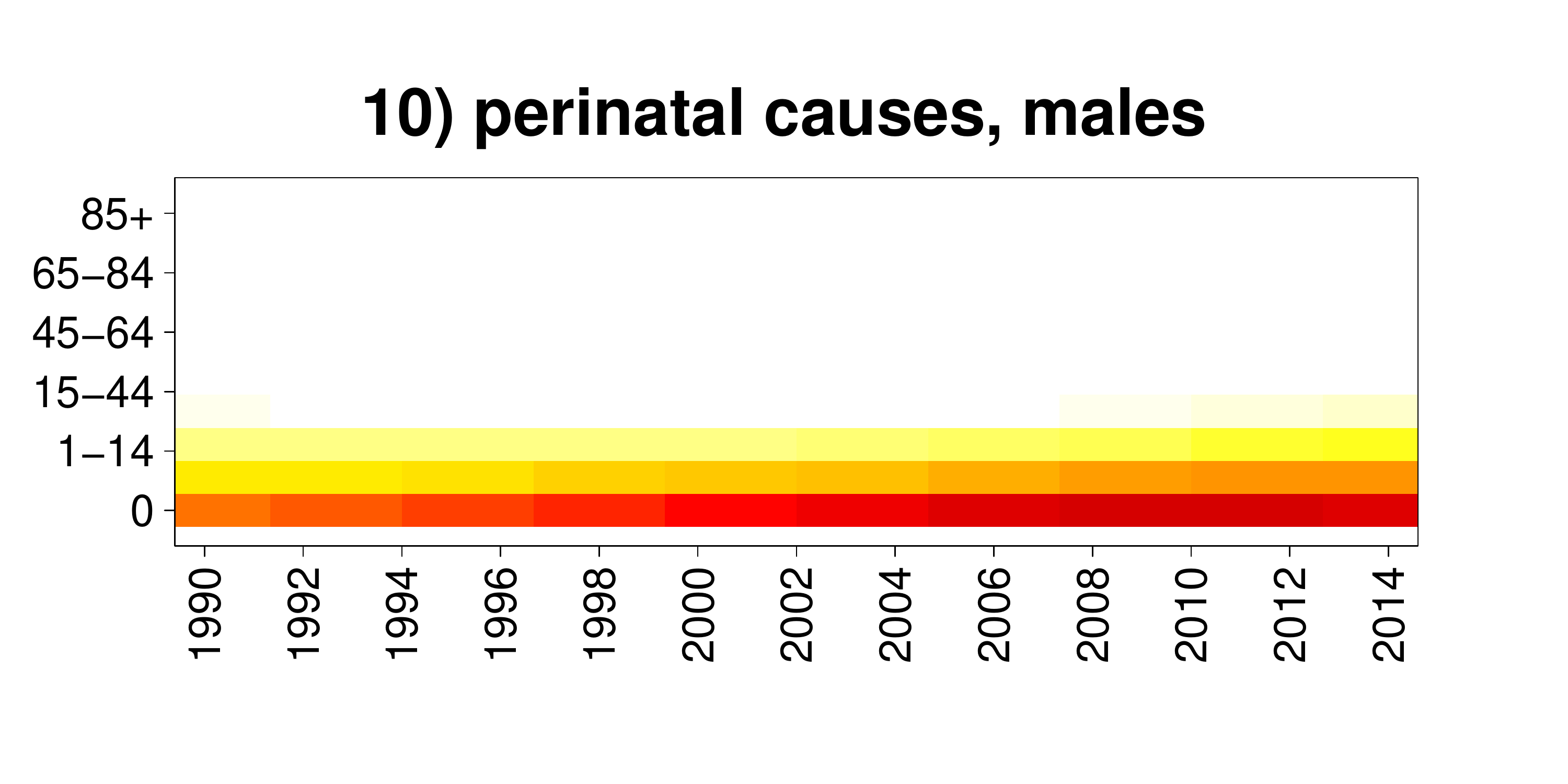}
\hfill
\includegraphics[width=.32\linewidth]{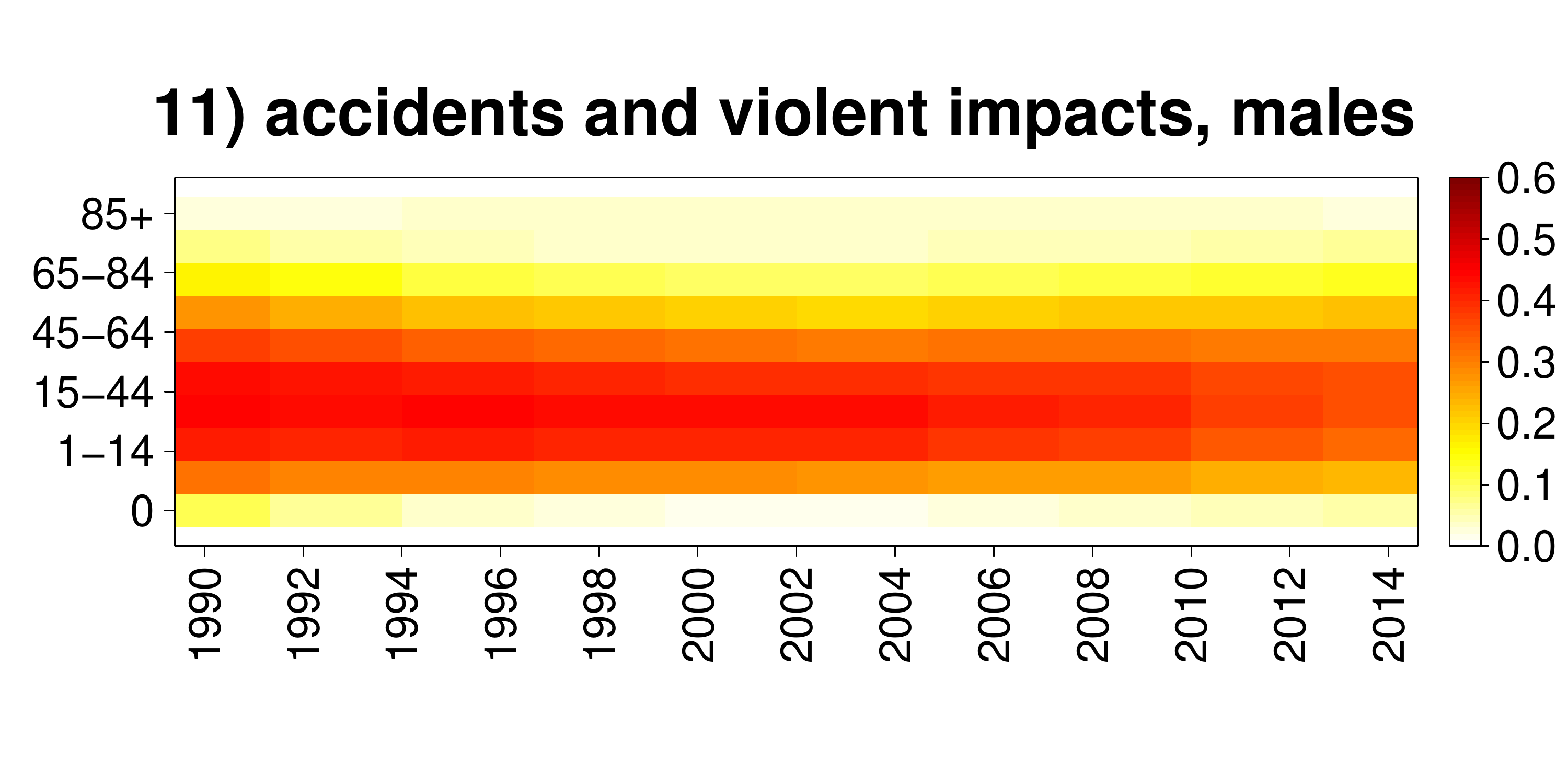}
\hfill
\includegraphics[width=.32\linewidth]{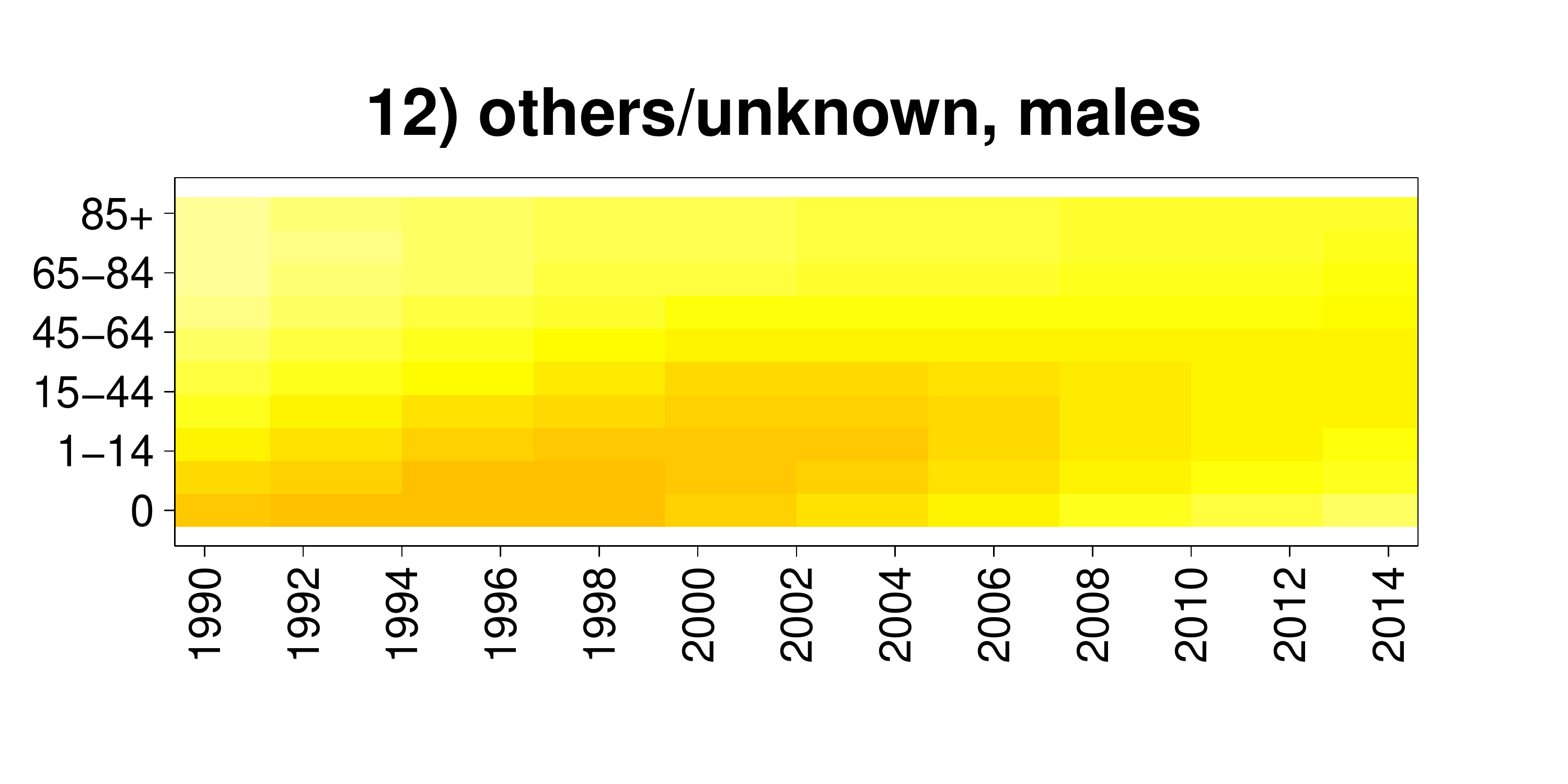}
\\[-0.5cm]
\includegraphics[width=.32\linewidth]{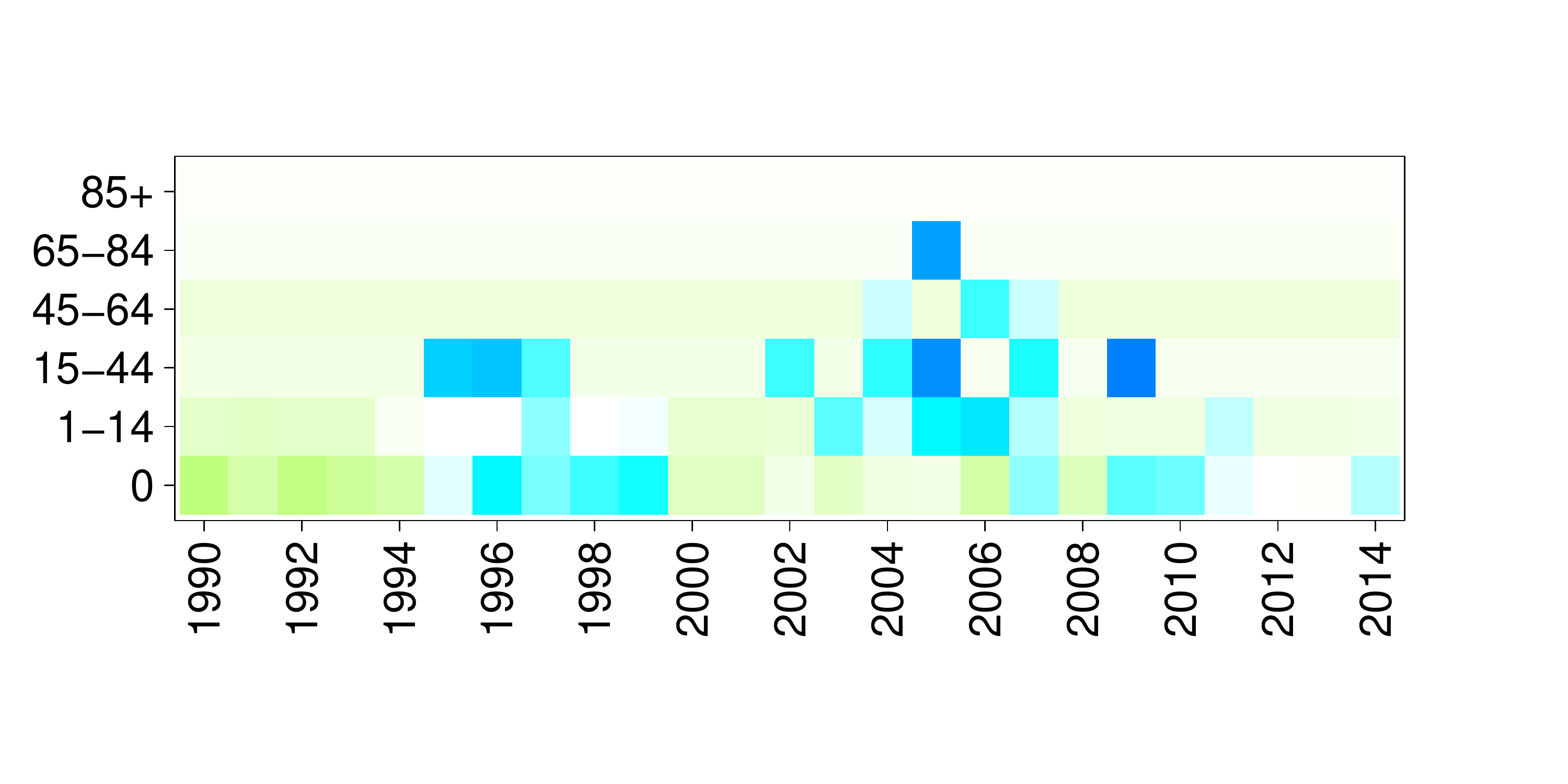}
\hfill
\includegraphics[width=.32\linewidth]{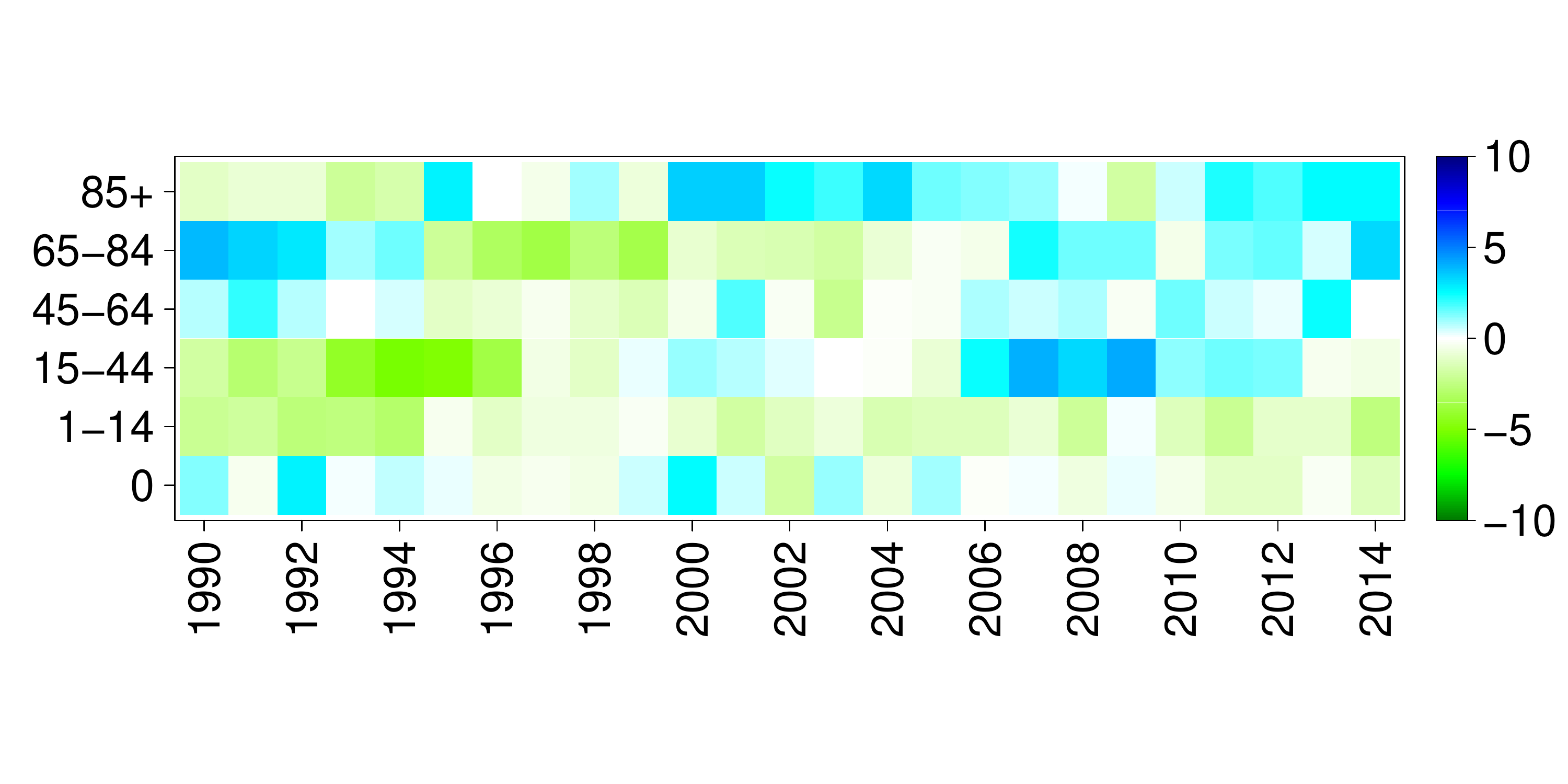}
\hfill
\includegraphics[width=.32\linewidth]{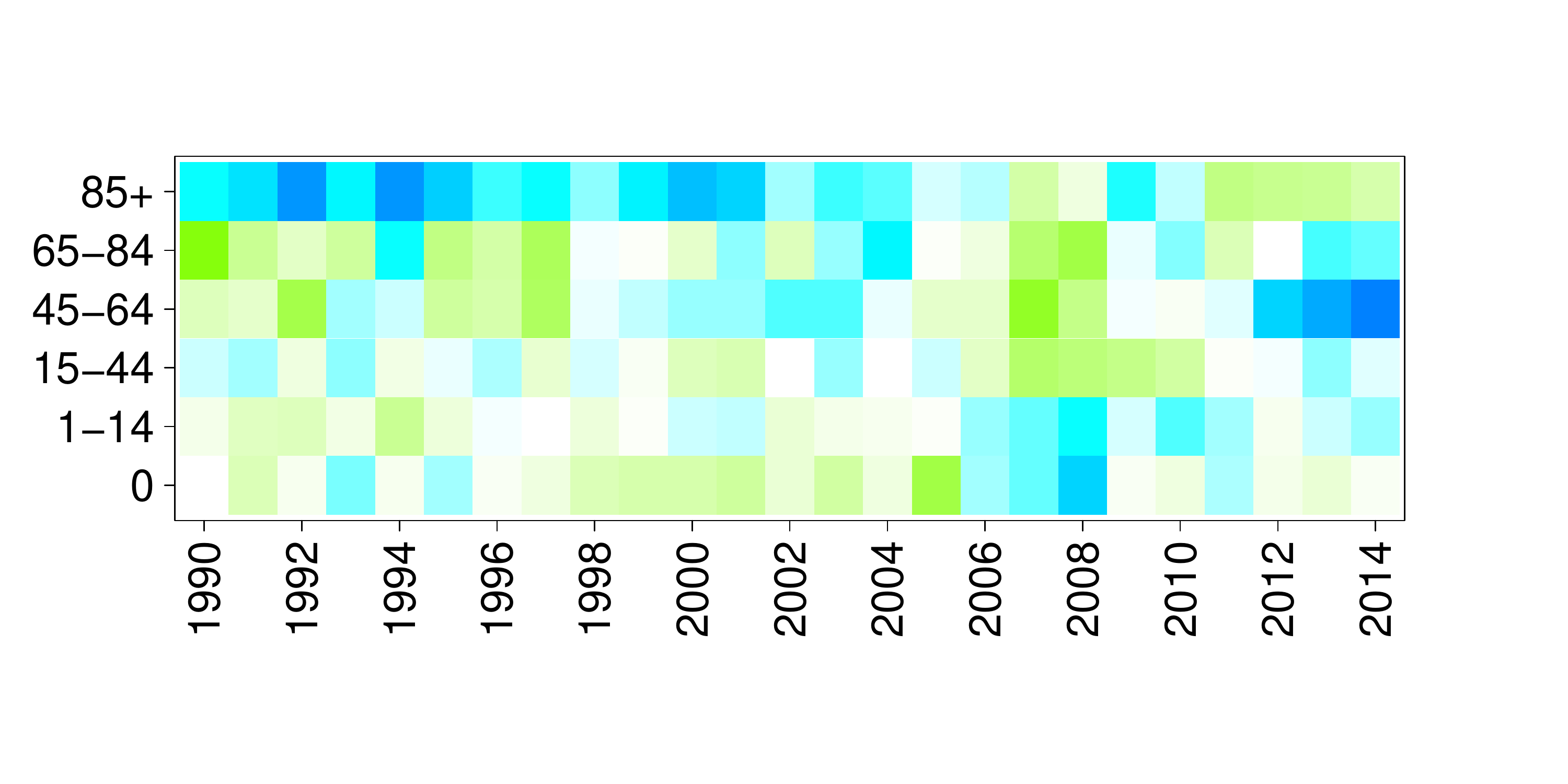}
\caption{\footnotesize
The odd rows illustrate the regression tree estimated probabilities $\theta^\text{tree}(k| \x)$
for males.
These plots all have the same scale given in the middle plot in each odd row.
The even rows show the corresponding Pearson's residuals
given by~\eqref{Equation: residuals}.
These plots all have the same scale given in the middle plot in each even row.
}
\label{Appendix, Figure: death causes, males}
\end{figure}
\begin{figure}
\centering
\includegraphics[width=.425\linewidth]{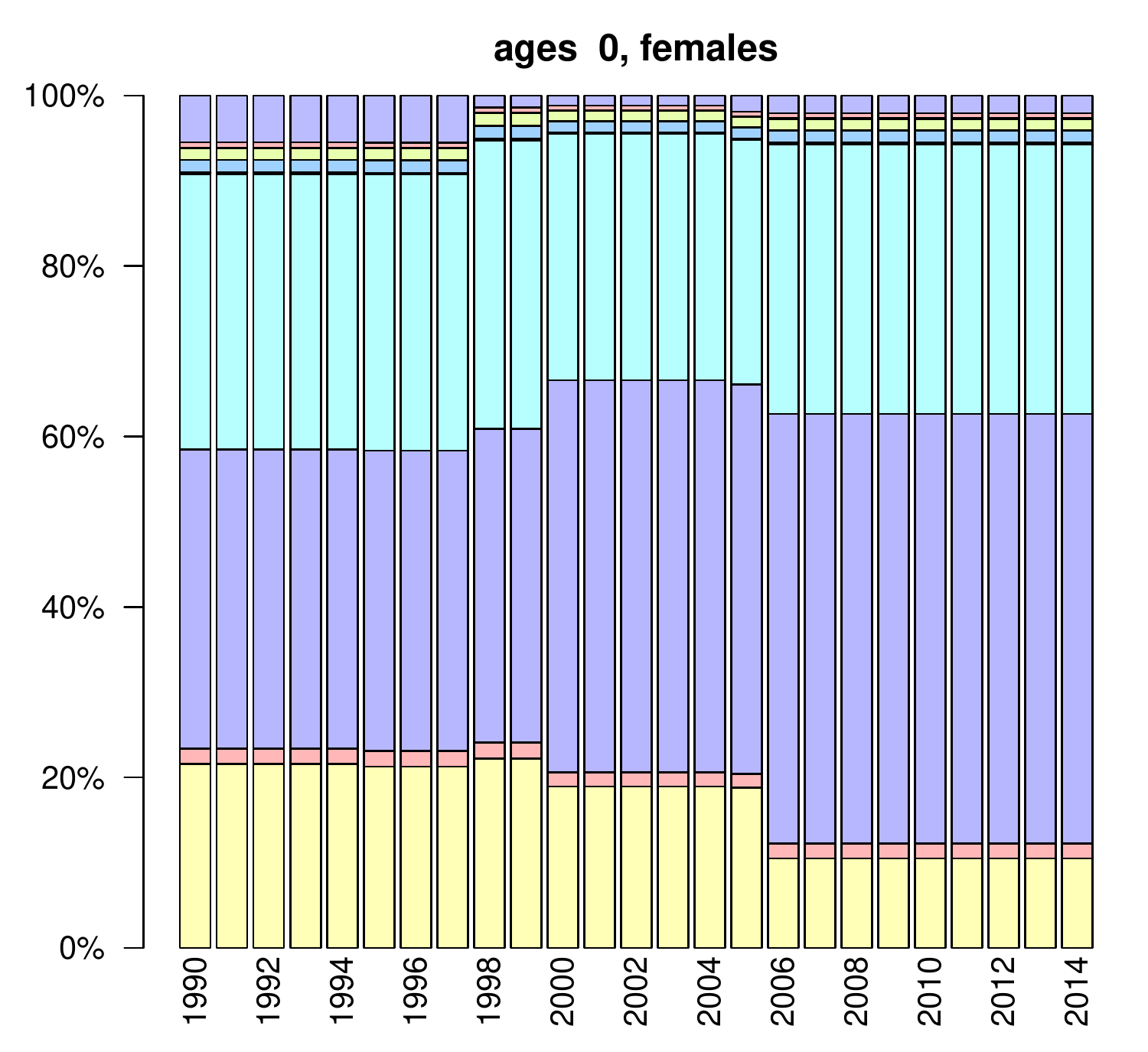}
\includegraphics[width=.425\linewidth]{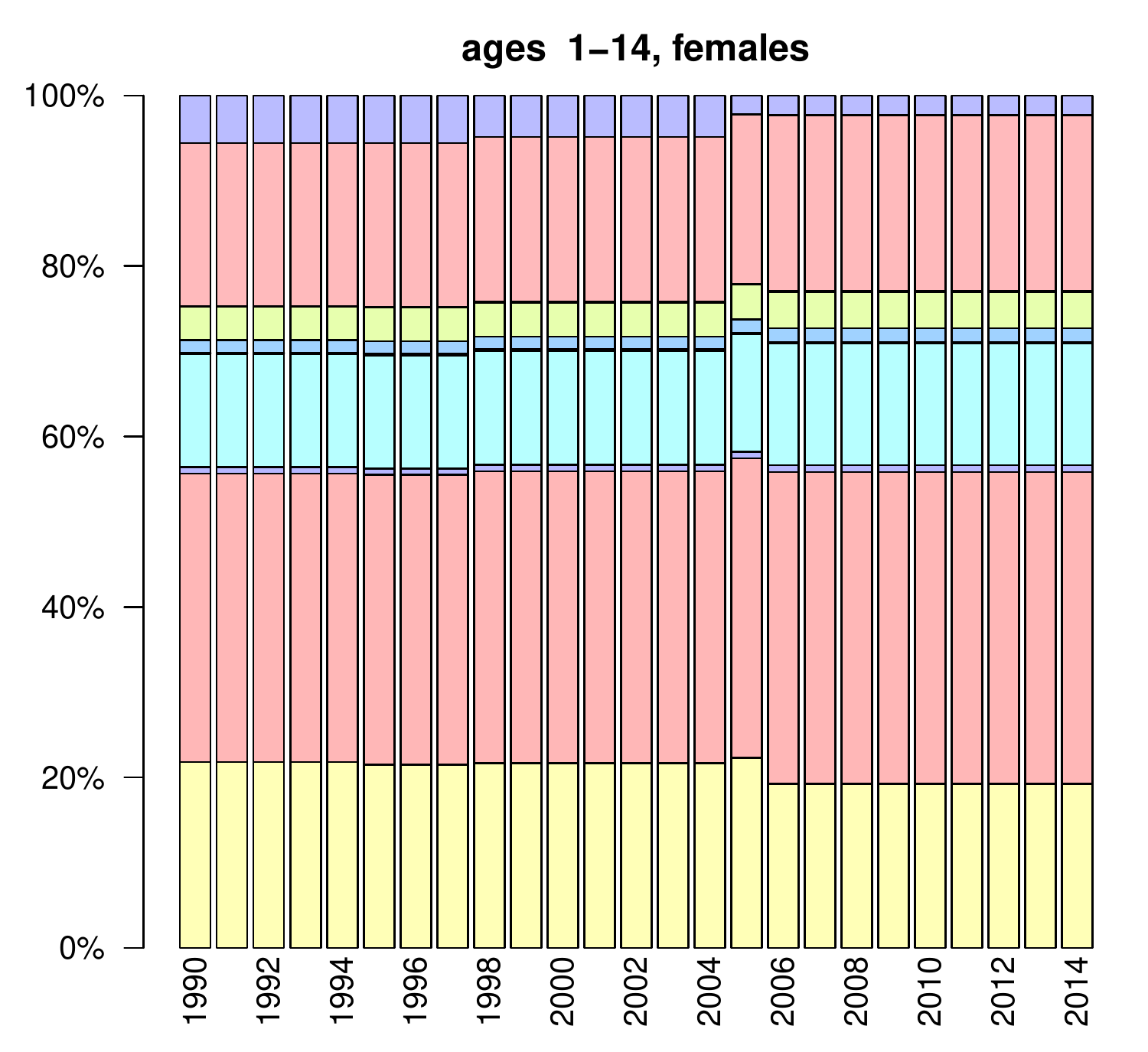}
\\
\includegraphics[width=.425\linewidth]{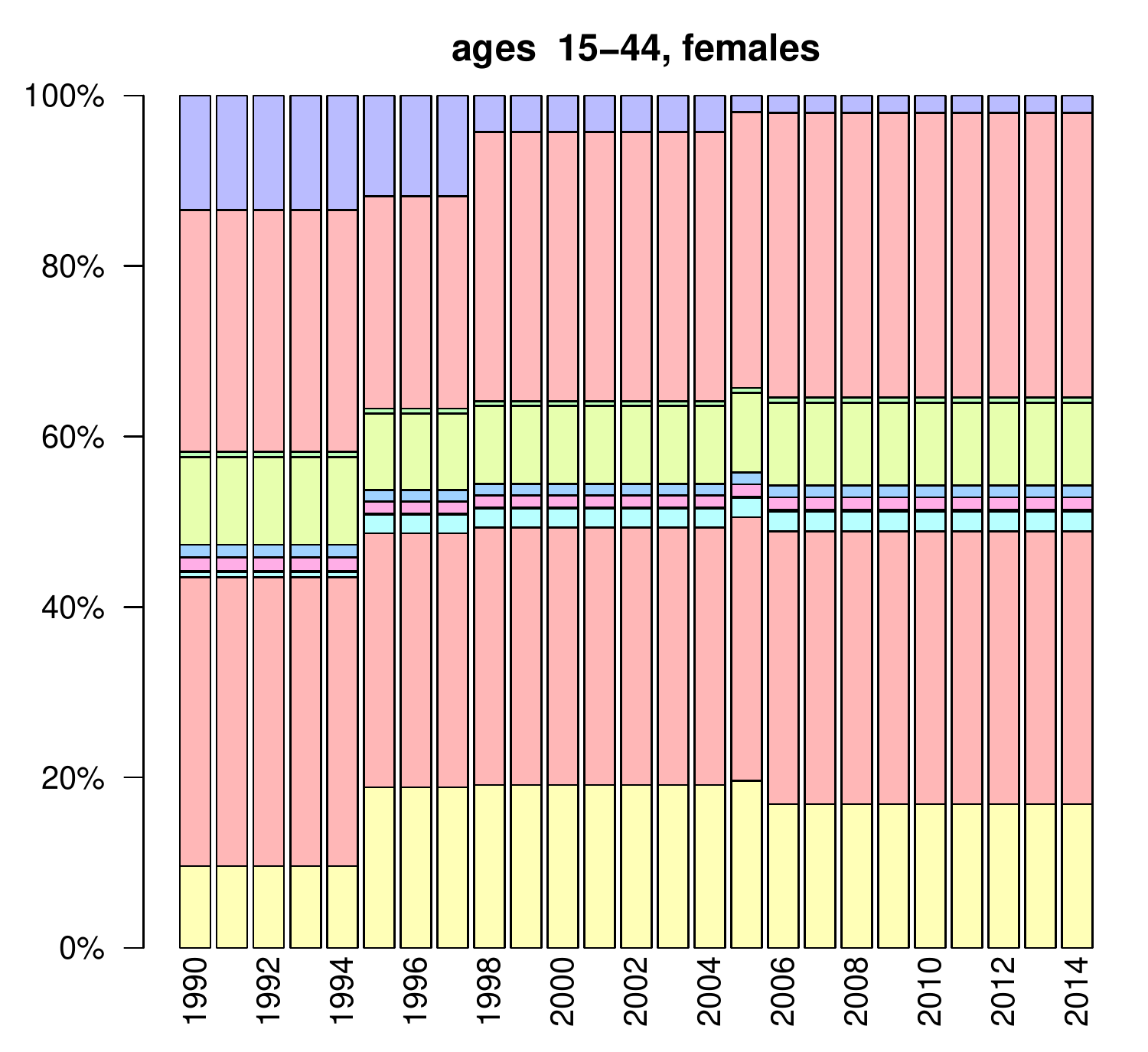}
\includegraphics[width=.425\linewidth]{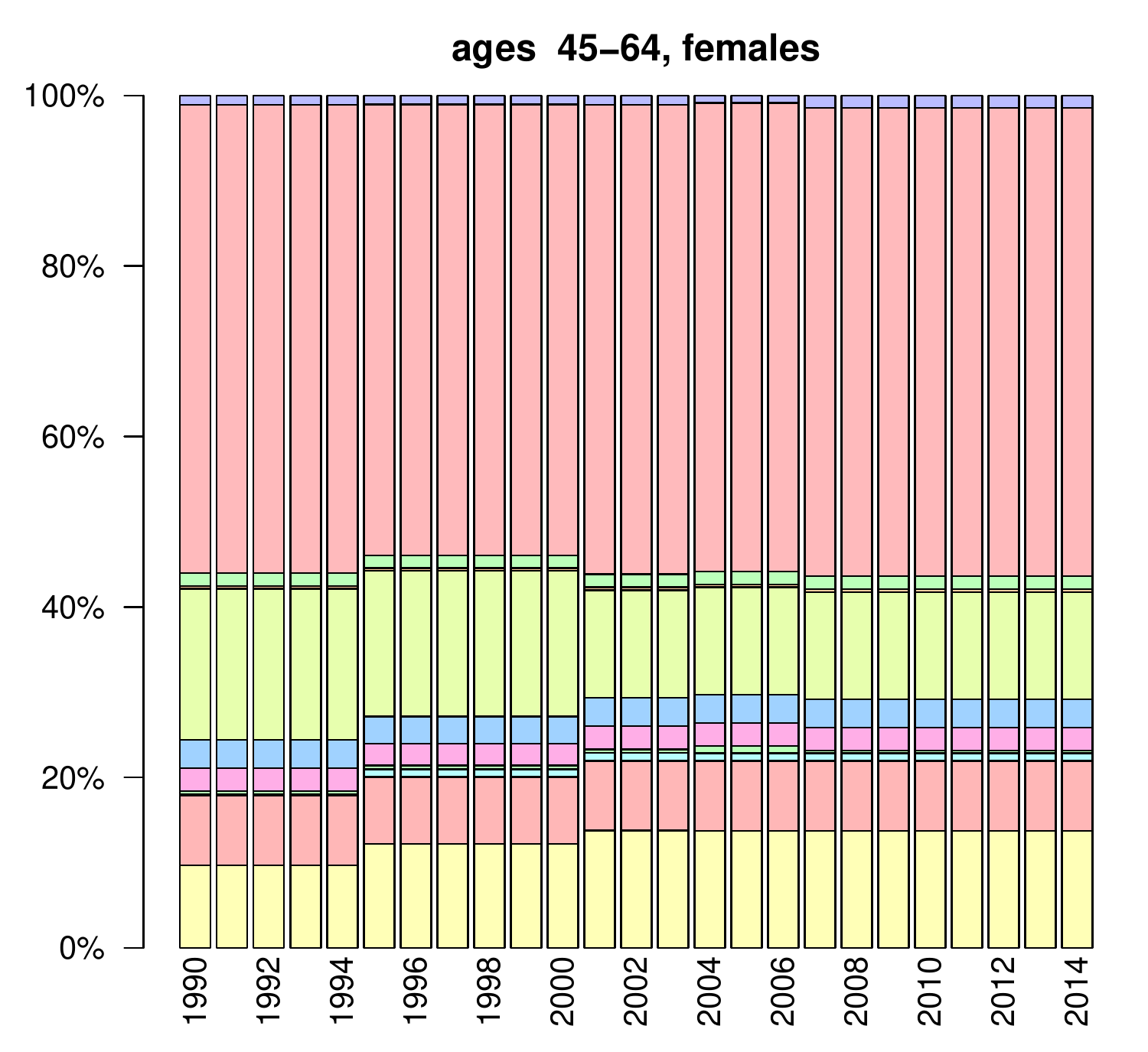}
\\
\includegraphics[width=.425\linewidth]{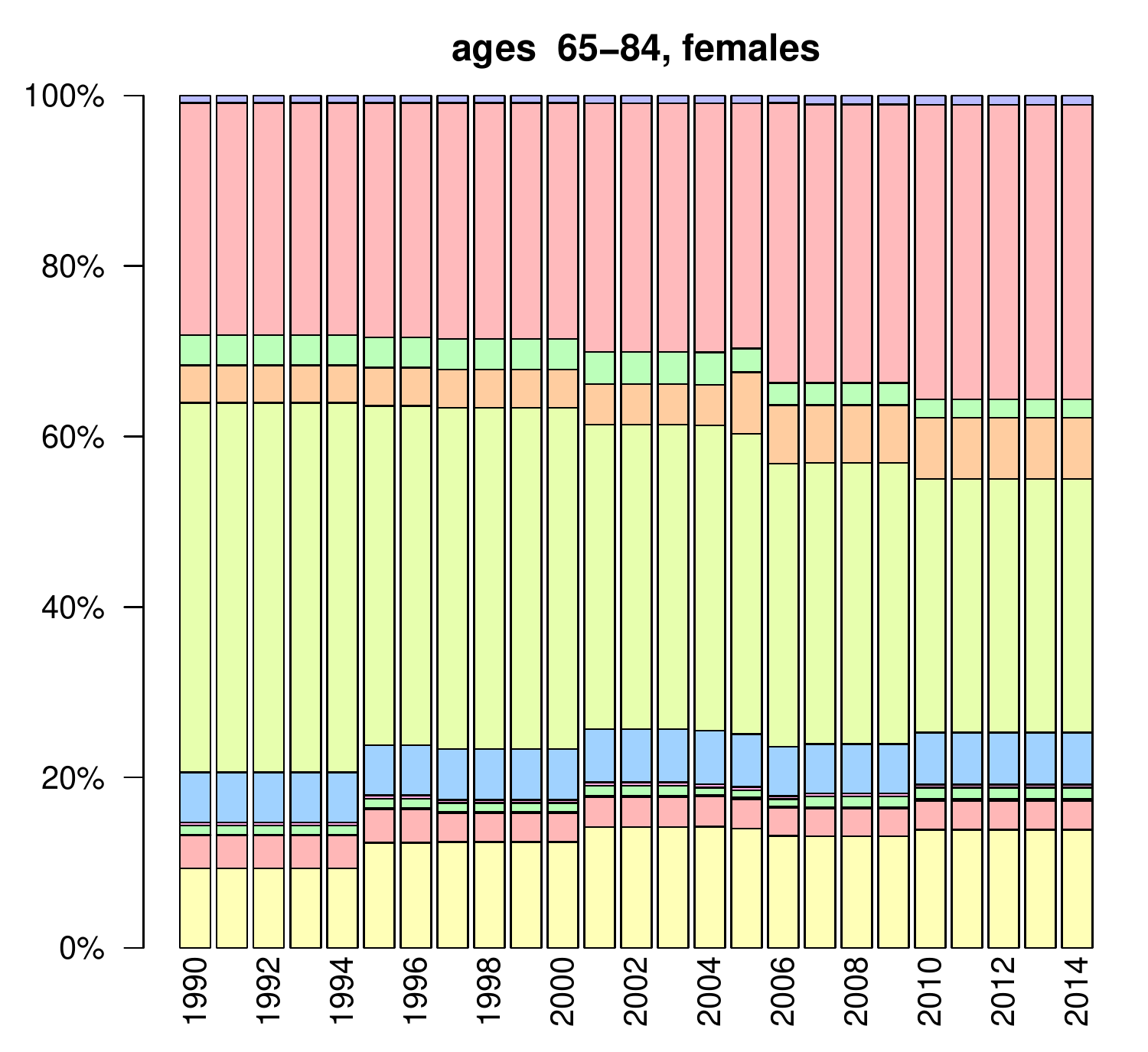}
\includegraphics[width=.425\linewidth]{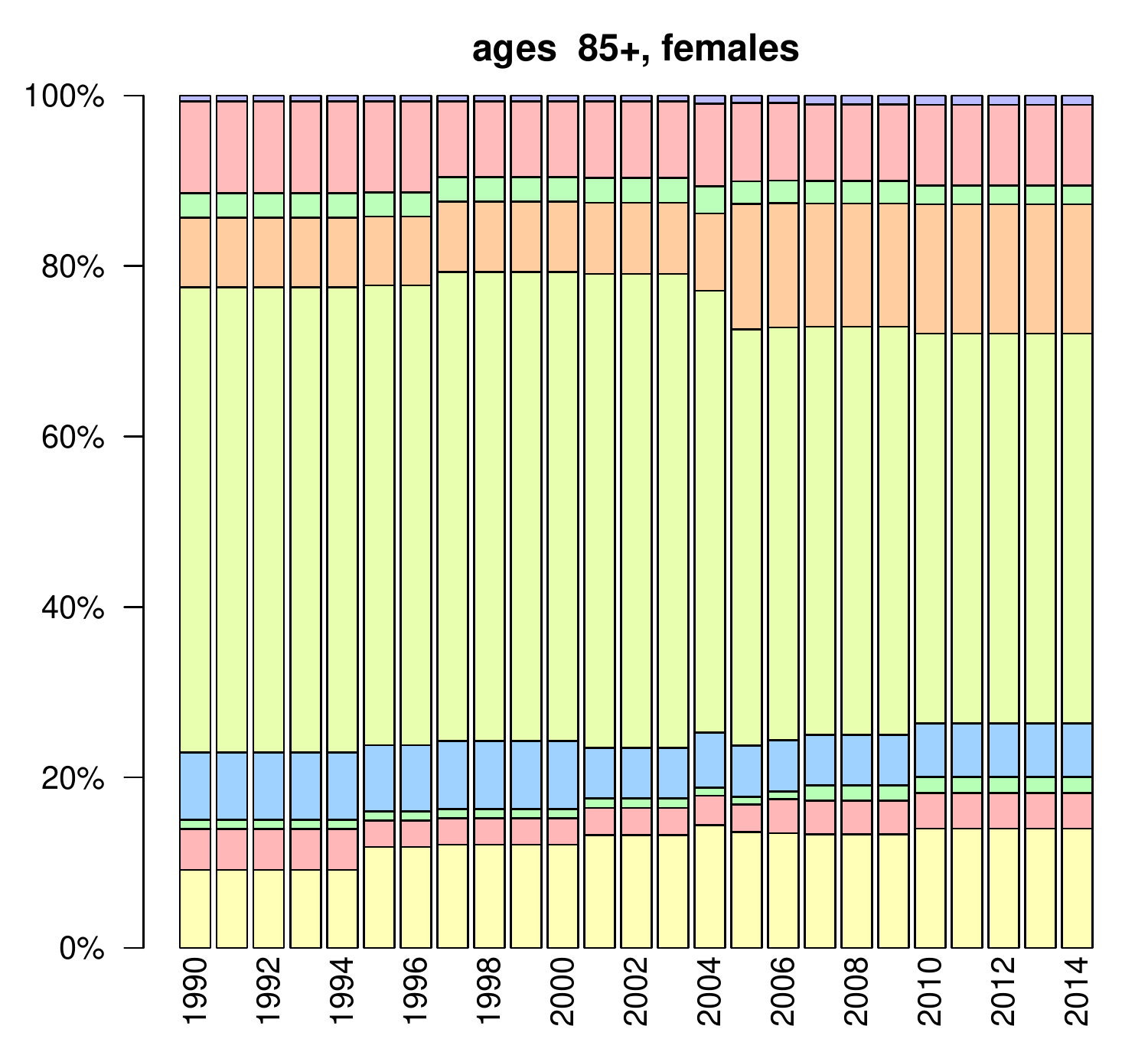}
\\
\includegraphics[width=0.75\linewidth]{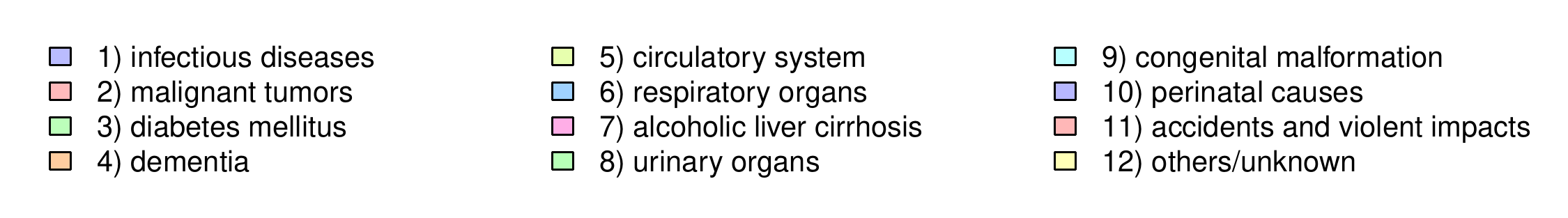}
\caption{\footnotesize
Regression tree estimated probabilities $\theta^\text{tree}(k| \x)$
for females and for the $12$ different causes of death considered.
}
\label{Appendix, Figure: death causes time, females}
\end{figure}
\begin{figure}
\centering
\includegraphics[width=.425\linewidth]{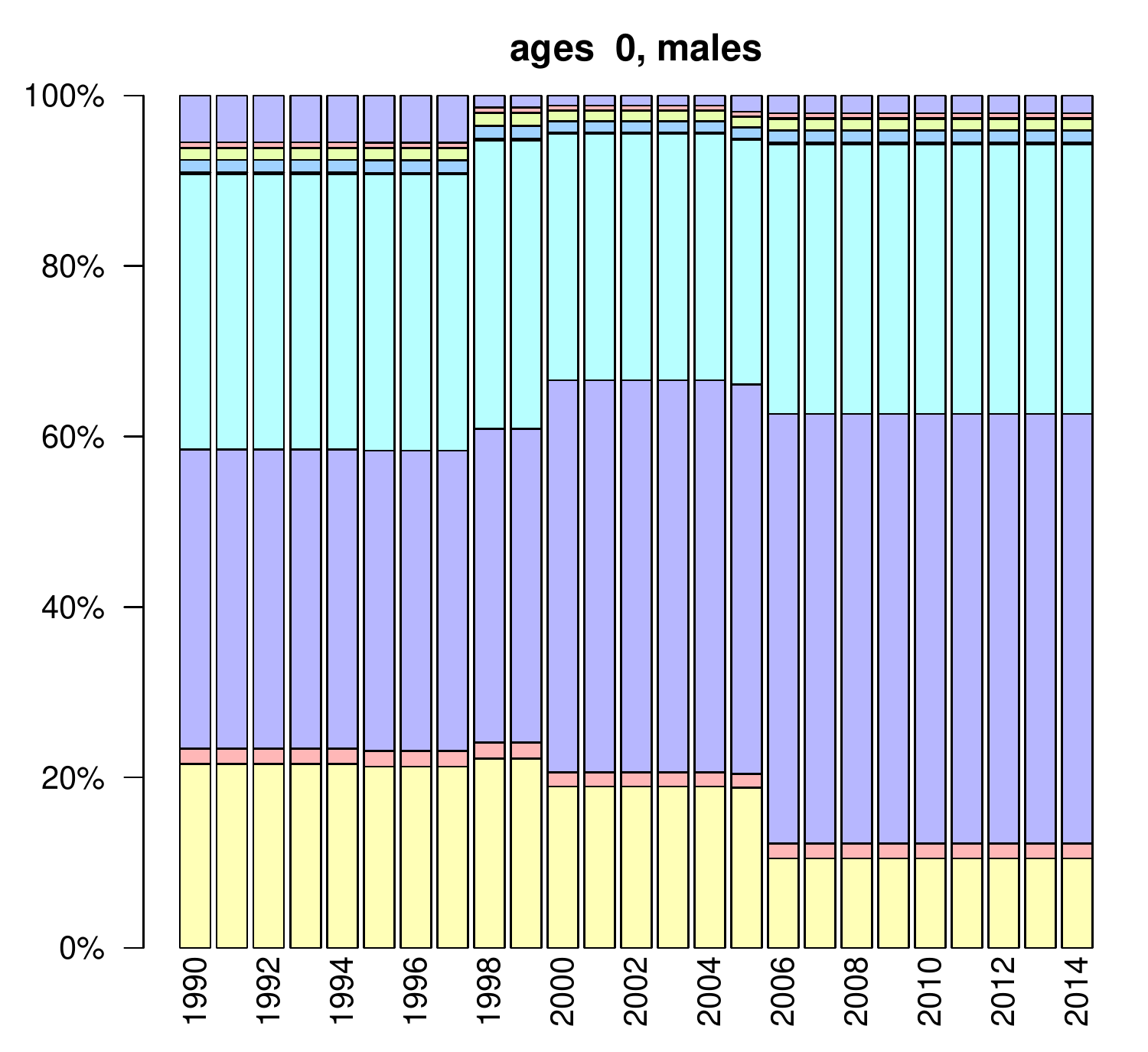}
\includegraphics[width=.425\linewidth]{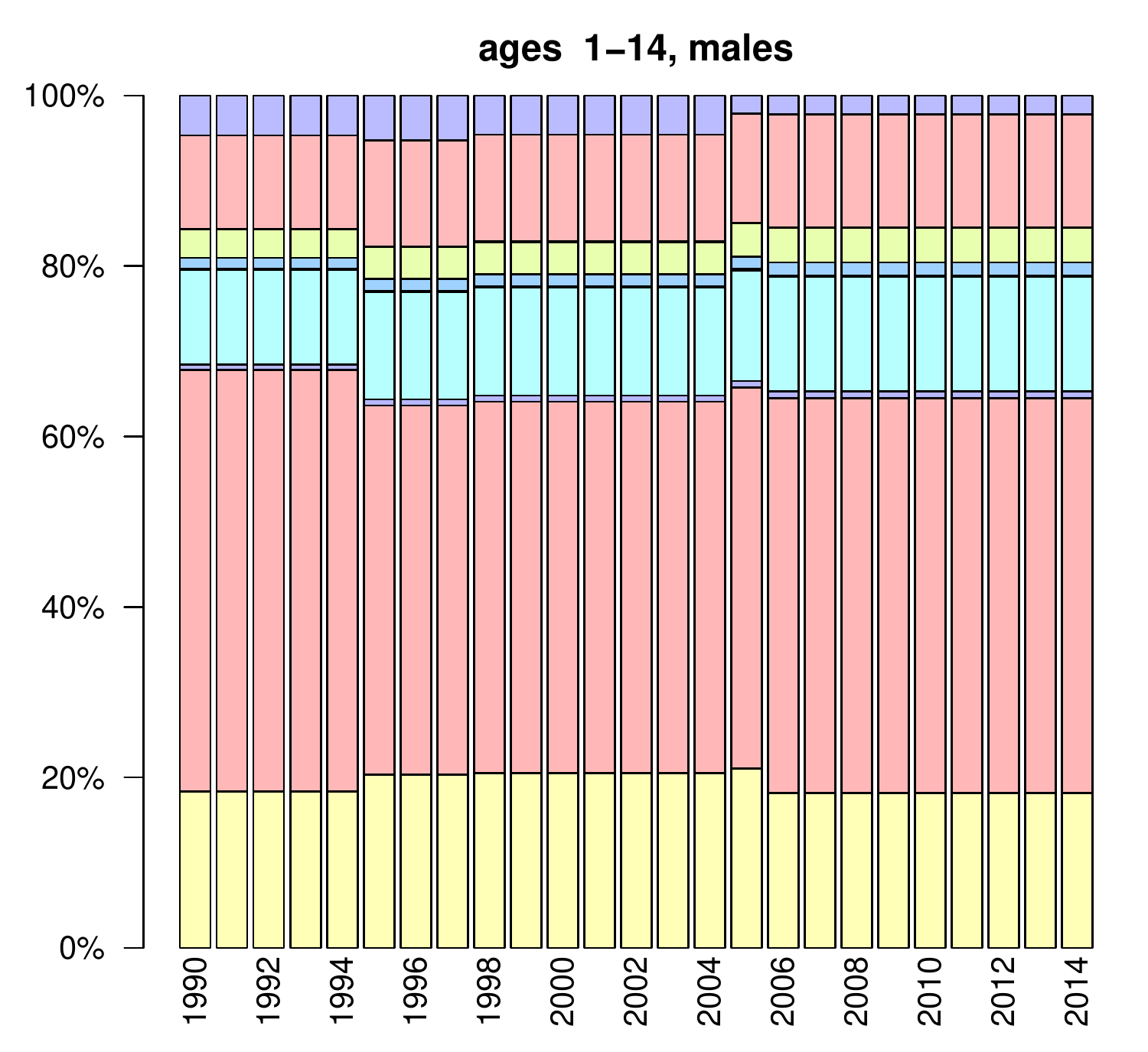}
\\
\includegraphics[width=.425\linewidth]{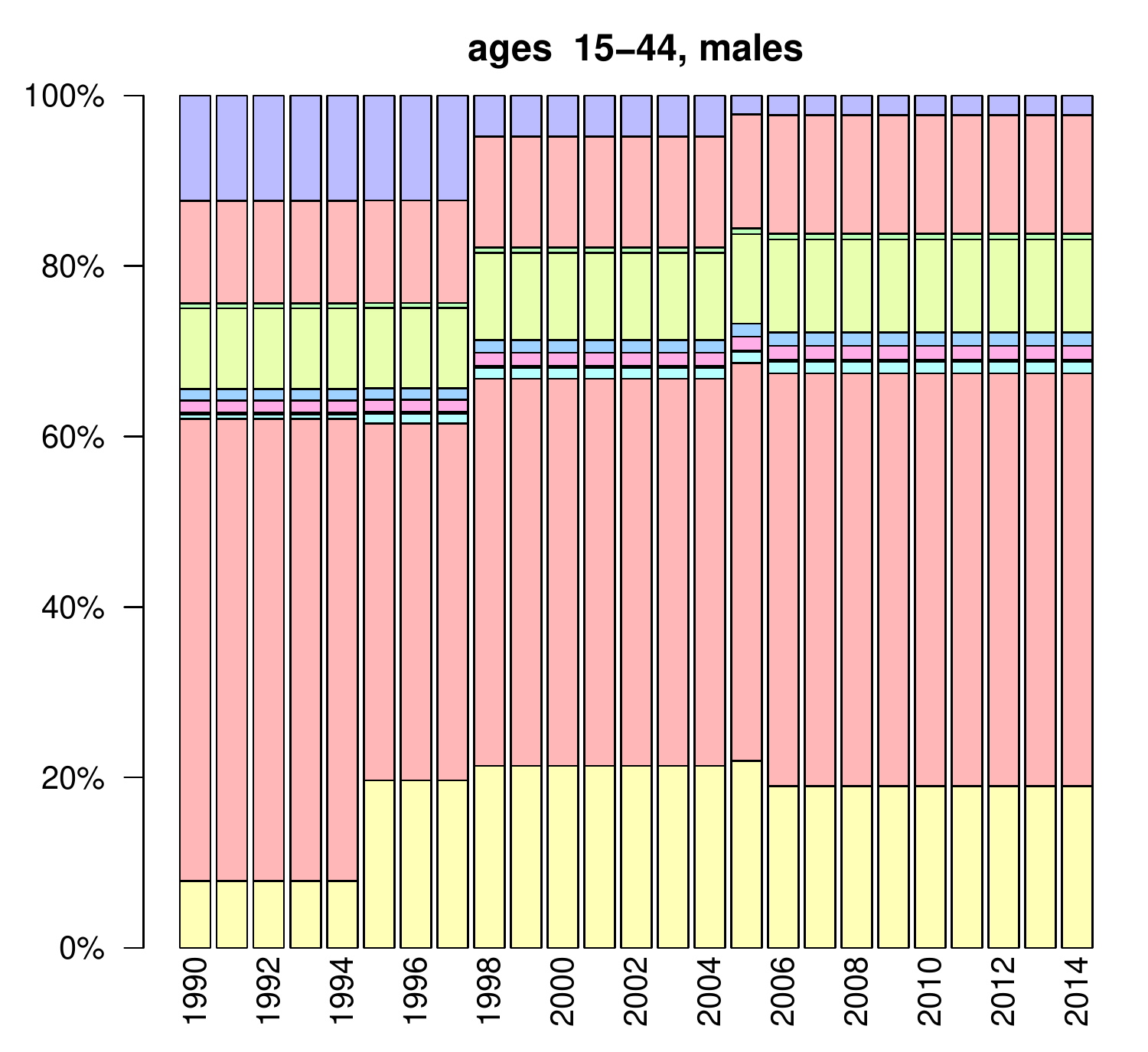}
\includegraphics[width=.425\linewidth]{./Figures/T4_M.pdf}
\\
\includegraphics[width=.425\linewidth]{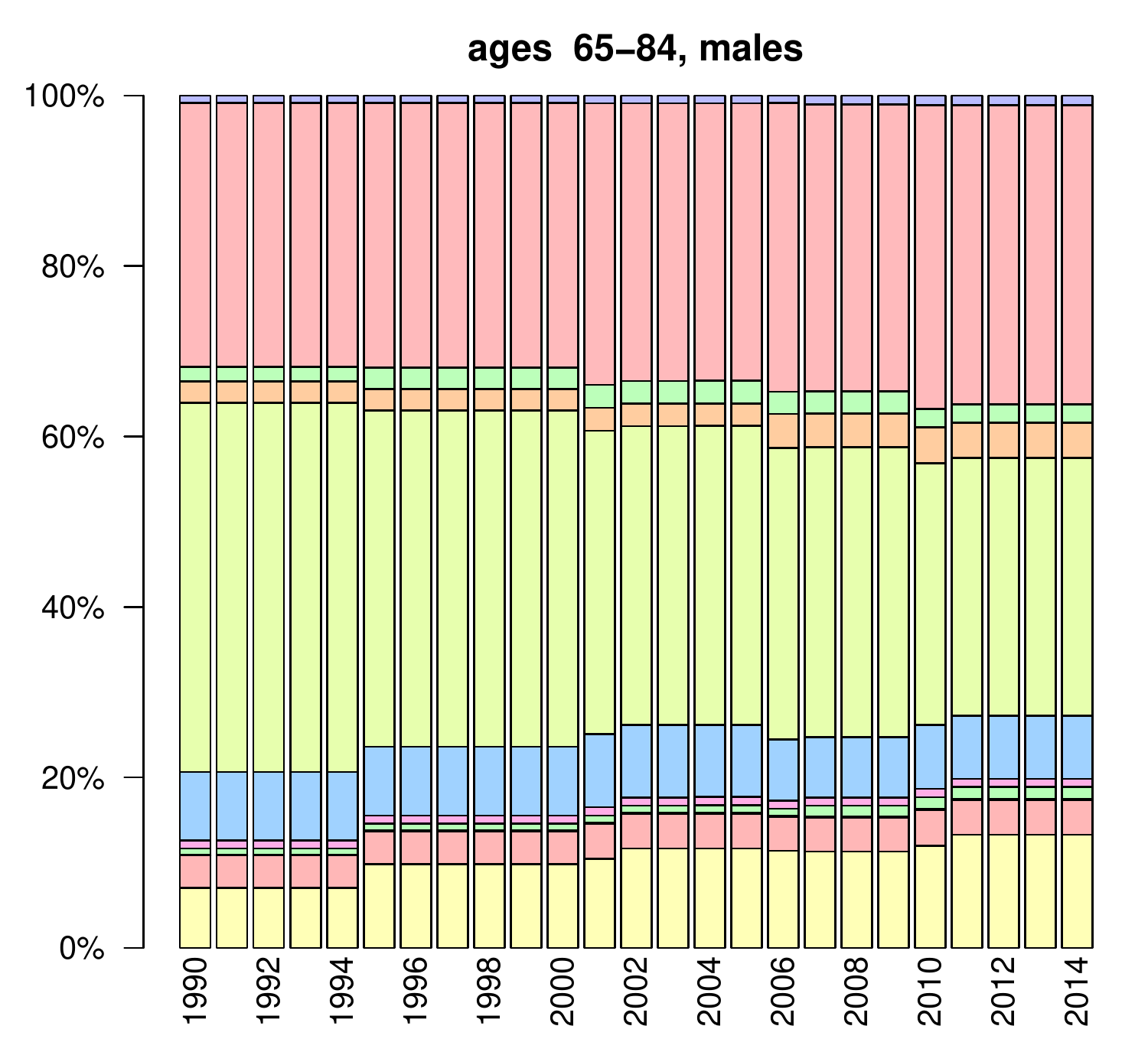}
\includegraphics[width=.425\linewidth]{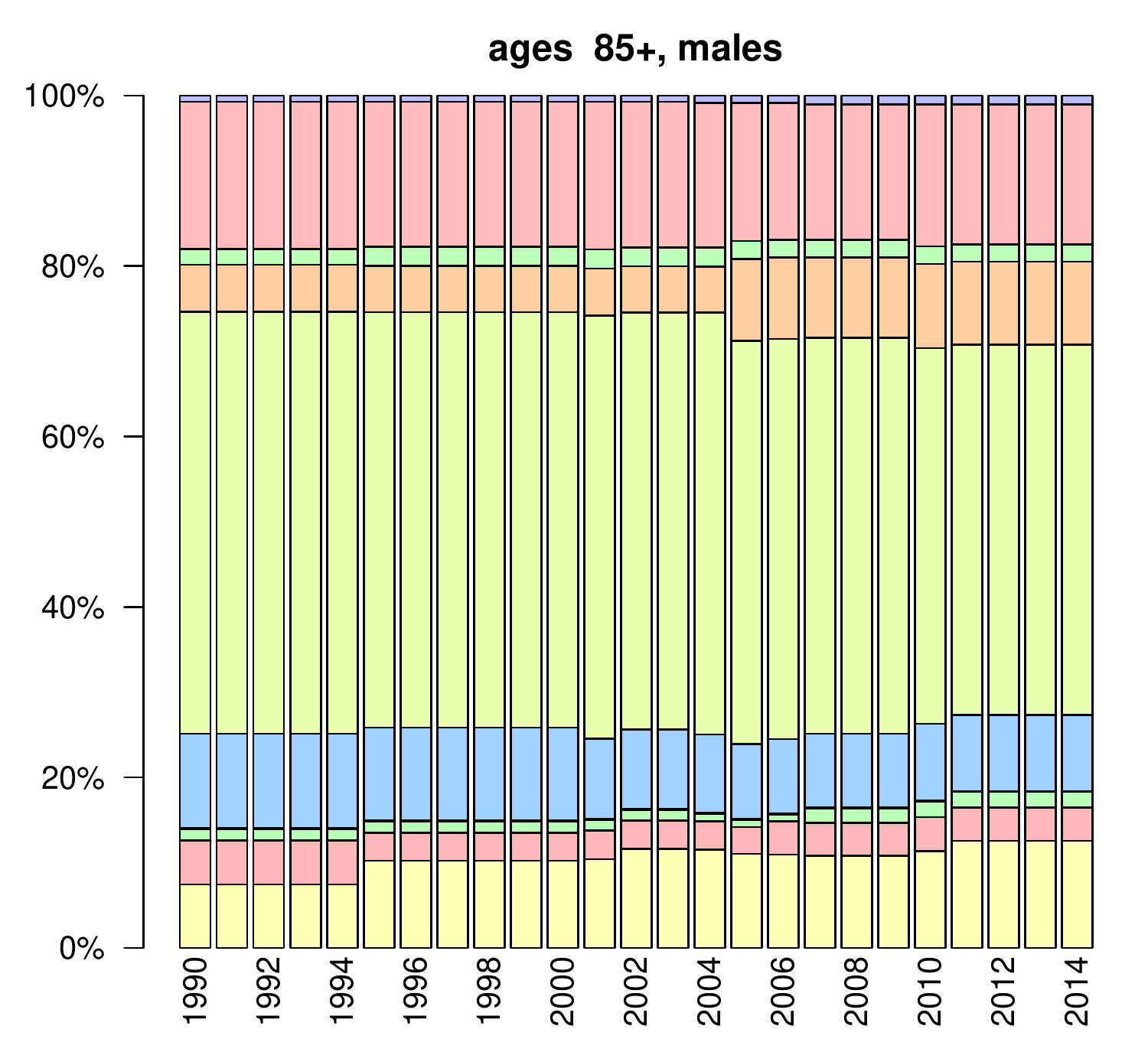}
\\
\includegraphics[width=0.75\linewidth]{./Figures/T_legend_horizontal.pdf}
\caption{\footnotesize
Regression tree estimated probabilities $\theta^\text{tree}(k| \x)$
for males and for the $12$ different causes of death considered.
}
\label{Appendix, Figure: death causes time, males}
\end{figure}

\end{document}